\pdfoutput=1
\documentclass[12pt]{article}
\usepackage[
top = 2.5cm, 
bottom = 2.5cm,  
left = 2.55cm,  
right = 2.55cm]{geometry}

\usepackage{subfigure}
\usepackage{caption}
\usepackage{latexsym}  
\usepackage{verbatim} 
\usepackage{float}
\usepackage{tikz}
\usetikzlibrary{matrix}
\usepackage{color}
\usepackage{graphicx,amssymb,amsfonts,amsmath,amssymb,amscd,amstext, mathrsfs}
\usepackage{graphicx} 
\usepackage{dsfont} 
\usepackage{xcolor}
\usepackage{amsmath}
\usepackage{empheq}
\usepackage{cite}

\usetikzlibrary{shapes.misc}

\usepackage[section]{placeins}

\numberwithin{equation}{section}

\newlength\dlf

\def\CT{\mathcal{T}}
\def\CS{\mathcal{S}}

\newcommand{\bw}{\begin{widetext}}
\newcommand{\ew}{\end{widetext}}
\newcommand{\bea}{\begin{eqnarray}}
\newcommand{\eea}{\end{eqnarray}}
\newcommand{\be}{\begin{equation}}
\newcommand{\ee}{\end{equation}}
\newcommand{\nn}{\nonumber}

\renewcommand{\bar}[1]{\overline{#1}}
\renewcommand{\tilde}[1]{\widetilde{#1}}
\renewcommand{\hat}[1]{\widehat{#1}}

\renewcommand{\cal}{\mathcal}
\newcommand{\CO}{\mathcal{O}}

\newcommand{\CN}{\mathcal{N}}

\newcommand{\CL}{\mathcal{L}}

\DeclareFontShape{OT1}{cmr}{mx}{n}{<->cmr10}{}
\newcommand{\titlefont}{\fontseries{mx}\selectfont}

\def\frac#1#2{{#1\over #2}}

\newcommand{\myRho}{{\mathfrak{r}}}

\usepackage{jheppub}

\usetikzlibrary{decorations.markings,positioning,decorations.pathmorphing}

\begin{document}

\begin{titlepage}

\begin{flushright} 
\end{flushright}

\begin{center} 

\vspace{0.35cm}

{\fontsize{21.5pt}{0pt}{\titlefont 
Nonperturbative Bounds on Scattering of\\
\vspace{0.3cm}
 Massive Scalar Particles in $d\ge 2$
}}

\vspace{1.6cm}  

{{Hongbin Chen$^1$, A. Liam Fitzpatrick$^1$,  Denis Karateev$^{2}$}}

\vspace{1cm} 

{{\it
$^1$Department of Physics, Boston University, 
Boston, MA  02215, USA
\\
\vspace{0.1cm}
$^2$D\'epartment de Physique Th\'eorique, Universit\'e de Gen\`eve,\\
24 quai Ernest-Ansermet, 1211 Gen\`eve 4, Switzerland
}}\\
\end{center}
\vspace{1.5cm}

{\noindent 
	We study two-to-two scattering amplitudes of a scalar particle of mass $m$. For simplicity, we assume the presence of $\mathbb{Z}_2$ symmetry and that the particle is $\mathbb{Z}_2$ odd. We consider two classes of amplitudes: the fully nonperturbative ones and effective field theory (EFT) ones with a cut-off scale $M$. Using the primal numerical method which allows us to impose full non-linear unitarity, we construct novel bounds on various observables in $2\leq d \leq 4 $ space-time dimensions for both classes of amplitudes. We show that our bounds are much stronger than the ones obtained by using linearized unitarity or positivity only. We discuss applications of our bounds to constraining EFTs. Finally, we compare our bounds to the amplitude in $\phi^4$ theory computed perturbatively at weak coupling, and find that they saturate the bounds. 
}

\end{titlepage}

\tableofcontents


\section{Introduction and summary of results}
\label{sec:intro}
 
The modern goal of the S-matrix bootstrap is to bound the space of allowed two-to-two scattering amplitudes non-perturbatively. 
In this paper, we will focus on the two-to-two scattering amplitude of two identical particles of mass $m$ with no bound states. We denote the scattering amplitude by $\CS(p_1, p_2, p_3, p_4)$ and its interacting part by $\CT(s,t,u)$, where $p_i$ are momenta of particles participating in the scattering and $s,t,u$ are the standard Mandelstam variables, satisfying  $s+t+u=4m^2$.   Our goal is to  find the space of all such scattering amplitudes allowed by the three constraints of crossing symmetry, analyticity, and partial amplitude unitarity: 
\begin{enumerate}
\item Crossing: $\CT(s,t,u)$ is invariant under permutations of $s,t,u$
\item Analyticity: $\CT(s,t,u)$ is an analytic function of $s,t,u$, except for physical branch cuts and poles.
\item Unitarity: the partial amplitudes $\CS_j(s)$ satisfy $|\CS_j(s)|^2 \leq1$ for $s> 4m^2 $ and all spins $j=0,2,4,\ldots$ 
\end{enumerate}
The partial amplitudes $\CS_j(s)$ are defined as integral transformations of $\mathcal{T}(s,t,u)$: 
\begin{equation}
	\label{eq:PA_general_d}
	\CS_j(s) \equiv 1 + \frac{i \CT_j(s)}{\CN_d(s)}, \qquad
	\CT_j(s) \equiv \int_{-1}^{+1} d \cos \theta \  \mu_{d,j}(\cos \theta)\CT(s,t(\cos \theta), u(\cos \theta)),
\end{equation} 
where $\mu_{d,j}$ and  $\CN_d$  are given in  (\ref{eq:measure}) and  (\ref{eq:PhaseSpaceFactor}). 
Further details of our conventions for scattering amplitudes, partial amplitudes, and unitarity are given in section \ref{sec:partial_amplitudes}.

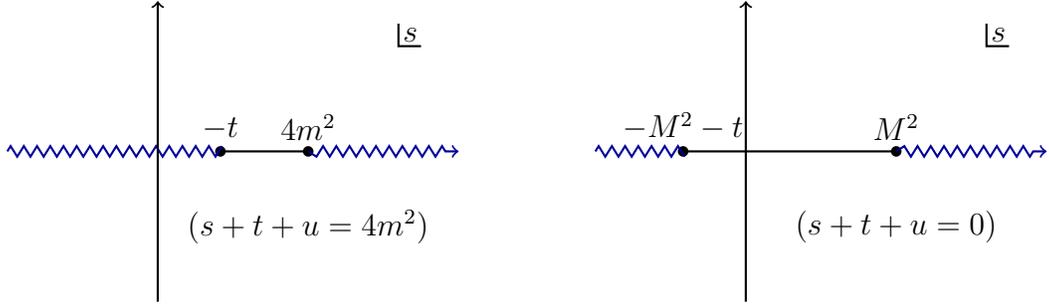
\begin{figure}[th!]
	\begin{center}

		\begin{tikzpicture}[thick]
			
			\node (label) at (2,-1) {$(s+t+u=4m^2)$};
			
			\draw (3.2,1.7) -- (3.2,1.4) -- (3.5,1.4) ;
			\node (label) at (3.36,1.55) {$s$};
			
			\fill(5/6, 0) circle (2pt) node[above] (l_branch) {$-t$} (2, 0) circle (2pt) node[above] (r_branch) {$4m^2$};
			
			
	    	\draw	[->,decorate,decoration={zigzag,segment length=5,amplitude=2,pre=lineto,pre length=0,post=lineto,post length=3}, blue!60!black]  (-2, 0) -- (l_branch.south) (r_branch.south) -- (4, 0) node [above left, black] {};
			
			\draw(l_branch.south) -- (r_branch.south);
			\draw[->] (0, -2) -- (0, 2) node[below left=0.1] {};
			
		\end{tikzpicture}
\qquad        \qquad	
		\begin{tikzpicture}[thick]
		
		\node (label) at (2,-1) {$(s+t+u=0)$};
		
		\draw (3.2,1.7) -- (3.2,1.4) -- (3.5,1.4) ;
		\node (label) at (3.36,1.55) {$s$};
		
		\fill(-5/6, 0) circle (2pt) node[above] (l_branch) {$-M^2-t$} (2, 0) circle (2pt) node[above] (r_branch) {$M^2$};
		
		\draw	[->,decorate,decoration={zigzag,segment length=5,amplitude=2,pre=lineto,pre length=0,post=lineto,post length=3}, blue!60!black]  (-2, 0) -- (l_branch.south) (r_branch.south) -- (4, 0) node [above left, black] {};
	
		\draw(l_branch.south) -- (r_branch.south);
		\draw[->] (0, -2) -- (0, 2) node[below left=0.1] {};
		
	\end{tikzpicture}
		\caption{Analytic structure in the $s$ complex plane for a fixed value of $t$ of two classes of amplitudes considered in the literature: the left plot is for the nonperturbative amplitude and the right one is for the EFT amplitude. Here $m$ is the mass of the particle and $M$ is the EFT ``cut-off''. It is also assumed that $m\ll M$, thus we set $m=0$ in this case. No poles are present due to the assumed presence of $\mathbb{Z}_2$ symmetry.}
		\label{fig:classes of amplitudes}
	\end{center}
\end{figure}

Regarding the analytic structure (on the $s$ complex plane for a fixed value of $t$), all the amplitudes considered in the literature can be split into two classes. In this paper, we refer to them as the \textbf{nonperturbative amplitudes} and the \textbf{EFT amplitudes}. The nonperturbative amplitudes have a branch cut starting at the two-particle threshold $4m^2$ together with the one related by the $s-u$ crossing. The EFT amplitudes instead have a branch cut starting from some ``cut-off'' scale $M$ together with the one related by the $s-u$ crossing. It is also assumed that the mass of the particle $m$ is much smaller than the ``cut-off'' scale $M$, namely $m\ll M$. In the EFT amplitudes, we can set $m=0$. Both classes of amplitudes are purely real on the horizontal axis between the two cuts. Notice that they could have poles in this region, however, we assume in this paper the absence of such poles (say, by imposing a $\mathbb{Z}_2$ symmetry and that the particle is $\mathbb{Z}_2$ odd). The analytic structure of nonperturbative and EFT amplitudes is depicted in figure \ref{fig:classes of amplitudes}.
The nonperturbative amplitudes were studied in various contexts in \cite{Paulos:2016fap, Paulos:2016but, Paulos:2017fhb,Doroud:2018szp, He:2018uxa, Cordova:2018uop, Guerrieri:2018uew, Paulos:2018fym, Homrich:2019cbt, EliasMiro:2019kyf, Cordova:2019lot, Bercini:2019vme, Gabai:2019ryw,Bose:2020shm,Bose:2020cod,Correia:2020xtr,Kruczenski:2020ujw, Guerrieri:2020bto,Hebbar:2020ukp,Karateev:2019ymz,Karateev:2020axc,Guerrieri:2020kcs,Tourkine:2021fqh,Guerrieri:2021ivu,He:2021eqn,EliasMiro:2021nul,Guerrieri:2021tak,Chen:2021pgx,Cordova:2022pbl,Albert:2022oes,Sinha:2020win,Chowdhury:2021ynh,Karateev:2022jdb}.\footnote{For an overview of recent results and discussion of some future directions, see \cite{Kruczenski:2022lot}.} The EFT amplitudes were studied extensively in various contexts in  \cite{Bellazzini:2020cot,Bellazzini:2021oaj,Caron-Huot:2020cmc,Caron-Huot:2021rmr,Davighi:2021osh,deRham:2021fpu,Henriksson:2021ymi,Henriksson:2022oeu,Caron-Huot:2022ugt,Tolley:2020gtv,deRham:2021bll,deRham:2022hpx,Chiang:2022ltp,Caron-Huot:2022jli}.

Using crossing, analyticity, and unitarity, one can bound the space of allowed scattering amplitudes using various methods to be discussed shortly.  The final output is a set of allowed values for some physical observables that we get to choose.  Ideally, we would like to choose these physical observables to be both intuitive and complete.  For instance, given a specific model Lagrangian, it should be clear how to compute these observables, and moreover, the full set of them should uniquely fix the full two-to-two scattering amplitude.  We will choose a  complete set of parameters/observables describing the scattering amplitudes in section \ref{sec:NonPerturbativeObservables}.

Let us now discuss tools that allow us to impose the constraints 1 - 3. 
Imposing any of these constraints 1, 2, or 3 individually is fairly straightforward; the main challenge is to impose them all simultaneously.  Looking only at $\CT(s,t,u)$ in Mandelstam variables obscures unitarity.  Projecting it onto partial amplitudes makes unitarity transparent, but obscures analyticity and crossing. This is because the full unitarity constraint $|\CS_j(s)|^2 \leq1$ is nonlinear in $\CT_j(s)$, namely
  \begin{equation}
	\textrm{full unitarity:}\quad |\CT_j |^2 \le  (\textrm{Im} \CT_j )(2 \CN_d),
	\label{eq:FullUnitarity}
\end{equation}
thus, it is difficult to keep track of the analytic properties of the underlying object $\CT(s,t,u)$ while studying this constraint. A weaker constraint could be obtained from \eqref{eq:FullUnitarity}. Dropping the square of the real part of $\mathcal{T}_j$ on the left-hand side of \eqref{eq:FullUnitarity} does not invalidate the inequality and leads to
\begin{equation}
	\label{eq:linear_unitarity}
	\textrm{linearized unitarity:}\quad	0 \le   \textrm{Im} \CT_j  \le 2  \CN_d.
\end{equation}
We refer to \eqref{eq:linear_unitarity} as linearized unitarity. The left-hand side of the linearized unitarity constraint is the so-called positivity constraint:
\begin{equation}
\textrm{positivity:}\quad0 \le \textrm{Im} \CT_j.
\label{eq:DispersionUnitarity}
\end{equation}

The most known tool for obtaining bounds on the amplitudes uses dispersion relations and combines them with positivity \eqref{eq:DispersionUnitarity}. There are various implementation of this tool, see for example \cite{Bellazzini:2020cot,Caron-Huot:2020cmc,Tolley:2020gtv}. In modern literature, this was mostly applied in the context of EFT amplitudes. For completeness of the discussion, we review the machinery of \cite{Caron-Huot:2020cmc} in the context of nonperturbative amplitudes and derive a set of rigorous positivity bounds in section \ref{sec:analytic_d>2}. The second tool which allows to impose linearized unitarity \eqref{eq:linear_unitarity} was recently developed in \cite{Chiang:2022ltp}. It was applied in the context of EFT amplitudes and led to stronger bounds than the ones obtained using positivity only.
Finally, there is a third class of numerical methods that allows us to impose the full non-linear unitarity \eqref{eq:FullUnitarity}. The methods from this class can be divided into two groups, the ``primal" ones \cite{Paulos:2017fhb,Homrich:2019cbt} and the ``dual'' ones \cite{Lopez:1976zs,He:2021eqn,EliasMiro:2021nul,Guerrieri:2021tak}.

In this paper, we will use the primal numerical method of \cite{Paulos:2017fhb,Homrich:2019cbt} to construct novel bounds on both the nonperturbative and EFT amplitudes. This is done in section \ref{sec:numeric_bounds}. The primal numerical method will be explained in section \ref{sec:PNA}. We will present our bounds on nonperturbative amplitudes in various number of space-time dimensions $2<d\leq 4$ (including fractional dimension) in section \ref{sec:NPA}. The results are given in figures \ref{fig:PA0_dPA0_Extrapolated} and \ref{fig:crossing_symmetric} -- \ref{fig:FL4}.  The numerical bounds on EFT amplitudes in $d=4$ space-time dimensions are given in section \ref{sec:EFTA}, see figures \ref{fig:EFT_0} and \ref{fig:EFT_2}. The method of \cite{Paulos:2017fhb,Homrich:2019cbt} can be easily modified to impose only linearized unitarity (or even only positivity). We will demonstrate on two examples (one for nonperturbative amplitudes and one for EFT amplitudes) that the bounds obtained using full non-linear unitarity are much stronger than the ones of linearized unitarity (or positivity).

We make publicly available all the numerical data obtained in section \ref{sec:numeric_bounds}. It allows to re-construct all the bounds and also to extract scattering and partial amplitudes on the boundary of allowed regions. The data can be downloaded from: \href{https://zenodo.org/record/6891946#.Ytwnmi8Roe0}{https://zenodo.org/record/6891946\#.Ytwnmi8Roe0}.

One could ask when the class of EFT amplitudes (recall the right part of figure \ref{fig:classes of amplitudes}) is relevant for physics. For instance, scattering amplitudes of truly massless particles such as pions, photons, or gravitons contain branch cuts all the way to the origin due to the presence of $\log(-s)$ terms.  Thus, the class of EFT amplitudes defined above does not seem to be relevant for describing this situation. One can argue however that the log terms are negligible at least in weakly coupled theories such as the general relativity describing gravitons. This is also true in theories where $m\ll M$ is a controlled approximation. For the most part, we will not discuss further the relevance of the EFT amplitudes for physics  and simply view them for now as some mathematical objects which must obey a set of constraints 1 - 3.

In section \ref{sec:EFTs}, we will show how the bounds obtained in section \ref{sec:numeric_bounds} can be used to bound effective field theory on a particular example of (pseudo-)Goldstone bosons. Finally in section \ref{sec:phi4}, we compare the perturbative two-particle amplitude of $\phi^4$ theory to our numerical bounds.

As a warm-up in appendix \ref{sec:2dwarmup}, aimed at the reader without much previous background in the numeric S-matrix bootstrap, we discuss all the above aspects in $d=2$ where there are significant technical simplifications but no fundamental difference in how the method works compared to $d>2$. Here, one can develop a much better intuition for how the three constraints of crossing, analyticity, and unitarity combine together.  Throughout this $d=2$ section, we refer to and compare to the analogous $d>2$  results obtained in the main text.  As a bonus, we will explain how to obtain numerical bounds of \cite{Paulos:2017fhb,Chen:2021pgx} purely analytically, following a method from \cite{EliasMiro:2019kyf}. Particularly, it is interesting to see that by only considering the S-matrix at the crossing symmetric point and its second derivative, one can obtain the full  S-matrix of the Sinh-Gordon model simply from unitarity and analyticity. 

In appendix \ref{app:limit_d2}, we discuss the $d\rightarrow 2$ limit of the general $d$ formulas given in section \ref{sec:partial_amplitudes}. In appendix \ref{app:polynomials}, we provide the connection between our observables defined in section \ref{sec:NonPerturbativeObservables} and the ones of \cite{Caron-Huot:2020cmc}.

\section{Scattering and partial amplitudes}
\label{sec:partial_amplitudes}
The goal of this section is to define unitarity constraints for scattering amplitudes of scalar particles in general number of dimensions $d>2$. 
The discussion below is an elaborated version of section 2.2 in \cite{Karateev:2019ymz}.

We denote by $\mathcal{S}(p_1,p_2,p_3,p_4)$ the scattering amplitude and by $\mathcal{T}(s,t,u)$ its non-trivial (interacting) part. The Mandelstam variables $s,t$ and $u$ obey the standard relation
\begin{equation}
	\label{eq:constraint}
	s+t+u=4m^2.
\end{equation}
The partial amplitude is denoted by $\mathcal{S}_j(s)$, where $j=0,2,4,\ldots$ is the total spin. It is related to the interacting part of the scattering amplitude as
\begin{equation}
\label{eq:Sj}
\mathcal{S}_j(s) = 1 + \frac{i}{\mathcal{N}_d(s)}\times \mathcal{T}_j(s),
\end{equation}
where $\mathcal{T}_j(s)$ is defined by
\begin{equation}
\label{eq:Tj}
\mathcal{T}_j(s) \equiv \mathbf{\Pi}_j\left[ \mathcal{T}\left(s,t(x),u(x)\right)\right].
\end{equation}
Here $\mathbf{\Pi}_j$ is the integral transform (projector) defined as
\begin{equation}
\label{eq:projector}
\mathbf{\Pi}_j\left[ f(x)\right] \equiv 
\int_{-1}^{+1}dx \,\mu_{d,\,j}(x)\, f(x)
\end{equation}
with the measure  $\mu_{d,\,j}(x)$ defined as
\begin{equation}
\label{eq:measure}
\mu_{d,\,j}(x) =  \frac{j!\,\Gamma\left(\frac{d-3}{2}\right)}{4\,\pi^{(d-1)/2}\Gamma(d-3+j)}
\times
(1-x^2)^{\frac{d-4}{2}} C_j^{(d-3)/2}(x).
\end{equation}
The function $\mathcal{N}_d(s)$ in \eqref{eq:Sj} is given by
\begin{equation}
\label{eq:PhaseSpaceFactor}
\mathcal{N}_d(s)\equiv 2^{d-1} \sqrt{s}\;\left(s-4m^2\right)^{(3-d)/2}.
\end{equation}
The object $C_j^{(d-3)/2}(x)$ in the measure \eqref{eq:measure} is the Gegenbauer polynomial and $d$ denotes the number of space-time dimensions. The variable $x$ is defined as $x\equiv\cos\theta$, where $\theta$ is the scattering angle. The Mandelstam variables $t$ and $u$ are related to $x$ as
\begin{equation}
\label{eq:mandelstam_variables_angle}
t = - \frac{s-4m^2}{2}\,(1-x),\quad
u = - \frac{s-4m^2}{2}\,(1+x).
\end{equation}
The relation \eqref{eq:Tj} can be inverted as follows
\begin{equation}
\label{eq:inverted_Tj}
\mathcal{T}(s,x) = a_{d}\times \sum_{j=0,2,4,\dots}(2j+d-3)
C_j^{(d-3)/2}(x)
\mathcal{T}_j(s),
\end{equation}
where the coefficient $a_{d}$ is defined as
\begin{equation}
	\label{eq:coefficients_ad}
a_{d}\equiv
(4\pi)^{(d-3)/2}\Gamma\left(\frac{d-3}{2}\right).
\end{equation}
These definitions are identical to the ones of section 3 in \cite{Paulos:2017fhb} and section 2.3 of \cite{Correia:2020xtr}.

Unitarity imposes the following simple constraint on the partial amplitudes
\begin{equation}
\label{eq:unitarity_1}
\left| \mathcal{S}_j(s) \right | \leq 1
\end{equation}
for the physical range of the Mandelstam variable $s$, namely $s\geq 4m^2$ and all the spins, namely $j=0,2,4,\cdots.$
The unitarity constraint \eqref{eq:unitarity_1} can be written in the semipositive-definite form as
\begin{equation}
\label{eq:unitarity_2}
\begin{pmatrix}
1 & 1\\
1 & 1
\end{pmatrix}+
\mathcal{N}_d(s)^{-1}\times
\begin{pmatrix}
0 & -i\mathcal{T}^*_j(s)\\
i\mathcal{T}_j(s) & 0
\end{pmatrix}\succeq 0.
\end{equation}
An equivalent form can be written as
\begin{equation}
	\label{eq:unitarity_3}
	\begin{pmatrix}
		1 & 0\\
		0 & 0
	\end{pmatrix}+
	\begin{pmatrix}
		-\frac{1}{2} \mathcal{N}_d^{-1} \text{Im} \mathcal{T}_j &  \mathcal{N}_d^{-1/2}\text{Re} \mathcal{T}_j\\
		\mathcal{N}_d^{-1/2}\text{Re} \mathcal{T}_j &  2\text{Im} \mathcal{T}_j
	\end{pmatrix}\succeq 0.
\end{equation}
The equivalence between \eqref{eq:unitarity_1}, \eqref{eq:unitarity_2} and \eqref{eq:unitarity_3} can be seen by simply taking into account \eqref{eq:Sj} and evaluating the determinant of \eqref{eq:unitarity_2} and \eqref{eq:unitarity_3}.

\section{Observables}\label{sec:NonPerturbativeObservables}

In this section, we will define several equivalent sets of parameters (called observables) which describe the interacting part of the scattering amplitude $\mathcal{T}(s,t)$. In general, each set has an infinite number of entries.\footnote{Notice that if there is a Lagrangian description of the theory, these entries can be computed in terms of a finite number of ``couplings'' which describe this Lagrangian.}  We start by discussing observables of nonperturbative amplitudes in section \ref{sec:NP_amplitudes_observables}.  In section \ref{sec:EFT_amplitudes_observables}, we will discuss observable of EFT amplitudes. Recall, that the definition of nonperturbative and EFT amplitudes was given in the beginning of section \ref{sec:intro}.

\subsection{Nonperturbative amplitudes} 
\label{sec:NP_amplitudes_observables}
One way to describe the amplitude is to evaluate it and its derivatives at some particular values $s_0$, $t_0$, and $u_0$, namely
\begin{equation}
	\label{eq:one_way}
	\partial_s^k\partial_t^l \mathcal{T}(s_0,t_0,u_0).
\end{equation}
The values $s_0$, $t_0$, and $u_0$ are convenient to choose in the Mandelstam region defined as
\begin{equation}
	\label{eq:range}
	0 \leq s_0\leq 4m^2,\qquad
	0 \leq t_0\leq 4m^2,\qquad
	0 \leq u_0\leq 4m^2,
\end{equation}
because in this region the amplitude is purely real.\footnote{Recall the analytic structure of the amplitude given in the left part of figure \ref{fig:classes of amplitudes}. The amplitude is purely real on the horizontal axis between the two branch cuts.} There are two choices of the parameters $s_0$, $t_0$ and $u_0$ we will use in this paper. The first choice is the crossing symmetric point, where we defined the observables $\lambda_{k,l}$ as
\begin{equation}
	\label{eq:description_1}
	\lambda_{k, l} \equiv \frac{m^{d-4+2(k+l)}}{k!l!} \partial_s^k\partial_t^l \mathcal{T}(4m^2/3,\, 4m^2/3,\, 4m^2/3).
\end{equation}
Using crossing symmetry, one can easily derive the following identities:\footnote{This can be done in Mathematica as follows. First,  define $\mathcal{T}(s,t,u)$ as  $\mathcal{T}(s,t,u)=f(s,t,u)+f(s,u,t)+f(u,s,t)+f(u,t,s)+f(t,u,s)+f(t,s,u)$, so that it is crossing symmetric. Then Taylor expand $\mathcal{T}(s,t,u)$ at the crossing symmetric point, and ask Mathematica to find the relationships between the Taylor coefficients, for example, via using the \texttt{Eliminate} function. This is how we got equation \eqref{eq:lambdaRelation}, and similarly \eqref{eq:LambdaRelation}.   }
\begin{align}\label{eq:lambdaRelation}
\lambda_{i,j}&=\lambda_{j,i},\quad \lambda_{0,k} =0 \quad \text{ for odd $k$}, \\
  \lambda_{1,1}&=\lambda_{2,0}, \quad \lambda_{1,3}=2 \lambda_{4,0} = 2\lambda_{2,2}/3,\quad 2 \lambda_{1,4}=\lambda_{3,2}, \quad
  \lambda_{5,1}=3 \lambda_{6,0} = 3(2 \lambda_{2,4}-\lambda_{3,3})/5, \cdots, \nonumber
  \end{align}
where we have presented all the identities up to $k+l\le 6$. In terms of the observables defined in  \eqref{eq:description_1}, the amplitude has the following series representation
\begin{equation}
	\label{eq:representation_CS}
	m^{d-4}\mathcal{T}(s,t,u)= \sum_{k,l =0}^\infty\lambda_{k,l}\;m^{-2(k+l)}(s-4m^2/3)^k(t-4m^2/3)^l,
\end{equation}
which is valid in the Mandelstam triangle
\begin{equation}
	0 \leq s\leq 4m^2,\qquad
	0 \leq t\leq 4m^2,\qquad
	0 \leq u\leq 4m^2.
\end{equation}
Due to \eqref{eq:lambdaRelation}, the above representation is automatically fully crossing invariant. For further details about how \eqref{eq:representation_CS} is related to the method in the literature that writes $\mathcal{T}$ in terms of symmetric polynomials, see appendix \ref{app:polynomials}.

The second choice is the forward limit, where we defined the observables $\Lambda_{k,l}$ as  
\begin{equation}
	\label{eq:description_2}
	\Lambda_{k, l} \equiv \frac{m^{d-4+2(k+l)}}{k!l!} \partial_s^k\partial_t^l \mathcal{T}(2m^2,\, 0,\, 2m^2).
\end{equation}
Similarly, using the fact that $\mathcal{T}$ is crossing symmetric, we can derive the following identities 
\begin{align}\label{eq:LambdaRelation}
  \Lambda_{k,0}&=0\quad \textrm{for odd $k$},\nonumber\\
  \Lambda_{1,1}&=\Lambda_{2,0},\quad \Lambda_{1,2}=\Lambda_{2,1},\quad \Lambda_{4,0}=\Lambda_{2,2}-\Lambda_{1,3}=\Lambda_{3,1}/2,\\\Lambda_{4,1}&=\Lambda_{2,3}-\Lambda_{1,4}=\Lambda_{3,2}/2,\quad
  \Lambda_{5,1}=3\Lambda_{6,0}=3(2\Lambda_{4,2}-\Lambda_{3,3})/5,\cdots,
\nonumber
\end{align}
where we again included identities up to $k+l\le 6$. Analogously to \eqref{eq:representation_CS}, there exist the following representation
\begin{equation}\label{eq:representation_FL}
	m^{d-4}\mathcal{T}(s,t) = \sum_{k,l =0}^\infty\Lambda_{k,l}\;m^{-2(k+l)}(s-2m^2)^k t^l
\end{equation}
valid in the region defined by \eqref{eq:range}. Due to \eqref{eq:LambdaRelation}, it is automatically $s$-$u$ invariant.

Another way to describe the amplitude is to use the partial amplitude $\mathcal{T}_j(s)$, where $j=0,2,4,\ldots$ is the spin as reviewed in section \ref{sec:partial_amplitudes}. Analogously to \eqref{eq:one_way}, we can then define the following set of parameters
\begin{equation}
	\label{eq:description_3}
	\tau_{j;k}
	\equiv \frac{m^{d-4+2k}}{k!}\partial_s^k\mathcal{T}_j(2m^2).
\end{equation}
The point $s_0=2m^2$ is also chosen from the range \eqref{eq:range} in this paper. Notice that \eqref{eq:description_3} are real due to the same reason as in \eqref{eq:description_1} and \eqref{eq:description_2}.

To summarize, we will be working in this paper with three equivalent (infinite dimensional) sets of observables
\begin{equation}
	\lambda_{k, l},\qquad
	\Lambda_{k, l},\qquad
	\tau_{j;k}.
\end{equation} 
There are other sets of observables discussed in the literature such as arks proposed in \cite{Bellazzini:2020cot} and scattering lengths. We will not construct bounds on them in this work.

\subsection{EFT amplitudes} 
\label{sec:EFT_amplitudes_observables}
Analogously to \eqref{eq:description_1}, we can define observables for the EFT amplitudes
\begin{equation}
	\label{eq:description_EFT}
	\lambda^\text{EFT}_{k, l} \equiv \frac{M^{d-4+2(k+l)}}{k!l!} \partial_s^k\partial_t^l \mathcal{T}(0,\, 0,\, 0).
\end{equation}
The observables in \eqref{eq:description_EFT} obey the same relations as in \eqref{eq:lambdaRelation}. Analogously to \eqref{eq:representation_CS}, we also have the series representation of the EFT amplitude which reads as
\begin{equation}
	\label{eq:observables_EFT}
	M^{d-4}\mathcal{T}(s,t,u)= \sum_{k,l =0}^\infty\lambda^\text{EFT}_{k,l}\;M^{-2(k+l)}s^kt^l.
\end{equation}
It is valid in the Mandelstam triangle
\begin{equation}
	0 \leq s\leq M^2,\qquad
	0 \leq t\leq  M^2,\qquad
	0 \leq u\leq M^2.
\end{equation}

\section{Positivity bounds for nonperturbative amplitudes}
\label{sec:analytic_d>2}

In this section, we will derive bounds on the observables defined in section \ref{sec:NP_amplitudes_observables} using dispersion relations and the positivity constraint \eqref{eq:DispersionUnitarity}. We refer to these bounds as the positivity bounds.
We will use the formalism of \cite{Caron-Huot:2020cmc} applied in the context of nonperturbative amplitudes which have $m\neq 0$ and their branch cut starts at $4m^2$. 


In section \ref{sec:dispersion_relations}, we will briefly review the formalism of \cite{Caron-Huot:2020cmc}. In section \ref{sec:SemiAnalyticBondsCS}, we will use it to bound the observables $\lambda_{k,l}$ defined in \eqref{eq:description_1} at the crossing symmetric point, and in section \ref{sec:SemiAnalyticBondsFL}, we will use it to bound the observables $\Lambda_{k,l}$ defined in \eqref{eq:description_2} for the forward limit.

\subsection{Dispersion relations}
\label{sec:dispersion_relations}

Consider the following function on the $s'$ complex plane
\begin{equation}
	\label{eq:function_F}
	F(s')\equiv\frac{\mathcal{T}(s',t)}{(s'-s_0)(s'-s_1)(s'-s_2)},
\end{equation}
where $t$, $s_0$, $s_1$ and $s_2$ should be seen as some fixed parameters. Due to the Froissart bound on the interacting part of the scattering amplitude $\mathcal{T}(s',t)$, this function decays at infinity as follows
\begin{equation}
	\label{eq:decay_infty}
	\lim_{s'\rightarrow \infty} s' F(s') = 0.
\end{equation}
On the $s'$ complex plane, the function $F(s')$ has two branch cuts inherited from $\mathcal{T}(s',t)$, see the left part of figure \ref{fig:classes of amplitudes}. It also has three simple poles at positions $s_0$, $s_1$ and $s_2$. In the rest of the $s'$ complex plane, the function $F(s')$ is analytic.

Since the function $F(s')$ is analytic in most of the complex plane, we have
\begin{equation}
	\oint_\gamma \frac{ds'}{2\pi i} F(s') = 0,
\end{equation}
where the contour $\gamma$ is picked as in the left part of figure \ref{fig:contour_integrals}, namely it does not contain any poles or branch cuts inside. Enlarging the contour to infinity and dropping the arcs at infinity due to \eqref{eq:decay_infty}, we obtain the right part of figure \ref{fig:contour_integrals}.  Using the Cauchy theorem to evaluate the integrals at the simple poles, we obtain the following formula (called the dispersion relation),
\begin{align}
	\label{eq:master}
	&\underset{s'=s_0,\,s_1,\,s_2}{\text{Res}} \left[\frac{\mathcal{T}(s',t)}{(s'-s_0)(s'-s_1)(s'-s_2)}\right] 
	=\\
	&\int_{-\infty}^{-t}\frac{ds'}{2\pi i}\, \frac{\text{Disc}_{s'}\mathcal{T}(s',t)}{(s'-s_0)(s'-s_1)(s'-s_2)}+
	\int_{4m^2}^{+\infty}\frac{ds'}{2\pi i}\, \frac{\text{Disc}_{s'}\mathcal{T}(s',t)}{(s'-s_0)(s'-s_1)(s'-s_2)},\nn
\end{align}
where the integrals are along the branch cuts.
The parameters $(s_0, s_1, s_2)$ can be chosen at our will.

\tikzset{cross/.style={cross out, draw=black, fill=none, minimum size=2*(#1-\pgflinewidth), inner sep=0pt, outer sep=0pt}, cross/.default={3pt}}

\def\xr{2.4}
\def\yr{2}

\def\xrl{3.5}
\def\yrl{2}

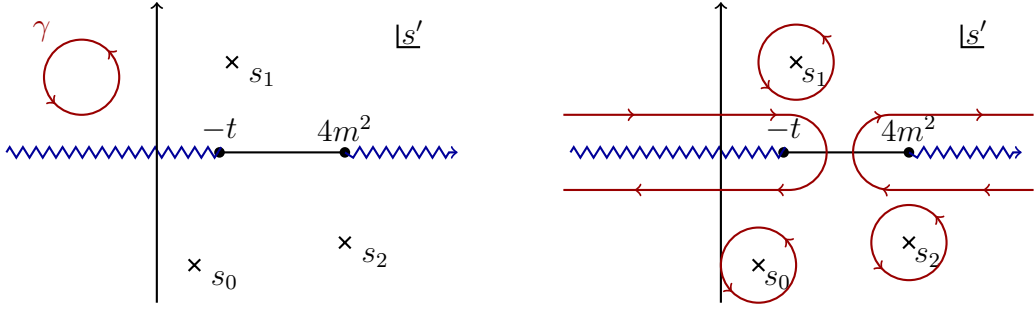
\begin{figure}[t]
	\begin{center}
		
				\begin{tikzpicture}[thick]
			\draw node[cross] at (0.5,-1.5) {} ;
			\draw node (label) at (0.9,-1.7) {$s_0$};
			
			\draw node[cross] at (1,1.2) {} ;
			\draw node (label) at (1.4,1) {$s_1$};
			
			\draw node[cross] at (2.5,-1.2) {} ;
			\draw node (label) at (2.9,-1.4) {$s_2$};
			
			\draw (3.2,1.7) -- (3.2,1.4) -- (3.5,1.4) ;
			\node (label) at (3.4,1.6) {$s'$};
			
			\fill(5/6, 0) circle (2pt) node[above] (l_branch) {$-t$} (5/2, 0) circle (2pt) node[above] (r_branch) {$4m^2$};
			
			
			\draw	[->,decorate,decoration={zigzag,segment length=5,amplitude=2,pre=lineto,pre length=0,post=lineto,post length=3}, blue!60!black]  (-2, 0) -- (l_branch.south) (r_branch.south) -- (4, 0) node [above left, black] {};
			
			\draw(l_branch.south) -- (r_branch.south);
			\draw[->] (0, -2) -- (0, 2) node[below left=0.1] {};

  \draw[red!60!black,decoration={markings,mark=between positions 0.125 and 0.875 step 0.5 with \arrow{>}},postaction={decorate}] (-1,1) circle (0.5) node [below right=-0.9 and -0.8] {$\gamma$};

			\end{tikzpicture}
		\qquad	\quad
		\begin{tikzpicture}[thick]
	
   \draw node[cross] at (0.5,-1.5) {} ;
   \draw node (label) at (0.75,-1.7) {$s_0$};
   
   \draw node[cross] at (1,1.2) {} ;
   \draw node (label) at (1.25,1) {$s_1$};
   
    \draw node[cross] at (2.5,-1.2) {} ;
   \draw node (label) at (2.75,-1.4) {$s_2$};

	\draw (3.2,1.7) -- (3.2,1.4) -- (3.5,1.4) ;
	\node (label) at (3.4,1.6) {$s'$};
	
	\fill(5/6, 0) circle (2pt) node[above] (l_branch) {$-t$} (5/2, 0) circle (2pt) node[above] (r_branch) {$4m^2$};
	
	
	\draw	[->,decorate,decoration={zigzag,segment length=5,amplitude=2,pre=lineto,pre length=0,post=lineto,post length=3}, blue!60!black]  (-2, 0) -- (l_branch.south) (r_branch.south) -- (4, 0) node [above left, black] {};
	
	\draw(l_branch.south) -- (r_branch.south);
	\draw[->] (0, -2) -- (0, 2) node[below left=0.1] {};
	
  \draw[red!60!black,decoration={markings,mark=between positions 0.125 and 0.875 step 0.5 with \arrow{>}},postaction={decorate}] (0.5,-1.5) circle (0.5);
  
    \draw[red!60!black,decoration={markings,mark=between positions 0.125 and 0.875 step 0.5 with \arrow{>}},postaction={decorate}] (1,1.2) circle (0.5);
    
        \draw[red!60!black,decoration={markings,mark=between positions 0.125 and 0.875 step 0.5 with \arrow{>}},postaction={decorate}] (2.5,-1.2) circle (0.5);
	
	\draw[xshift=50,red!60!black,decoration={markings,mark=between positions 0.125 and 0.875 step 0.25 with \arrow{>}},postaction={decorate}] (\xr,-\yr/4) -- (\yr/4,-\yr/4) arc (-90:-270:\yr/4) (\yr/4,\yr/4) -- (\xr,\yr/4) ;
	
  \draw[xshift=40,red!60!black,decoration={markings,mark=between positions 0.125 and 0.875 step 0.25 with \arrow{>}},postaction={decorate}] (-\xrl,\yrl/4) -- (-\yrl/4,\yrl/4) arc (90:-90:\yrl/4) (-\yrl/4,-\yrl/4) -- (-\xrl,-\yrl/4);

\end{tikzpicture}
		\caption{Contour integrals of the function $F(s')$ defined in equation \eqref{eq:function_F}. }
		\label{fig:contour_integrals}
	\end{center}
\end{figure}

The discontinuity of the amplitude is defined as
\begin{equation}
	\begin{aligned}
		\text{Disc}_s\mathcal{T}(s,t) 
		&\equiv \mathcal{T}(s+i\epsilon,t) - \mathcal{T}(s-i\epsilon,t)\\
		&= \mathcal{T}(s+i\epsilon,t) - \left(\mathcal{T}\left(s+i\epsilon,t\right)\right)^* \\
		&= 2i\,\text{Im} \mathcal{T}(s+i\epsilon,t),
	\end{aligned}
\end{equation}
where the second line holds on the real axis only.
Plugging the $su$-crossing equation \begin{equation}
	\label{eq:su_crossing}
	\mathcal{T}(s,t) = \mathcal{T}(4m^2-s-t,\;t)
\end{equation}
into the first term on the right-hand side of \eqref{eq:master}, and performing straightforward manipulations, we obtain 
\begin{multline}
	\int_{-\infty}^{-t}\frac{ds'}{2\pi i}\,
	\frac{\text{Disc}_s\mathcal{T}(s',t)}{(s'-s_0)(s'-s_1)(s'-s_2)} =\\ 
	\int_{4m^2}^{+\infty}\frac{d\xi}{2\pi i}\, \frac{\text{Disc}_\xi\mathcal{T}(\xi,t)}{(\xi+s_0+t-4m^2)(\xi+s_1+t-4m^2)(\xi+s_2+t-4m^2)},
\end{multline}
where we have introduced the following change of variables
\begin{equation}
	\xi \equiv 4m^2-s'-t.
\end{equation}
One can use the above result to bring the main formula \eqref{eq:master} into the following form \begin{multline}
	\label{eq:dispersion}
	\underset{s'=s_0,\,s_1, s_2}{\text{Res}} \left[\frac{\mathcal{T}(s',t)}{(s'-s_0)(s'-s_1)(s'-s_2)}\right] =
	\int_{4m^2}^{+\infty}\frac{ds'}{\pi}\,\text{Im}\mathcal{T}(s',t)
	\Bigg(
	\frac{1}{(s'-s_0)(s'-s_1)(s'-s_2)}\\
	+\frac{1}{(s'+s_0+t-4m^2)(s'+s_1+t-4m^2)(s'+s_2+t-4m^2)}
	\Bigg).
\end{multline}

Decomposing the scattering amplitude on the RHS of the above equation with the help of  \eqref{eq:mandelstam_variables_angle} - \eqref{eq:coefficients_ad}, we get
\begin{equation}
	\label{eq:dispersion_final}
	\underset{s'=s_0,\,s_1, s_2}{\text{Res}} \left[\frac{\mathcal{T}(s',t)}{(s'-s_0)(s'-s_1)(s'-s_2)}\right] =
	\sum_{j=0,2,4,\dots} 	\int_{4m^2}^{+\infty}\frac{ds'}{\pi}\, \text{Im}\mathcal{T}_j(s') h_j(s'),
\end{equation}
where the function $h_j(s')$ is a purely kinematic object defined as
\begin{align}
	&h_j(s') \equiv a_d\times
	(2j+d-3) 
	C_j^{(d-3)/2}\left(1+\frac{2t}{s'-4m^2}\right)\times\\
	&\bigg(
	\frac{1}{(s'-s_0)(s'-s_1)(s'-s_2)}+
	\frac{1}{(s'+s_0+t-4m^2)(s'+s_1+t-4m^2)(s'+s_2+t-4m^2)}
	\bigg)\nonumber
\end{align}
and the coefficient $a_d$ is defined in equation \eqref{eq:coefficients_ad}.

It is convenient to define the following angular bracket notation for the integration with the imaginary part of the partial amplitude and summation over spins
\begin{equation}
	\label{eq:averages}
	\langle h_j(s')\rangle \equiv 
	a_d\, m^{d-6}\times 
	\sum_{j=0,2,4,\dots}  (2j+d-3)C_j^{(d-3)/2}(1)	\int_{4m^2}^{+\infty}\frac{ds'}{\pi}\text{Im}\mathcal{T}_j(s')
	h_j(s').
\end{equation}
Note that the extra factor of $C_j^{(d-3)/2}(1)$ is introduced in the above definition in order to simplify the expressions later on. We will refer to $\langle h_j(s')\rangle $ as the ``average'' of the function $h_j(s') $. Note that the factor of $m^{d-6}$ is inserted such that $\langle h_j(s')\rangle $ has the same mass dimension as $h_j(s')$.

The dispersion relation written in the form \eqref{eq:dispersion_final} together with the positivity constraint
\begin{equation}
	\label{eq:simplified_unitarity}
	\text{Im}\mathcal{T}_j(s) \geq 0
\end{equation}
and the definition of ``averages'' \eqref{eq:averages} will be the key ingredients for constructing bounds on various observables.

\subsection{Bounds on $\lambda_{k,l}$}\label{sec:SemiAnalyticBondsCS}
In this subsection, we bound the observables $\lambda_{k,l}$ defined in \eqref{eq:description_1} at the crossing symmetric point. We choose the points $s_0$, $s_1$ and $s_2$ as follows
\begin{equation}
	s_0 = s, \qquad s_1=4m^2/3,\qquad s_2=8m^2/3-t.
\end{equation}
This choice maintains the $st$ crossing symmetry. Plugging them into \eqref{eq:dispersion_final}, we obtain the following expression
\begin{align}
	\label{eq:dispersion_n2_1}
	&\underset{s'=\{s,\; 2m^2,\; 2m^2-t\}}{\text{Res}} \left[\frac{m^{d-6}\mathcal{T}(s',t)}{(s'-s)(s'-4m^2/3)(s'+t-8m^2/3)}\right] =\\
	&\left\langle\frac{3 \left(4 m^2-2 s'-t\right)}{\left(s'-s\right) (s'-4m^2/3) \left(3 s'+3 t-8 m^2\right) \left(s'+s+t-4m^2\right)}
	\frac{C_j^{(d-3)/2}\left(1+\frac{2t}{s'-4m^2}\right)}{C_j^{(d-3)/2}(1)}
	\right\rangle,\nonumber
\end{align}
where the angular bracket is defined in \eqref{eq:averages}. 

In order to evaluate the left-hand side (LHS) of the expression \eqref{eq:dispersion_n2_1}, we use the polynomial representation of the amplitude \eqref{eq:representation_CS} in the vicinity of the crossing symmetric point. Plugging \eqref{eq:representation_CS} into the LHS of \eqref{eq:dispersion_n2_1}, and expanding around $s=4m^2/3$ and $t=4m^2/3$, we get
\begin{multline}
	m^{6}\times\textrm{LHS}=	\lambda _{2,0}+\lambda _{2,1}\hat{t}+ \left(\hat{t}^2 \lambda _{4,1}+\hat{t} \lambda _{4,0}\right)\hat{s}\\
	+\left(\lambda _{4,0}+\hat{t}^2 \left(\lambda _{4,2}-2 \lambda _{6,0}\right)+\hat{t} \lambda _{4,1}\right)\hat{s}^2  +2  \lambda _{4,0}\hat{t}^2+\cdots,
\end{multline}
where we have defined
\begin{equation}
	\hat{s}\equiv m^{-2}s-4/3,\qquad
	\hat{t}\equiv m^{-2}t-4/3.
\end{equation}

Similarly we can expand the right-hand side (RHS)  of the equation \eqref{eq:dispersion_n2_1} around $s=4m^2/3$ and $t=4m^2/3$. Equating the coefficients in front of the same $\hat{s}^k \hat{t}^l$ terms, we get the following relations 
\begin{align}
	\lambda_{2,0}\label{eq:lambdaAverage20}&=\left\langle\frac{54m^6\mathcal{F}(d,j,s')}{\left(3 s^{\prime}-4m^2\right)^{3}}\right\rangle, \\
	\lambda_{2,1} &=\left\langle\frac{108m^8 j (d+j-3)\mathcal{F}(d+2,j-1,s') }{(d-2) \left(s'-4m^2\right) \left(3 s'-4m^2\right)^3}  -\frac{243m^8 \mathcal{F}(d,j,s')}{\left(4m^2-3 s'\right)^4}\right\rangle,\label{eq:lambdaAverage21}\\
	\lambda_{4,0}\label{eq:lambdaAverage401}&=\left\langle\frac{486m^{10}\mathcal{F}(d,j,s')}{\left(3 s^{\prime}-4m^2\right)^{5}}\right\rangle,
\end{align}

\begin{align}
	\lambda_{4,0}=&\left\langle\frac { 27m^{10} \Gamma (d/2 - 1 ) } {2 ( 3 s ^ { \prime } - 4 m^2) ^ { 5 } } \left[ \frac{36\mathcal{F}(d,j,s')}{\Gamma \left(d/2-1\right)}+\frac{j(d+j-3)\left(3 s^{\prime}-4m^2\right)^2}{\left(s^{\prime}-4m^2\right)^{2}}\right.\right.\\
	&\times\left(\left.\left.\frac{(j-1)(d+j-2)\mathcal{F}(d+4,j-2,s')}{\Gamma(d/2+1)}-\frac{9\left(s^{\prime}-4m^2\right)\mathcal{F}(d+2,j-1,s')}{\Gamma(d/2)\left(3 s^{\prime}-4m^2\right)}\right)\right]\right\rangle\nonumber,
\end{align}
where we have defined
\begin{equation}
	\begin{aligned}
		\mathcal{F}(d,j,s)&\equiv { }_{2} F_{1}\left(-j, d+j-3 ; \frac{d-2}{2} ;-\frac{4m^2}{3\left(s-4m^2\right)}\right)\\
		&=\frac{j!}{(d-3)_j}C_j^{(d-3)/2}\left(\frac{4m^2-3s'}{3 (4m^2-s')}\right).
	\end{aligned}
\end{equation}
The factor $(d-3)_j$ in the above equation is the  Pochhammer symbol.
Note that we have two expressions for $\lambda_{4,0}$, and subtracting one from the other we get a null constraint
\begin{align}\label{eq:lambdaNull}
	0=&\bigg\langle\frac { 27m^{10} j(d+j-3)} {(d-2) ( 3 s ^ { \prime } - 4 m^2) ^4 \left(s^{\prime}-4m^2\right)^{2}}\times\left(-9\left(s^{\prime}-4m^2\right)\mathcal{F}(d+2,j-1,s') \right.\\
	&+\left.	2(j-1)(1+(j-2)/d)\left(3 s^{\prime}-4m^2\right)\mathcal{F}(d+4,j-2,s')\right)\bigg\rangle\nonumber.
\end{align}
One can expand \eqref{eq:dispersion_n2_1} to higher orders to find more null constraints. For example, there is one null constraint coming from two different expressions of $\lambda_{4,1}$, and another one from two different expressions of $\lambda_{6,0}$. 

\paragraph{Analytic bounds}
The definition of ``averages'' in equation \eqref{eq:averages} involves an integration over the region $s'\in[4m^2,\infty )$ and a summation over spins $j$. Using the positivity condition \eqref{eq:simplified_unitarity}, we can derive various bounds. 

If we look at the equations \eqref{eq:lambdaAverage20} and \eqref{eq:lambdaAverage401}, we notice that the integrands in the averages are non-negative function,\footnote{Note that $\mathcal{F}(d,j,s)$ is positive for the $d>2, j\ge 0$ and $s> 4m^2$.
	This can be shown by writing it as a sum of manifestly positive terms as follows
	\begin{equation}
		\mathcal{F}(d,j,s) = \sum_{i=0}^j\frac{ (j-i+1)_i (d+j-3)_i}{i! \left(\frac{d}{2}-1\right)_i}\left(\frac{4 m^{2}}{3\left(s-4 m^{2}\right)}\right)^{i}.
	\end{equation}
} thus we immediately conclude that
\begin{equation}
	\lambda_{2,0}\ge 0 \quad \textrm{and} \quad \lambda_{4,0}\ge 0.
\end{equation}
Notice that the expressions in the angular brackets of $\lambda_{2,0}$ and $\lambda_{4,0}$ only differ by a factor of $(3s'-4m^2)^2$ in the denominator (and a constant factor), we can easily show that 
\begin{equation}
	0\le \frac{\lambda_{4,0}}{ \lambda_{2,0}}\le \frac{9}{64}.
\end{equation}
Here we have used the fact that 
\begin{equation}\label{eq:AverageInequality1}
	\int_{4m^2}^\infty ds' \frac{q(s')}{(3s'-4m^2)^n}\leq
	\frac{1}{8m^2}\int_{4m^2}^\infty ds' \frac{q(s')}{(3s'-4m^2)^{n-1}},
\end{equation}
which holds for any function $q(s')$ obeying $q(s')\ge0$ in the  integration region.

For $\lambda_{2,1}$, notice that the first term in the angular bracket is also non-negative (for $d>2$, $j\ge 0$ and $s\ge 4m^2$), therefore, we have 
\begin{equation}
	\lambda_{2,1}\ge-\left\langle \frac{243m^8\mathcal{F}(d,j,s')}{\left(4m^2-3 s'\right)^4}\right\rangle\ge -\frac{9}{16}\lambda_{2,0},
\end{equation}
where we have used \eqref{eq:AverageInequality1} again to relate it to $\lambda_{2,0}$. In other words, we have obtained the lower bound
\begin{equation}\label{eq:lambda21LowerBound}
	-\frac{9}{16}\leq\frac{\lambda_{2,1}}{\lambda_{2,0}}.
\end{equation}

\paragraph{Numerical bounds}
One can also construct bounds on $\lambda_{2,1}$  numerically \cite{Caron-Huot:2020cmc}. In order to do that let us denote the expressions in the angular brackets in equations \eqref{eq:lambdaAverage20}, \eqref{eq:lambdaAverage21}, and \eqref{eq:lambdaNull} by $L_{2,0}(j,s')$, $L_{2,1}(j,s')$ and $N_{4,0}(j,s')$ respectively. Then these equations can be written in a compact form as follows
\begin{equation}
	\label{eq:defs_1}
	\lambda_{2,0}=\langle L_{2,0}(j,s')\rangle,\qquad \lambda_{2,1}=\langle L_{2,1}(j,s')\rangle, \qquad 0=\langle N_{4,0}(j,s')\rangle.
\end{equation}
One can then define the following simple optimization problem
\begin{equation}
	\label{eq:optimization_1}
	\begin{cases}
		\textrm{minimize} & A\\
		\textrm{subject to} & A\, L_{2,0}(j,s')-L_{2,1}(j,s')+c N_{4,0}(j,s')\ge0  \\
	\end{cases}
\end{equation} 
for all $s'\ge 4m^2$ and $j=0,2,4,\cdots$. Here, we are minimizing over any possible real value of $c$.  Taking the ``average'' of the second line in \eqref{eq:optimization_1} and using the definitions \eqref{eq:defs_1}, we then conclude that
\begin{equation}
	A \lambda_{2,0}-\lambda_{2,1}\geq 0.
\end{equation}
Thus, the optimization problem \eqref{eq:optimization_1} finds the upper bound on the ratio $\lambda_{2,1}/\lambda_{2,0}$.


We solved the optimization problem \eqref{eq:optimization_1} using the \texttt{LinearOptimization} function in Mathematica at a large number of different values of $s'$ and $j$ (from $j=0$ up to some value $j_\text{max}$) as the constraints. For $d=3$ and $d=4$, we found the optimal value of $A$ to be $A\approx 1.152$, which means that 
\begin{equation}
	\lambda_{2,1} \leq 1.152  \lambda_{2,0}.
\end{equation}

Similarly, one can find a lower bound on the ratio $\lambda_{2,1}/\lambda_{2,0}$ by solving the following optimization problem
\begin{equation}
	\label{eq:optimization_2}
	\begin{cases}
		\textrm{minimize} & B \\
		\textrm{subject to} & -B L_{2,0}\left(j, s^{\prime}\right)+L_{2,1}\left(j, s^{\prime}\right)+cN_{4,0}\left(j, s^{\prime}\right) \geq 0
	\end{cases}
\end{equation}
for all $\forall s^{\prime} \geq 4 m^{2}$ and $j=0,2,4, \cdots$.  Taking the ``average'' of the second line in \eqref{eq:optimization_2} and using the definitions \eqref{eq:defs_1} we conclude that
\begin{equation}
	-B \lambda_{2,0}+\lambda_{2,1}\geq 0.
\end{equation}
Thus, the optimization problem \eqref{eq:optimization_2} finds the lower bound on the ratio $\lambda_{2,1}/\lambda_{2,0}$.
By solving the optimization problem \eqref{eq:optimization_2}, we find numerically that the optimal value of $B$ reads as $B=-9/16$, which is precisely the result found already in \eqref{eq:lambda21LowerBound}.
Summarizing, for $d=3$ and $d=4$, we obtained the following two-sided bound.\footnote{The upper bound in \eqref{eq:lambdaTwoSided} stays the same if we also include the null constraint from $\lambda_{4,1}$. However, including also the null constraint coming from $\lambda_{6,0}$, we get a slightly better bound $\frac{\lambda_{2,1}}{\lambda_{2,0}} \leq 1.126$. We expect that more null constraints in the optimization problem will improve this bound, but their effects will be small, similar to the case considered in \cite{Caron-Huot:2020cmc}. Another comment about the upper bound in \eqref{eq:lambdaTwoSided} is that at the precision we are reporting, it does not depend on the spacetime dimension $d$. We have checked up to $d=26$ and the dependence on $d$ is very weak. The difference for different $d$ only shows up at the fifth digit after the decimal point in the cases that we considered.}
\begin{equation}
	-\frac{9}{16} \leq \frac{\lambda_{2,1}}{\lambda_{2,0}} \leq 1.152.
	\label{eq:lambdaTwoSided}
\end{equation}

We compare \eqref{eq:lambdaTwoSided} with the numerical S-matrix result in figures \ref{fig:crossing_symmetric3} and  \ref{fig:crossing_symmetric4}. We see that the numerical S-matrix bounds are much stronger for generic values of $\lambda_{2,0}$, but coincide with the positivity bounds for small enough values of $\lambda_{2,0}$. This is interesting in at least two aspects. First, this confirms the correctness of our numerical results in section \ref{sec:numeric_bounds}. Second, this also shows how including the full unitarity constraint (numerical S-matrix bootstrap) improves the result obtained using only positivity.

\subsection{Bounds on $\Lambda_{k,l}$}
\label{sec:SemiAnalyticBondsFL}

In this section, we bound the observables $\Lambda_{k,l}$ defined in \eqref{eq:description_2} in the forward limit. We choose the points $s_0$, $s_1$ and $s_2$ as follows
\begin{equation}
	s_0 = s, \qquad s_1=2m^2,\qquad s_2=2m^2-t.
\end{equation}
This choice maintains the $su$ symmetry.
Plugging them into \eqref{eq:dispersion_final}, we obtain
\begin{multline}
	\label{eq:dispersion_n2_2}
	\underset{s'=\{s,\; 2m^2,\; 2m^2-t\}}{\text{Res}} \left[\frac{m^{d-6}\mathcal{T}(s',t)}{(s'-s)(s'-2m^2)(s'+t-2m^2)}\right] =\\
	\left\langle
	\frac{(2s'+t-4m^2)}{(s'-s)(s'-2m^2)(s'+t-2m^2)(s'+s+t-4m^2)}
	\frac{C_j^{(d-3)/2}\left(1+\frac{2t}{s'-4m^2}\right)}{C_j^{(d-3)/2}(1)}
	\right\rangle,
\end{multline}
where the definition of the angular bracket is given in \eqref{eq:averages}.

In order to evaluate the LHS of the expression \eqref{eq:dispersion_n2_2}, we use the polynomial representation of the amplitude \eqref{eq:representation_FL} in the forward limit. Plugging \eqref{eq:representation_FL} into the LHS of \eqref{eq:dispersion_n2_2}, and expanding around $s=2m^2$ and $t=0$, we get
\begin{multline}
	\label{eq:expansion_LHS}
	m^6\times \textrm{LHS}=\Lambda _{2,0}+\Lambda _{2,1}\tilde{t}+ \left(\Lambda _{2,2}-\Lambda _{4,0}\right)\tilde{t}^2\\ + \left(\Lambda _{4,0}+ \Lambda _{4,1}\tilde{t}+ \left(\Lambda _{4,2}-2 \Lambda _{6,0}\right)\tilde{t}^2\right)\tilde{s}^2+\left(\Lambda _{4,1}\tilde{t}^2 + \Lambda _{4,0}\tilde{t}\right)\tilde{s} +...,
\end{multline}
where we have defined
\begin{equation}
	\tilde{s}\equiv m^{-2}s-2,\qquad 
	\tilde{t}\equiv m^{-2}t.
\end{equation}
Similarly we can expand the RHS  of equation \eqref{eq:dispersion_n2_1} around $s=2m^2$ and $t=0$. Equating the coefficients in front of the same $\tilde{s}^k \tilde{t}^l$ terms we get the following relations 
\begin{small}
	\begin{align}
		\nn
		\Lambda_{2,0}&= \left\langle \frac{2m^6}{(s'-2m^2)^3} \right\rangle,\qquad
		\Lambda_{4,0}=\left\langle \frac{2m^{10}}{(s'-2m^2)^5} \right\rangle,\nonumber\\
		\label{eq:avereges}
		\Lambda_{2,1}&=-\left\langle m^8\frac{3(s'-4m^2)-\frac{4}{d-2}	\mathcal{J}_2(s'-2m^2)}{(s'-2m^2)^4(s'-4m^2)} \right\rangle,\\
		\Lambda_{2,2}-\Lambda_{4,0} &= 4m^{10}\left\langle \frac{
			(s'-4m^2)^2+
			\mathcal{J}_2(s'-2m^2) \left( \frac{4-5d}{2d(d-2)}(s'-4m^2)-\frac{2}{d}m^2\right)+
			\frac{1}{d(d-2)}\mathcal{J}_2^2(s'-2m^2)^2
		}{(s'-2m^2)^5(s'-4m^2)^2} \right\rangle.
		\nn
	\end{align}
\end{small}
Here we have defined the eigenvalue of the quadratic Casimir $\mathcal{J}_2$ as
\begin{equation}
	\mathcal{J}_2 \equiv j(j+d-3).
\end{equation}

\paragraph{Analytic bounds} Similar to the $\lambda_{k,l}$ case we considered in the last subsection, since integrands in the first two lines of  \eqref{eq:avereges}  are non-negative in the integration region $s\in[4m^2,\infty )$, 
we immediately conclude that
\begin{equation}\label{eq:Lambda1140}
	\Lambda_{2,0} \geq 0,\qquad
	\Lambda_{4,0} \geq 0.
\end{equation}

Let us now address $\Lambda_{2,1}$.  Using linearity of the average, we can rewrite \eqref{eq:avereges} as follows
\begin{equation}
	-\Lambda_{2,1} = 3\left\langle\frac{m^8}{(s'-2m^2)^4} \right\rangle-\frac{4}{d-2}
	\left\langle \frac{	\mathcal{J}_2 m^8}{(s'-2m^2)^3(s'-4m^2)} \right\rangle.
\end{equation}
Since the second term above is non-negative (for $d>2$ which is what we consider in this section), the following inequality follows
\begin{equation}
	-\Lambda_{2,1} \leq 3\left\langle\frac{m^8}{(s'-2m^2)^4} \right\rangle \leq  \frac{3\Lambda_{2,0}}{4}.
\end{equation}
In other words, we have obtained the lower bound
\begin{equation}\label{eq:Lambda21Lambda20}
	-\frac{3}{4}\leq\frac{\Lambda_{2,1}}{\Lambda_{2,0}}.
\end{equation}
To get the second inequality here, we used the following simple property of the integral
\begin{equation}
	\int_{4m^2}^\infty ds' \frac{q(s')}{(s'-2m^2)^n}\leq
	\frac{1}{2m^2}\int_{4m^2}^\infty ds' \frac{q(s')}{(s'-2m^2)^{n-1}},
\end{equation}
similar to the inequality in \eqref{eq:AverageInequality1}, which holds true for any function $q(s')$ obeying $q(s')\geq 0$ in the integration region. Using the above inequality, we can also show that 
\begin{equation}
	\label{eq:ab_Lambda}
	0\le  \frac{\Lambda_{4,0}}{\Lambda_{2,0}}\leq \frac{1}{4},
\end{equation}
where we also included the lower bound from \eqref{eq:Lambda1140} for completeness.
Further bounds can be derived from the last line of \eqref{eq:avereges}. The numerical procedure described in the previous subsection applied in this case does not improve the analytic bound \eqref{eq:ab_Lambda}.

\section{Numerical bounds in $d>2$}
\label{sec:numeric_bounds}
In order to put numerical bounds on observables defined in section \ref{sec:NonPerturbativeObservables}, we employ the primal numerical approach of \cite{Paulos:2017fhb,Homrich:2019cbt}. For a concise summary of this approach see sections 1 and 4.1 of \cite{Hebbar:2020ukp}.  We will briefly summarize this approach also in section \ref{sec:PNA}. There we start by explaining the primal numerical approach in the context of nonperturbative amplitudes and then indicate what changes if we want to apply it to EFT amplitudes. We show our bounds on nonperturbative amplitudes in section \ref{sec:NPA}. We show our bounds on EFT amplitudes in section \ref{sec:EFTA}.  Recall, that the definition of nonperturbative and EFT amplitudes was given in the beginning of section \ref{sec:intro}.

\subsection{Primal numerical approach}
\label{sec:PNA}
Analyticity and crossing symmetry are implemented by writing the following ansatz for the scattering amplitude 
\begin{multline}
\label{eq:ansatz}
m^{d-4}\mathcal{T}(s,t,u) = \sum_{a+b\leq N_\textrm{max}}
\alpha_{ab}\times\Big(
\myRho(s,s_0)^a\myRho(t,t_0)^b+\\
\myRho(s,s_0)^a\myRho(u,u_0)^b+
\myRho(t,t_0)^a\myRho(u,u_0)^b
\Big),
\end{multline}
where $\alpha_{ab}$ are the unknown real dimensionless coefficients symmetric in their indices $a,b$ and the $\myRho$-variable is defined as
\begin{equation}
\label{eq:rho}
\myRho(z;z_0) \equiv \frac{\sqrt{4m^2-z_0}-\sqrt{4m^2-z}}{\sqrt{4m^2-z_0}+\sqrt{4m^2-z}}.
\end{equation}
Here $z_0$ is some free parameters of our choice. In this paper, we make the following choice
\begin{equation}
s_0=t_0=u_0=4m^2/3.
\end{equation}
This choice does not affect the bounds given large enough values of $N_\textrm{max}$. The ansatz \eqref{eq:ansatz} can be used in $d>2$ dimensions. The parameter $N_\textrm{max}$ is introduced to make the sum finite. All the bounds presented below are constructed for $N_\textrm{max}=\{20, 22, 24, 26\}$ and then extrapolated to $N_\textrm{max}=\infty$ with a linear function fit $q + r/N_\textrm{max}$, where $q$ and $r$ are fit parameters. In our numerical procedure, we set $m=1$. All the dimensionless quantities we bound do not depend on this choice.

The coefficients $\alpha_{ab}$ can be related straightforwardly to the observables defined in section \ref{sec:NonPerturbativeObservables}. Consider for example the observables defined in equation \eqref{eq:description_1}. We can obtain the expression for $(\lambda_{0,0}, \lambda_{2,0}, \lambda_{2,1})$ in terms of $\alpha_{ab}$ by applying  \eqref{eq:description_1} to \eqref{eq:ansatz}. They read as
\begin{equation}
	\lambda_{0,0} = \alpha_{00},\quad
	\lambda_{2,0} = \frac{9}{256} \alpha_{01}+\frac{9}{512} \alpha_{02}-\frac{9}{1024} \alpha_{11},\quad
	\lambda_{2,1} = -\frac{405}{32768} \alpha_{01}+\ldots
\end{equation}
Solving these relations for $(\alpha_{00},\alpha_{01},\alpha_{02})$ and using the solution in \eqref{eq:ansatz}, we obtain the ansatz which depends on the following unknown parameters
\begin{equation}
	\label{eq:coefficients_ans}
	(\lambda_{0,0}, \lambda_{2,0}, \lambda_{2,1}, \alpha_{03}, \alpha_{04},\ldots, \alpha_{11}, \alpha_{12}, \ldots).
\end{equation}

We can now perform the integral transform of the ansatz \eqref{eq:ansatz} according to \eqref{eq:PA_general_d} to obtain the partial amplitudes. They will also depend on the set of unknown coefficients  \eqref{eq:coefficients_ans}. We impose full non-linear unitarity constraint in the form \eqref{eq:unitarity_3} for a discrete number of points $N_\text{grid}$ in $s$ and a finite number of spins $j=0,2,\ldots, L_\textrm{max}$. In practice, we use $N_\text{grid}=200$ and $L_\textrm{max}=N_\textrm{max}+10$ unless stated differently. The coefficients of the ansatz \eqref{eq:coefficients_ans} satisfying \eqref{eq:unitarity_3} are found by solving various numerical optimization problems using the SDPB software \cite{Simmons-Duffin:2015qma,Landry:2019qug}.

Let us briefly discuss the $d\rightarrow 2$ limit. In \cite{Chen:2021pgx}, we have obtained bounds in $d=2$. There, we used a much simpler ansatz
\begin{equation}
\label{eq:ansatz_d=2}
m^{-2}\mathcal{T}(s,t) =
\sum_{a=0}^{N_\textrm{max}}\beta_{a}\times\Big(
\myRho(s,s_0)^a+
\myRho(t,t_0)^a
\Big),
\end{equation}
where the $\beta_\alpha$s are the unknown real parameters. Clearly, the ansatz \eqref{eq:ansatz} contains a lot more unknown parameters compared to \eqref{eq:ansatz_d=2}. In the vicinity of $d=2$, we will need to use very large values of $L_\textrm{max}$ and $N_\text{grid}$ in order to numerically reproduce  \eqref{eq:ansatz_d=2} from \eqref{eq:ansatz}. Also, the numerics might become very sensitive to the precision of some numerical integrals we use internally. See appendix \ref{app:limit_d2} for further discussion. Concluding, we expect to have trouble when constructing bounds in the vicinity of $d=2$. In practice, we will be able to go as low as $d= 2.4$.

Using linearized unitarity \eqref{eq:linear_unitarity} instead of the full non-linear unitarity \eqref{eq:unitarity_3} in the primal numerical method is trivial. We simply need to replace the single condition \eqref{eq:unitarity_3} given by a 2 by 2 matrix with two conditions given by two 1 by 1 matrices
\begin{equation}
	\label{eq:lin_unit}
	\textrm{Im}  \CT_j\geq 0,\qquad
	2  \CN_d-\textrm{Im}  \CT_j\geq 0.
\end{equation}
Notice however that these do not constraint the purely real part of the ansatz \eqref{eq:ansatz} given by the coefficient $\alpha_{0,0}=\lambda_{0,0}$. During the optimization process, the coefficient $\alpha_{0,0}=\lambda_{0,0}$ will remain undetermined and will cause instabilities. The simplest way to deal with this issue is to set  $\lambda_{0,0}=0$.

To impose only positivity, we need to take into account the first condition in \eqref{eq:lin_unit} and drop the second one. Notice however that if we found one solution which satisfies $\textrm{Im}  \CT_j\geq 0$, we could obtain an infinite set of solutions by re-scaling the coefficients of the ansatz \eqref{eq:coefficients_ans}. In order to obtain a uniquely defined solution, we need to further fix one of the coefficients in  \eqref{eq:coefficients_ans}. For example, we can fix $\lambda_{2,0}$. The bounds we obtain then will be exactly of the form as in section \ref{sec:analytic_d>2}, more precisely, as in equation \eqref{eq:lambdaTwoSided}. The explicit values for the bounds obtained in section \ref{sec:analytic_d>2} and by using the primal numerical method will also coincide since the two approaches impose the same amount of constraints.

To apply the above method to EFT amplitudes (where the  branch cut is assumed to be absent below $s=M^2$ instead of $s=4m^2$), one should simply modify the analytic structures of the ansatz \eqref{eq:ansatz}. See figure \ref{fig:classes of amplitudes} for a visual representation of the analytic structure of both classes of amplitudes. The EFT ansatz reads as
\begin{multline}
	\label{eq:ansatz_EFT}
	M^{d-4}\mathcal{T}(s,t,u) = \sum_{a+b\leq N_\textrm{max}}
	\alpha_{ab}\times\Big(
	\myRho(s,s_0)^a\myRho(t,t_0)^b+\\
	\myRho(s,s_0)^a\myRho(u,u_0)^b+
	\myRho(t,t_0)^a\myRho(u,u_0)^b
	\Big),
\end{multline}
where the $\myRho$-variable is defined as
\begin{equation}
	\label{eq:rho_EFT}
	\myRho(z;z_0) \equiv \frac{\sqrt{M^2-z_0}-\sqrt{M^2-z}}{\sqrt{M^2-z_0}+\sqrt{M^2-z}}.
\end{equation}
The parameter $z_0$ can now be chosen as
\begin{equation}
	s_0=t_0=u_0=0.
\end{equation}
Recall that we also assume $m\ll M$ of the particles, thus we can set $m=0$ in \eqref{eq:constraint}, \eqref{eq:PhaseSpaceFactor} and \eqref{eq:mandelstam_variables_angle}. The observables we bound now are defined in \eqref{eq:description_EFT}. In our numerical procedure we set $M=1$. All the dimensionless quantities we bound do not depend on this choice.

\begin{figure}[tb!]
\begin{center}
\includegraphics[width=0.85\textwidth]{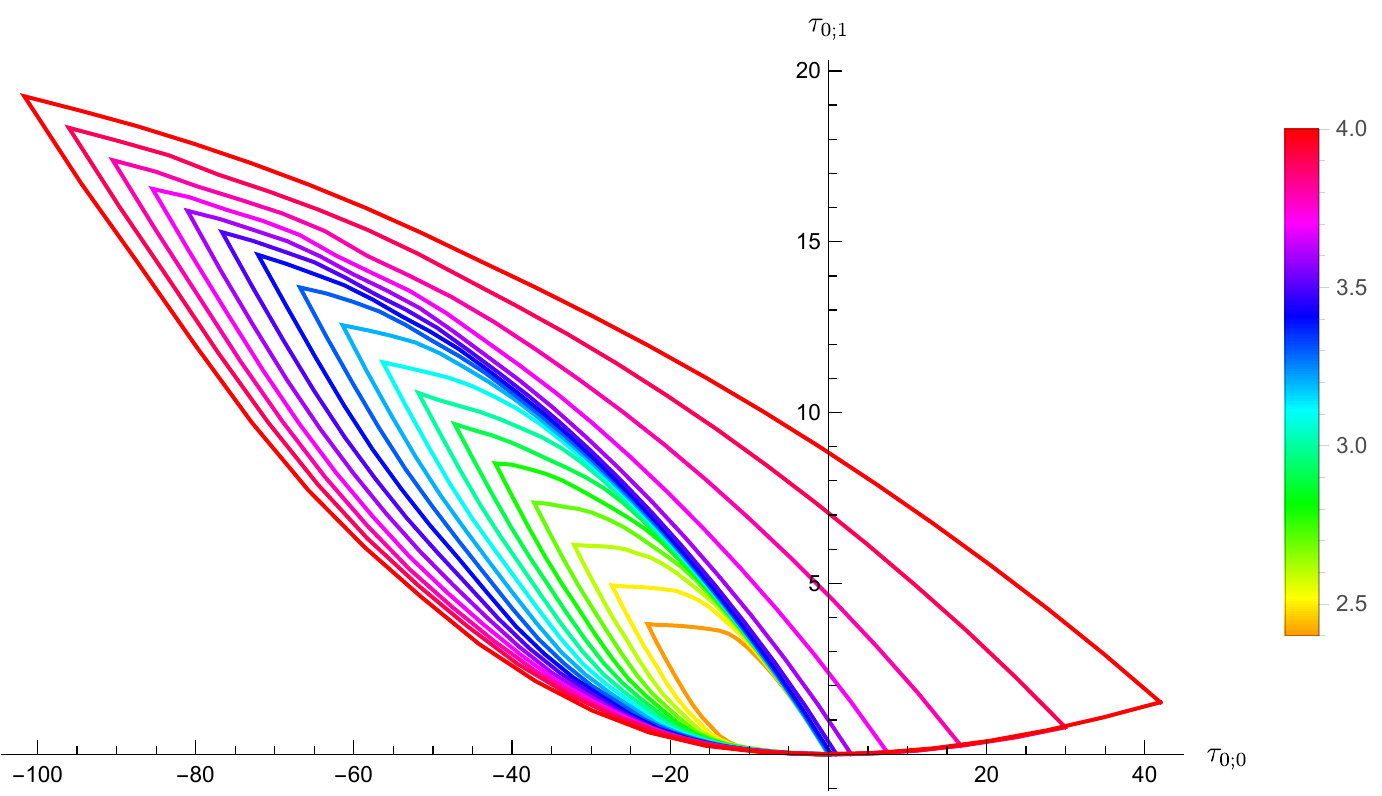}
\caption{Nonperturbative bound on the observables $(\tau_{0;0},\, \tau_{0;1})$ defined in \eqref{eq:description_3} for various spacetime dimensions $d\in[2.4,4]$. Different colors represent different $d$. For each value of $d$, the allowed region is inside the corresponding ``leaf'' shape. The plot is built with $L_\text{max}=50$.}
\label{fig:PA0_dPA0_Extrapolated}

\vspace{20mm}

\includegraphics[width=0.44\textwidth]{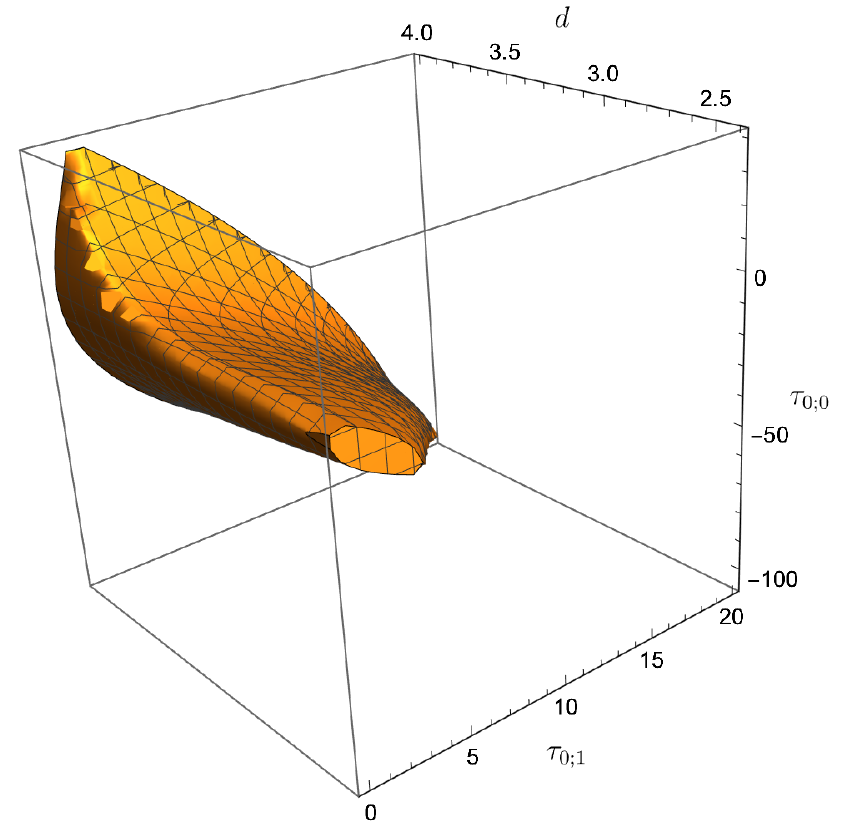}
\qquad
\includegraphics[width=0.45\textwidth]{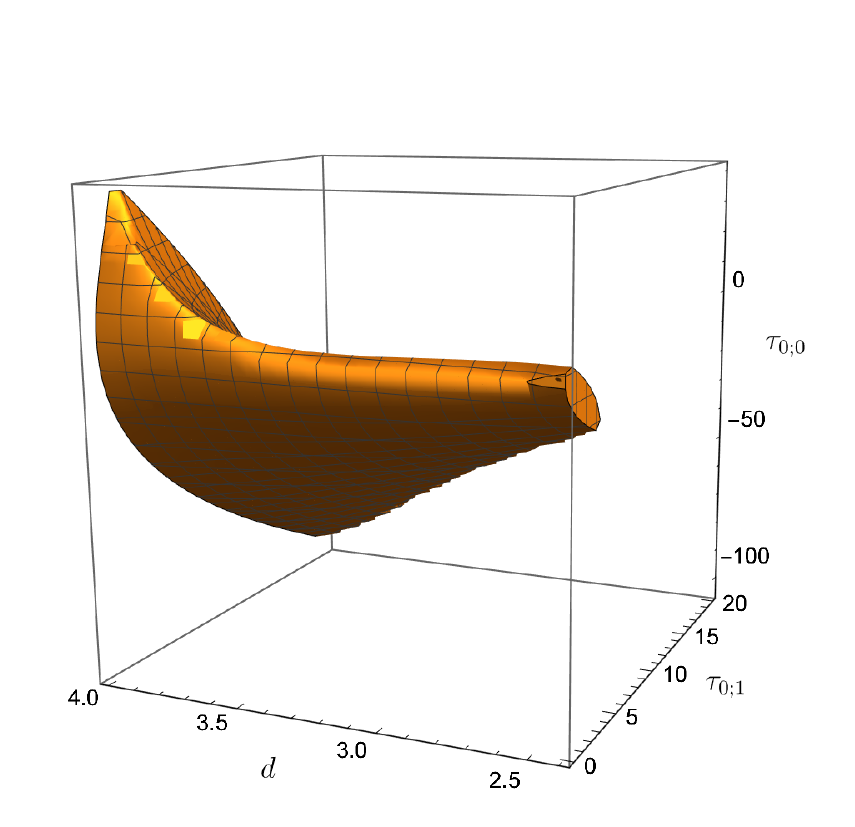}
\caption{Nonperturbative bound on the observables $(\tau_{0;0},\, \tau_{0;1})$ defined in \eqref{eq:description_3} as a function of the spacetime dimension $d$. The plot is built with $L_\text{max}=50$.}
\label{fig:3dplot}
\end{center}
\end{figure}

\subsection{Nonperturbative amplitudes}
\label{sec:NPA}

\begin{figure}[htb!]
\begin{center}
\includegraphics[width=0.41\textwidth]{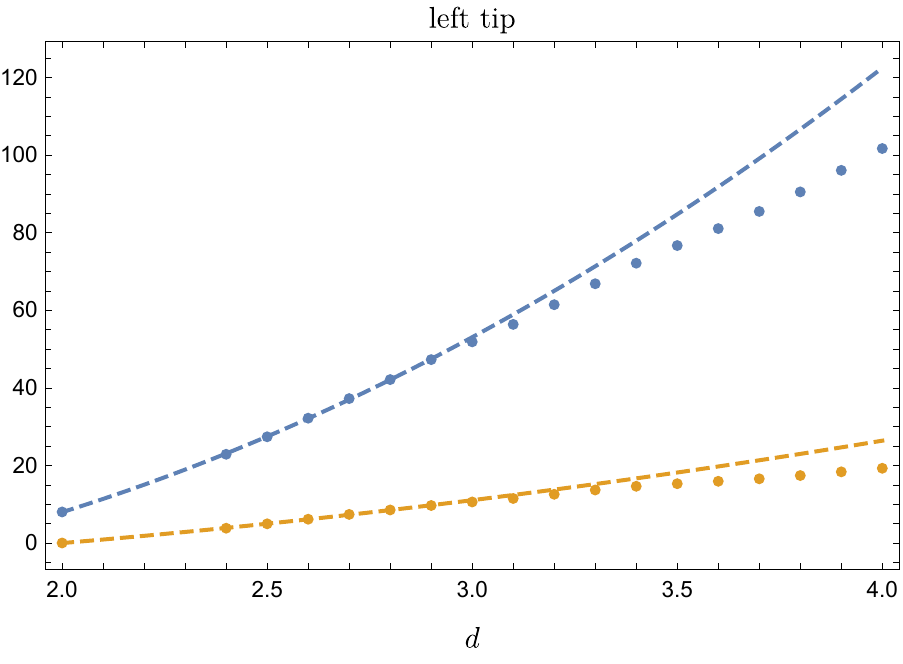}
\qquad
\includegraphics[width=0.5\textwidth]{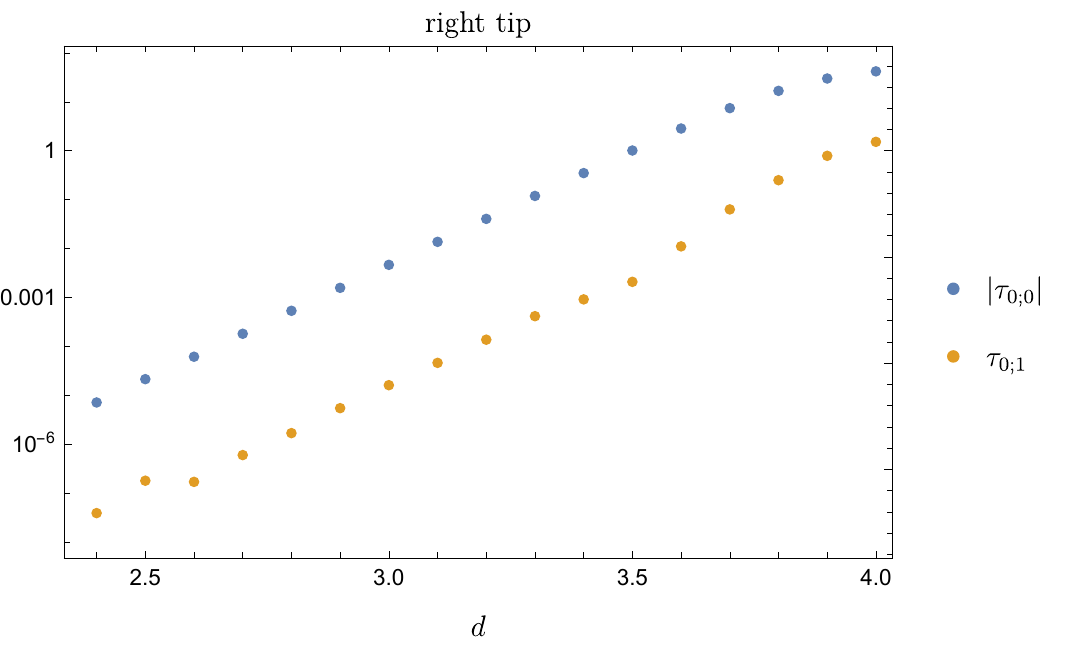}

\caption{Dependence of left and right tips of the allowed regions in figure \ref{fig:PA0_dPA0_Extrapolated} on the spacetime dimension $d$. In the left plot, we have also indicated the $d=2$ result from equation \eqref{eq:boundLambda0}, namely $(\tau_{0;0},\tau_{0;1})=(-8, 0)$. The dashed lines indicate how the $d>2$ results approach $d=2$ one. In the right plot, we see that the right tip in figure \ref{fig:PA0_dPA0_Extrapolated} approaches zero as we lower the spacetime dimension $d$.} 
\label{fig:PA0_dPA0_left_tip}

\vspace{14mm}

\includegraphics[width=0.45\textwidth]{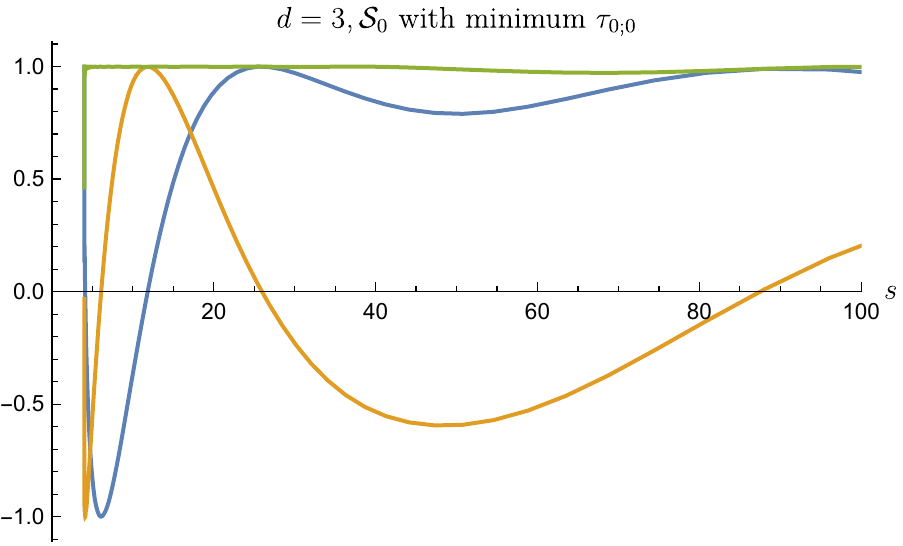}
\includegraphics[width=0.53\textwidth]{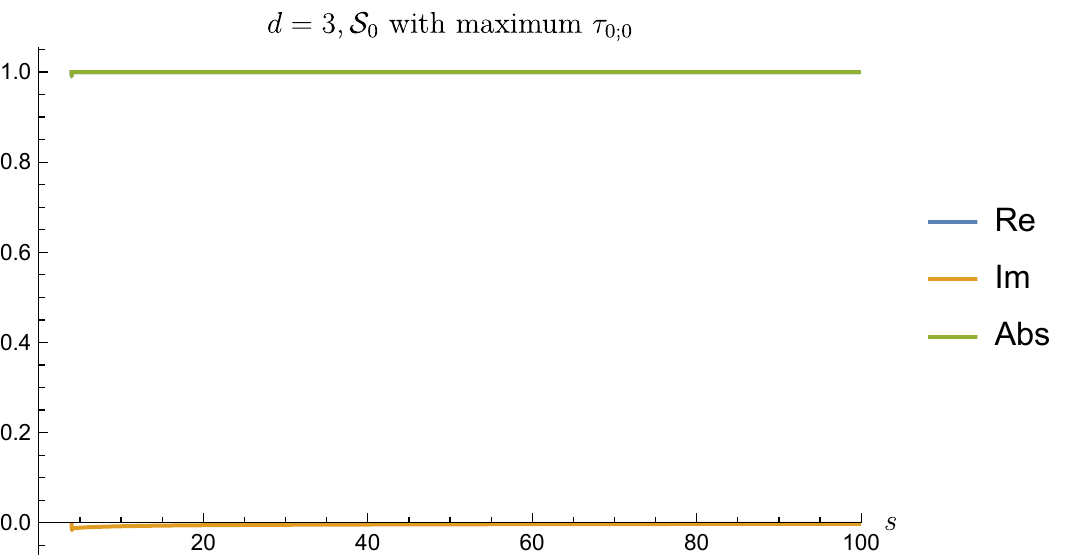}
\includegraphics[width=0.45\textwidth]{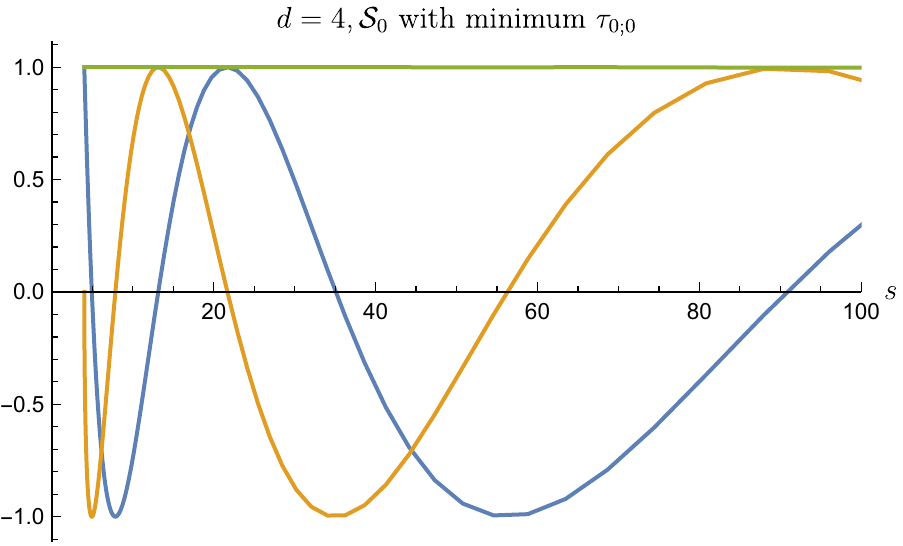}
\includegraphics[width=0.53\textwidth]{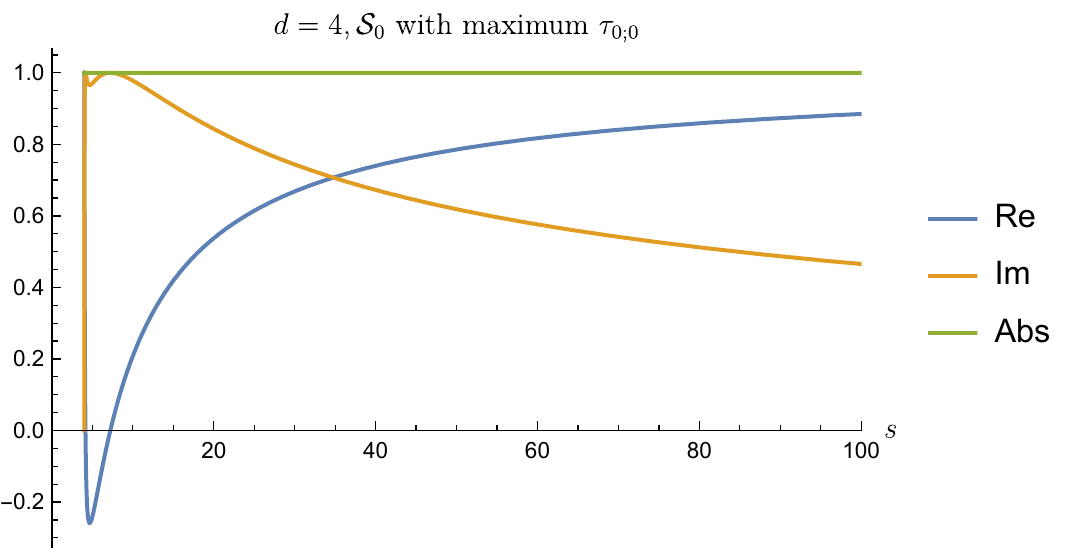}
\caption{Spin-zero partial amplitudes $\mathcal{S}_0$ at the tips of figure \ref{fig:PA0_dPA0_Extrapolated} for $d=3$ and $d=4$. We used $N_\text{max}=26$ and $L_\text{max}=50$ for these plots. }
\label{fig:partial_amplitudes}
\end{center}
\end{figure}

Let us now present our numerical results for the class of nonperturbative amplitudes. We start by showing the bounds on $(\tau_{0;0},\, \tau_{0;1})$ observables for various values of space-time dimensions $d$. They are given in figure \ref{fig:PA0_dPA0_Extrapolated}. For each value of $d$, the allowed region of the parameters is contained inside the corresponding leaf shape. Different dimensions are marked by different colors. In figure \ref{fig:3dplot}, we provide 3d plots that show how the bound on $(\tau_{0;0},\, \tau_{0;1})$ evolves as a function of $d$. The bounds on figure \ref{fig:PA0_dPA0_Extrapolated} have two well-pronounced tips. The position of the left tip steadily moves when increasing $d$. However, the right tip stays very close to the origin for $2<d\lesssim3.5$ and only starts visibly moving to the right around $d\gtrsim 3.5$. We present the positions of both tips as  functions of $d$ in figure \ref{fig:PA0_dPA0_left_tip}, where one can see more clearly their behavior. We also show how the tips of $d>2$ connect to the tips of $d=2$ which are at $(\tau_{0;0}, \tau_{0;1})=(-8,0)$ and $(\tau_{0;0}, \tau_{0;1})=(0,0)$ (for more details, see appendix \ref{sec:2dwarmup} or section 4 of \cite{Chen:2021pgx}).

On the boundary of the leaves in figure \ref{fig:PA0_dPA0_Extrapolated}, one can reconstruct numerically the scattering amplitude and all its partial amplitudes. As an example in figure \ref{fig:partial_amplitudes}, we provide the spin-zero partial amplitude at the tips of $d=3$ and $d=4$. All of these partial amplitudes saturate unitarity as can be seen from the plot.

Before proceeding with the rest of our numerical results, let us emphasize that all the numerical data and the description of how to use these data can be downloaded from: \href{https://zenodo.org/record/6891946#.Ytwnmi8Roe0}{https://zenodo.org/record/6891946\#.Ytwnmi8Roe0}. This data allows not only to reconstruct bounds like figures \ref{fig:PA0_dPA0_Extrapolated} and \ref{fig:3dplot}, but also to reconstruct the amplitudes and partial amplitudes on the boundary of the allowed region. Figure \ref{fig:partial_amplitudes} should be simply seen as an example.

\begin{figure}[H]
	\begin{center}
		\includegraphics[width=0.9\textwidth]{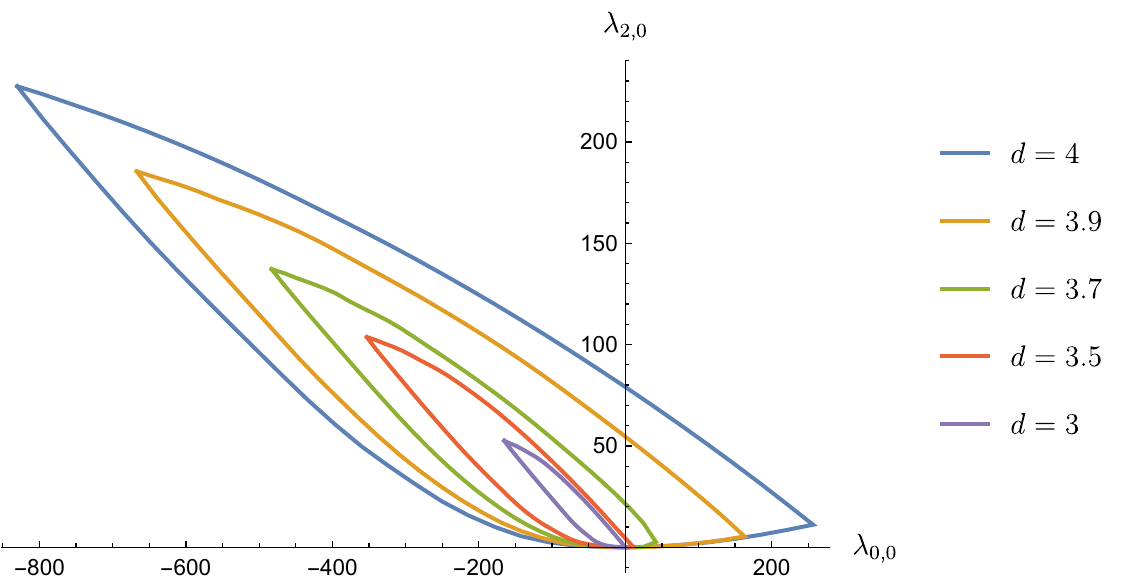}

		\caption{Nonperturbative bounds on the observables $(\lambda_{0,0}\, \lambda_{2,0})$ defined in \eqref{eq:description_1} at the crossing symmetric point obtained using full unitarity. For each value of $d$, the allowed region is inside the corresponding ``leaf'' shape.  }
		\label{fig:crossing_symmetric}
		
		\vspace{10mm}
		
		\includegraphics[width=0.9\textwidth]{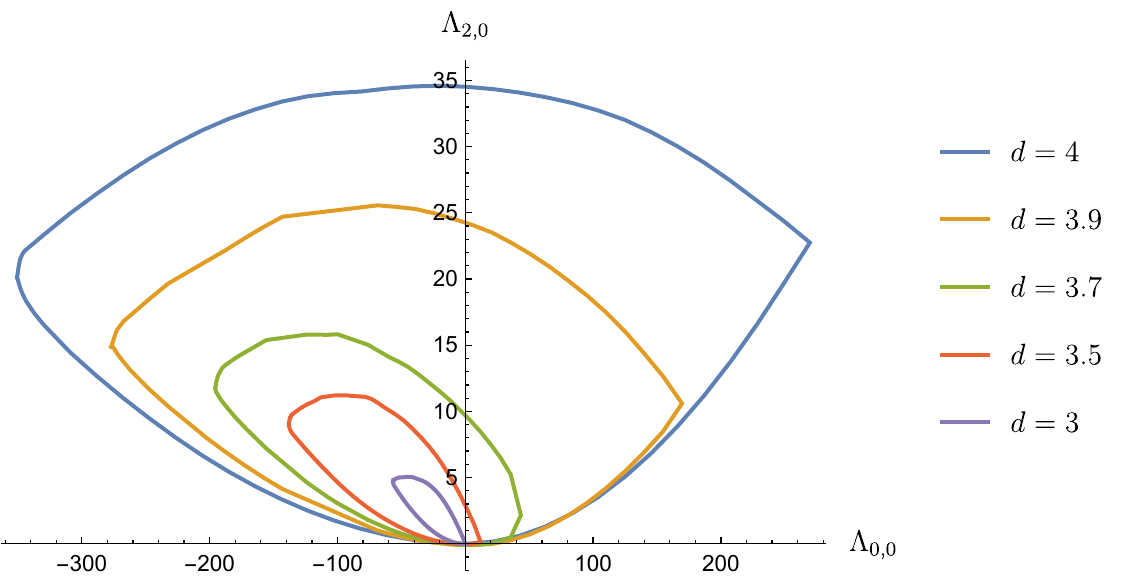}
		\caption{Nonperturbative bounds on the observables $(\Lambda_{0,0}\, \Lambda_{2,0})$  defined in \eqref{eq:description_2} in the forward limit obtained using full unitarity. For each value of $d$, the allowed region is inside the corresponding ``leaf'' shape. The wiggles in the upper boundaries of the bounds in $d=3.5$, $3.7$ and $3.9$ spacetime dimensions are due to errors in the extrapolation to $N_\text{max}=\infty$.}

		\label{fig:plot_forward}
	\end{center}
\end{figure}

The bounds on the observables $(\lambda_{0,0},\, \lambda_{2,0})$ at the crossing symmetric point and  $(\Lambda_{0,0},\, \Lambda_{2,0})$ in the forward limit for various spacetime dimensions $d$ are presented in figures \ref{fig:crossing_symmetric} and \ref{fig:plot_forward}, respectively. For each value of $d$, the allowed region is again inside the corresponding leaf. On both plots, one observes tips on the right that behave similarly to the ones of figure \ref{fig:PA0_dPA0_Extrapolated}. One can see some wiggles in figure \ref{fig:plot_forward} for $d=3.5$, $3.7$ and $3.9$, which are completely unphysical. They are present due to errors in the extrapolation to $N_\text{max}=\infty$.

\begin{figure}[H]
	\begin{center}
\includegraphics[width=0.9\textwidth]{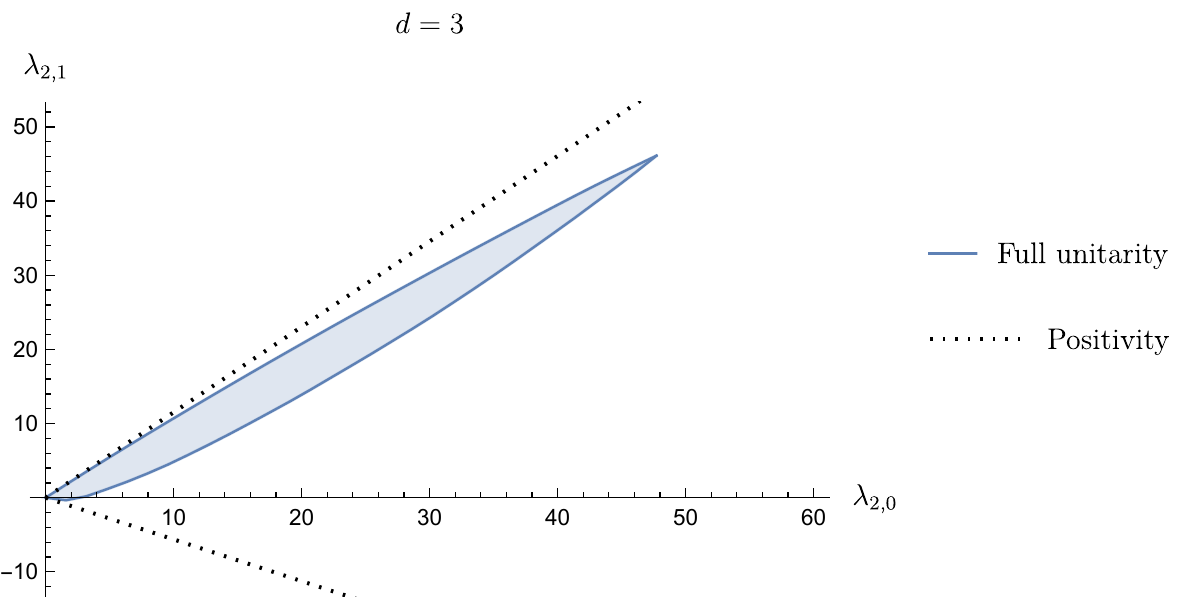}
		\caption{Nonperturbative bounds in $d=3$ on the observables $(\lambda_{2,0},\, \lambda_{2,1})$ defined in \eqref{eq:description_1} at the crossing symmetric point. The allowed region obtained using full unitarity is shaded in blue. The dashed lines represent the positivity bound derived in the last section, see \eqref{eq:lambdaTwoSided}. The allowed region is enclosed in the cone between the two dashed lines.  The positivity bound agrees with the full unitarity bounds at small $\lambda_{2,0}$, but overall is much weaker.}
		\label{fig:crossing_symmetric3}
		
		\vspace{8mm}
		
		\includegraphics[width=0.9\textwidth]{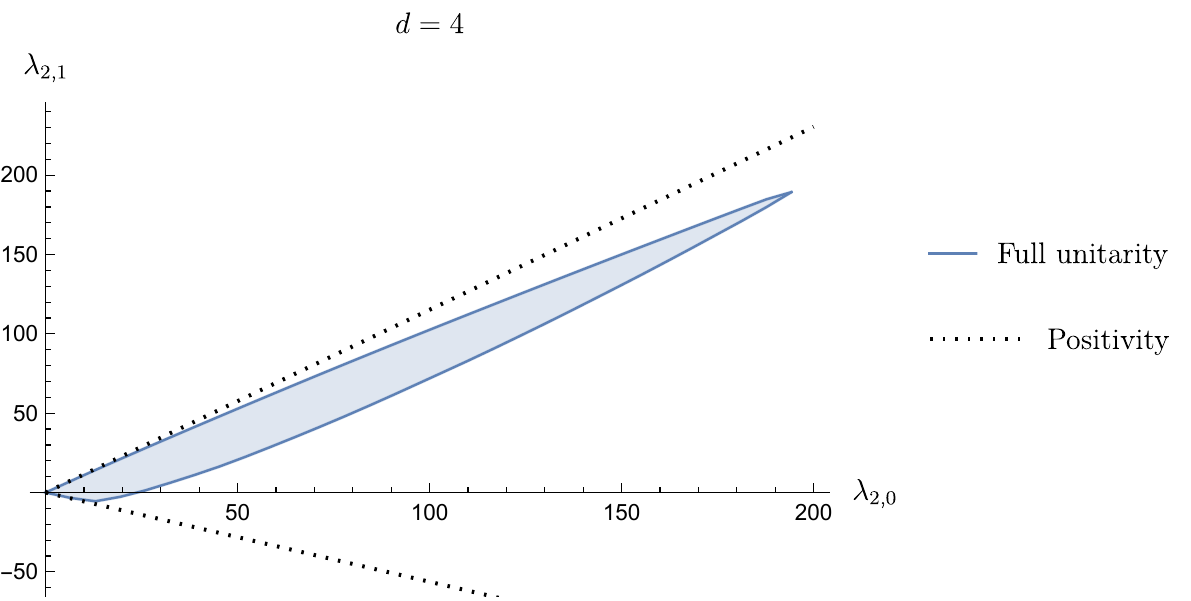}
		\caption{Nonperturbative bounds in $d=4$ on the observables $(\lambda_{2,0},\, \lambda_{2,1})$  defined in \eqref{eq:description_1} at the crossing symmetric point. The allowed region obtained using full unitarity is shaded in blue. The dashed lines represent the positivity bound derived in the last section, see \eqref{eq:lambdaTwoSided}. The allowed region is enclosed in the cone between the two dashed lines.  The positivity bound agrees with the full unitarity bounds at small $\lambda_{2,0}$, but overall is much weaker.}
		\label{fig:crossing_symmetric4}
	\end{center}
\end{figure}

\begin{figure}[H]
	\begin{center}
		\includegraphics[width=0.9\textwidth]{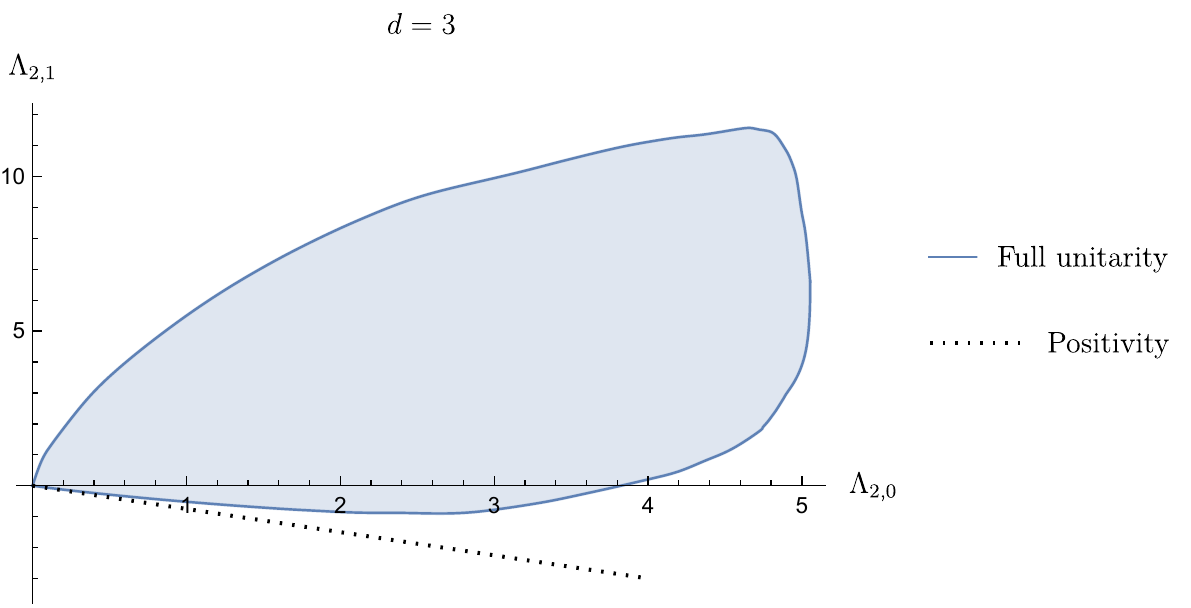}
		\caption{Nonperturbative bounds in $d=3$ on the observables $(\Lambda_{2,0},\, \Lambda_{2,1})$ defined in \eqref{eq:description_2} in the forward limit. The  ``whale shape'' allowed region  obtained using full unitarity is shaded in blue. This bound has only one kink at the origin. The dashed line represents the positivity bound given in \eqref{eq:Lambda21Lambda20}. The allowed region from positivity lies above the dashed line.  The positivity bound is much weaker than the full unitarity bound. }
		\label{fig:FL3}
		
		\vspace{15mm}
		
		\includegraphics[width=0.9\textwidth]{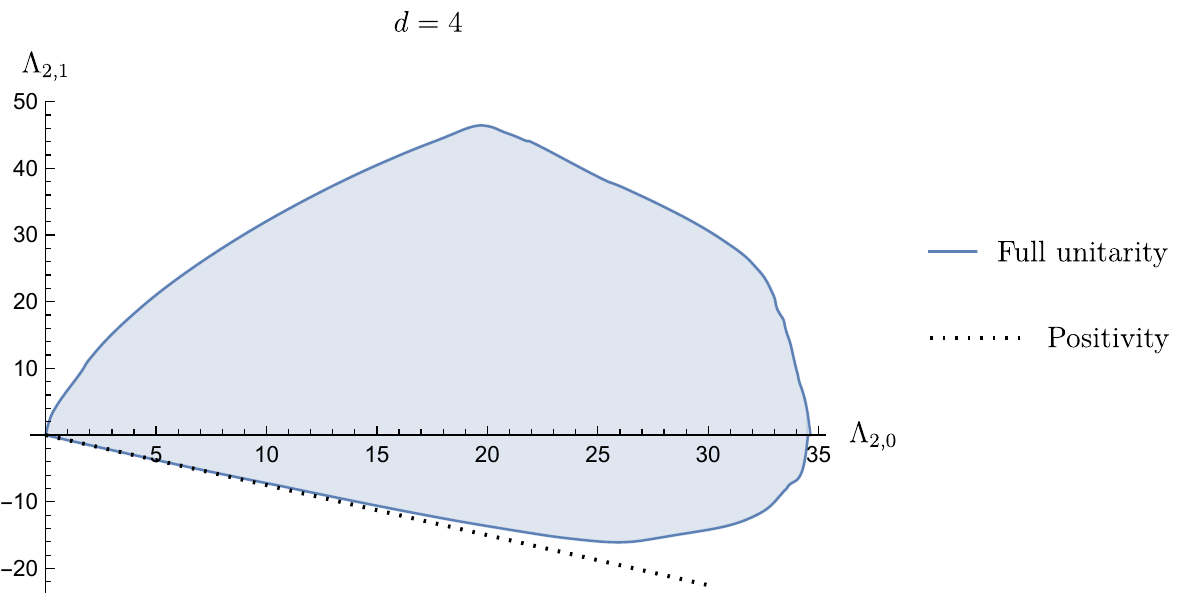}
	\caption{Nonperturbative bounds in $d=4$ on the observables $(\Lambda_{2,0},\, \Lambda_{2,1})$ defined in \eqref{eq:description_2} in the forward limit. The allowed region obtained using full unitarity is shaded in blue. This bound has two kinks, one at the origin and one on the upper edge. The dashed line represents the positivity bound given in \eqref{eq:Lambda21Lambda20}. The allowed region from positivity lies above the dashed line.  The positivity bound is much weaker than full unitarity bound.}
		\label{fig:FL4}
	\end{center}
\end{figure}

The bounds on $(\lambda_{2,0},\, \lambda_{2,1})$ in $d=3$ and $d=4$  are presented in figures \ref{fig:crossing_symmetric3} and \ref{fig:crossing_symmetric4}, respectively. The allowed region is shaded in blue. These have a very elongated shape with two kinks/tips on the very left and very right. Under closer inspection, one sees that the bottom boundaries are smooth. The bounds $(\Lambda_{2,0},\, \Lambda_{2,1})$ in $d=3$ and $d=4$ are presented in figures \ref{fig:FL3} and \ref{fig:FL4}, respectively. The allowed region is also shaded in blue. In $d=3$, the bound has a single kink (left tip), whereas, in $d=4$, the bound has two kinks. The kink on the upper boundary of the $d=4$ plot simply corresponds to the right tip in figure \ref{fig:crossing_symmetric4}. 

\begin{figure}[!htb]
	\begin{center}
		\includegraphics[width=0.9\textwidth]{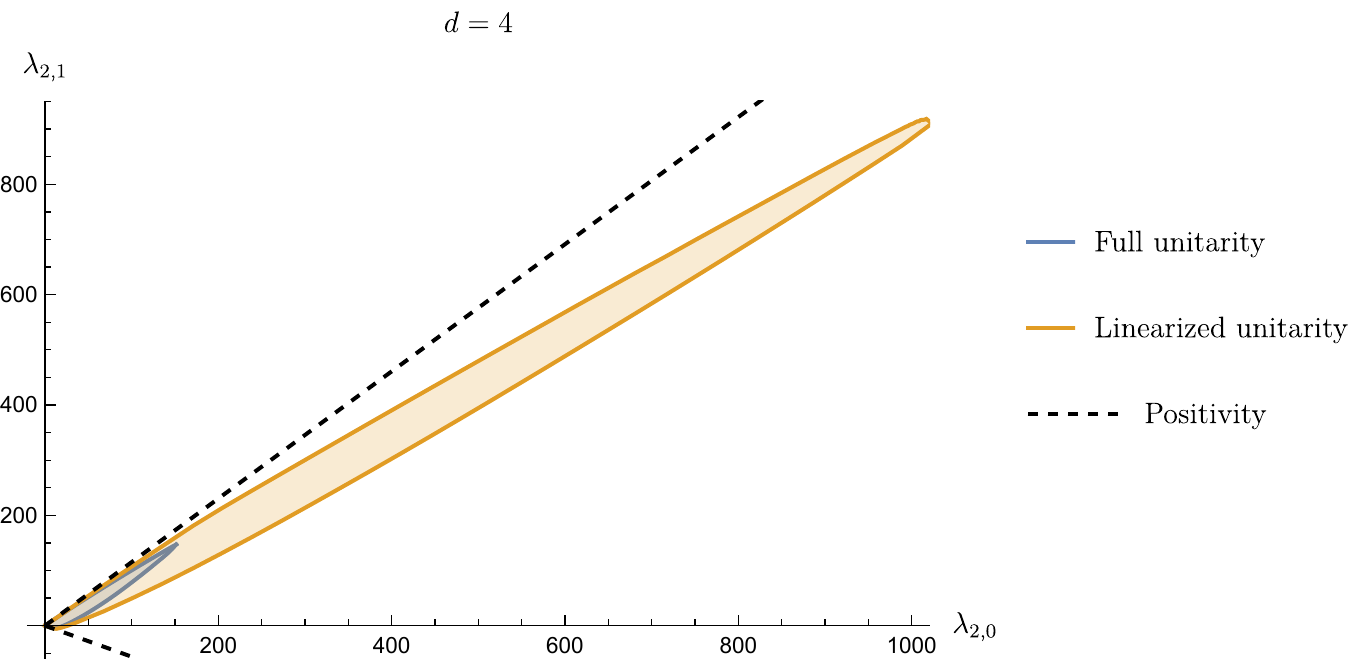}
		\caption{Nonperturbative bounds in $d=4$ on the observables $(\lambda_{2,0} , \lambda_{2,1})$ defined in \eqref{eq:description_1} at the crossing symmetric point with various amount of unitarity imposed. Black dashed lines indicate positivity bounds, with the allowed region lying in the cone between the two lines. The region shaded in yellow is the allowed region obtained using linearized unitarity only. The region shaded in blue is the allowed region obtained using full non-linear unitarity. The plot is constructed with $N_\textrm{max}=20$. Here, we do not perform  the extrapolation with $N_\textrm{max}$.}
		\label{fig:linearizedUnitarity_NP}
	\end{center}
\end{figure}

Let us now discuss how our bounds compare with the ones obtained by using only linearized unitarity or positivity. First of all, full non-linear unitarity gives bounds on the observables $\lambda_{0,0}$ and $\Lambda_{0,0}$, whereas linearized unitarity or positivity does not. Several bounds using positivity were obtained in section \ref{sec:analytic_d>2}. We found in section \ref{sec:analytic_d>2} analytically that $\lambda_{2,0}\geq 0$ and $\Lambda_{2,0}\geq 0$, which is consistent with figures \ref{fig:crossing_symmetric3} - \ref{fig:linearizedUnitarity_NP}. More interesting bounds given by equations \eqref{eq:lambdaTwoSided} and \eqref{eq:Lambda21Lambda20} are depicted by the black dashed lines in figures \ref{fig:crossing_symmetric3} - \ref{fig:linearizedUnitarity_NP}, we see that they are weaker than the bounds from full unitarity, but agree very well at small $\lambda_{2,0}$ or $\Lambda_{2,0}$, as expected. Finally, in figure \ref{fig:linearizedUnitarity_NP}, we display three bounds together obtained by using positivity only (black dashed line), linearized unitarity (yellow region), and full non-linear unitarity (blue region) for $d=4$. We can clearly see how much stronger the bounds become when we require more unitarity.

\begin{figure}[htb!]
	\begin{center}
		\includegraphics[width=0.60\textwidth]{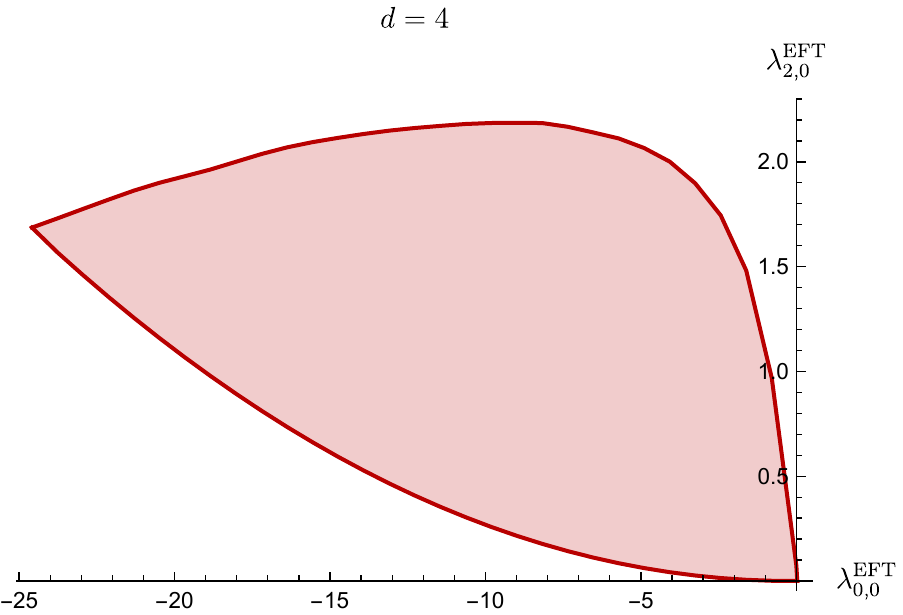}
		\caption{Bound in $d=4$ on the observables $(\lambda^\text{EFT}_{0,0}, \lambda^\text{EFT}_{2,0})$ of EFT amplitudes defined in \eqref{eq:description_EFT} using full non-linear unitarity. The allowed region is shaded in red.}
		\label{fig:EFT_0}
	\end{center}
\end{figure}

\subsection{EFT amplitudes}
\label{sec:EFTA}
Let us now address bounds on EFT amplitudes with $m=0$ and a branch cut starting from the cut-off scale $M$. Please recall that their observables are defined in \eqref{eq:description_EFT}. Here we do not perform a systematic study of EFT amplitudes in various space-time dimensions $d$ and focus instead on $d=4$ only. 

We start by bounding the observable $(\lambda^\text{EFT}_{0,0}, \lambda^\text{EFT}_{2,0})$. The result is given in figure \ref{fig:EFT_0}, the allowed region of parameters is shaded in red. Such type of bounds is inaccessible to techniques that use linearized unitarity or positivity as explained at the end of section \ref{sec:PNA}. An interesting point useful for the discussion in the next section is $\lambda^\text{EFT}_{0,0}=0$. For this particular value of $\lambda^\text{EFT}_{0,0}$, we obtain the bounds on $\lambda_{2,0}^\textrm{EFT}$ at different values of $N_\textrm{max}$ and extrapolate to $N_\textrm{max}=\infty$. We got 
\begin{equation}
	\label{eq:l20_EFT}
\textrm{If } \lambda^\text{EFT}_{0,0}=0:\qquad
0\leq\lambda^\text{EFT}_{2,0}\leq 0.073.
\end{equation}

\begin{figure}[t!]
	\begin{center}
	\includegraphics[width=0.65\textwidth]{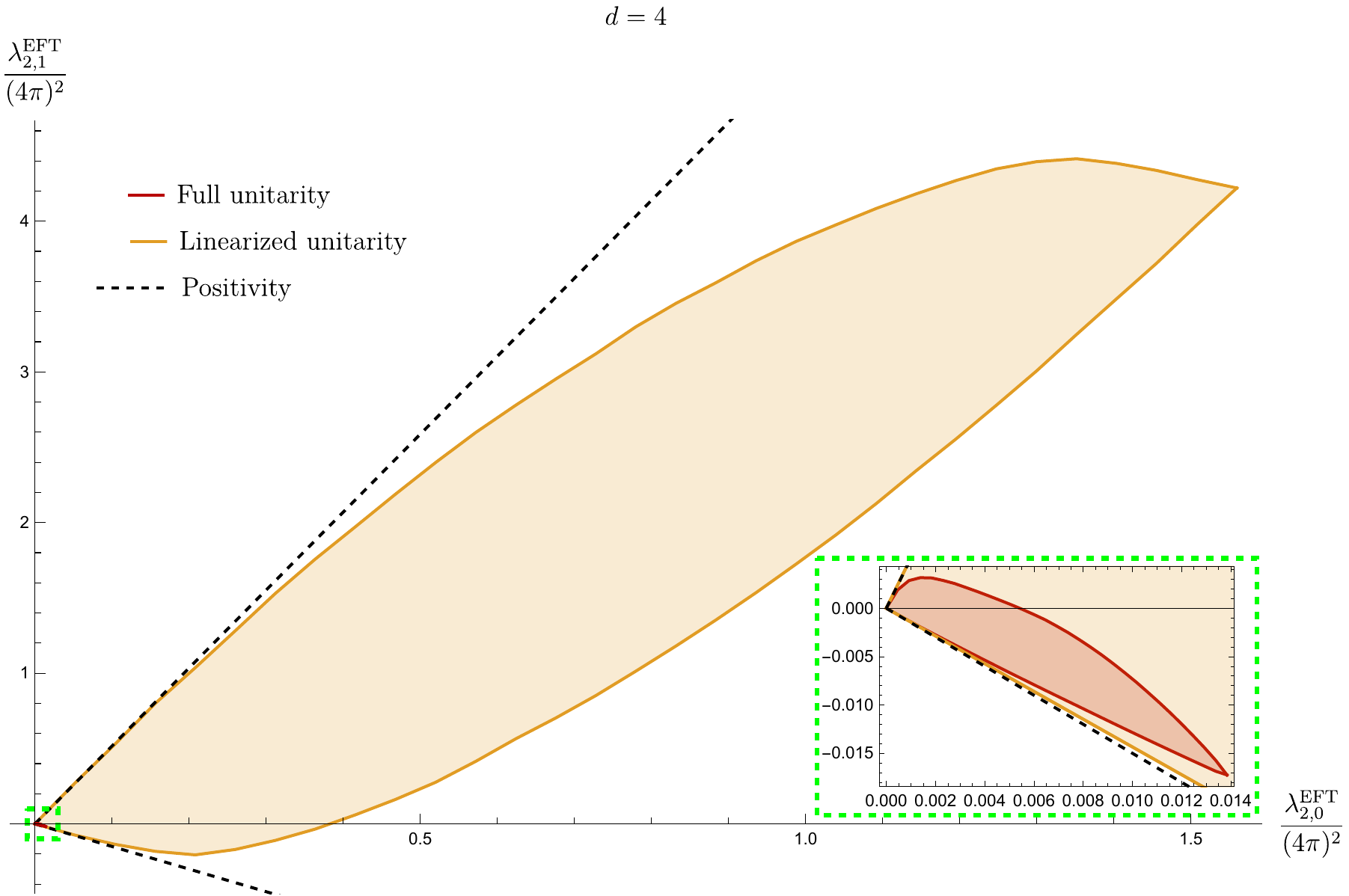}
		
		\caption{Bounds in $d=4$ on the observables $(\lambda^\text{EFT}_{2,0}, \lambda^\text{EFT}_{2,1})$ of EFT amplitudes defined in \eqref{eq:description_EFT}. The allowed region built with linearized unitarity is shaded in yellow. The allowed region built with full non-linear unitarity is shaded in red.  The black dashed lines are bounds obtained in \cite{Caron-Huot:2020cmc} using positivity only, which we summarized in \eqref{eq:positivity_bound_EFT}. Notice the re-scaling of the axes labels by $(4\pi)^2$ compared to all the other plots. This plot is in perfect agreement with figure 5 in \cite{Chiang:2022ltp}. The inset at the right corner is the zoomed version of the region near the origin. It will be presented again in figure \ref{fig:EFT_2}. }
		\label{fig:EFT_1}
	\end{center}
\end{figure}

\begin{figure}[!htb]
\centering
\includegraphics[width=0.8\textwidth]{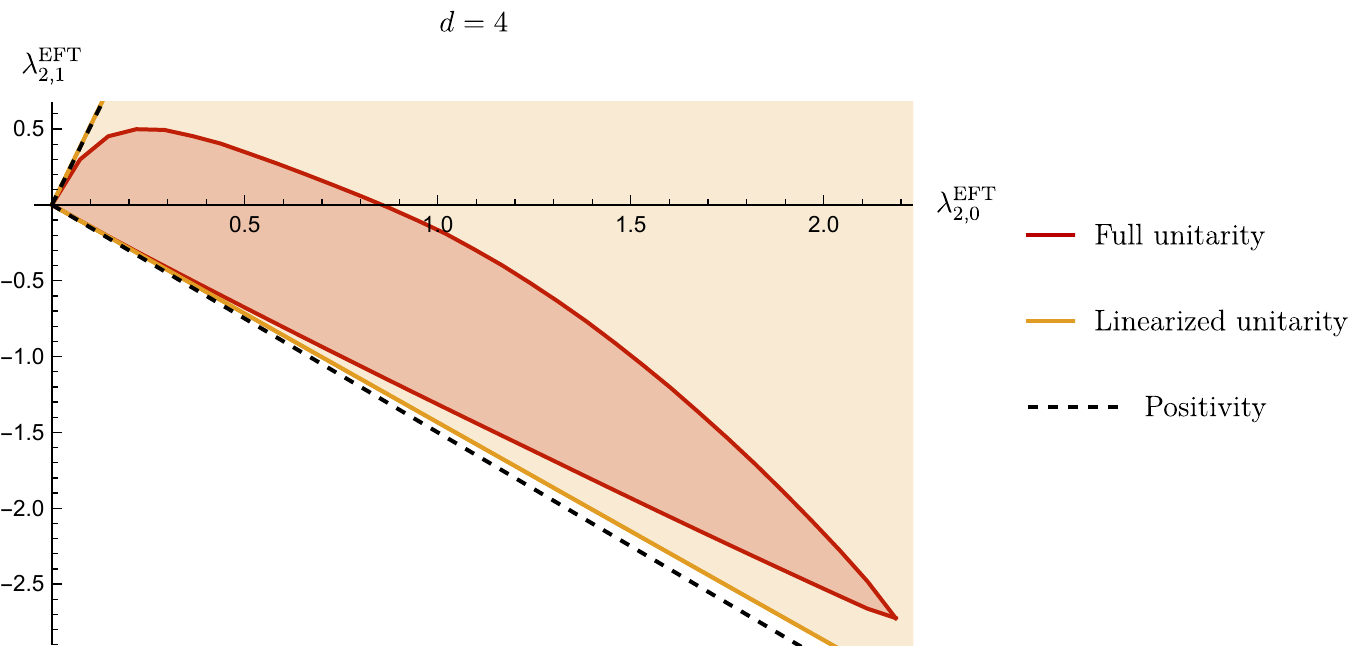}
		\caption{Zoomed version of figure \ref{fig:EFT_1} around the origin. The black dashed lines are bounds obtained in \cite{Caron-Huot:2020cmc} using positivity only, which we summarized in   \eqref{eq:positivity_bound_EFT}. The yellow region is obtained with linearized unitarity. The red region is obtained with full non-linear unitarity. Notice that we do not include the $(4\pi)^2$ re-scaling of the axes labels compared to figure \ref{fig:EFT_1}. }
		\label{fig:EFT_2}
\end{figure}

The bound on $(\lambda^\text{EFT}_{2,0}, \lambda^\text{EFT}_{2,1})$ is given in figures \ref{fig:EFT_1} and \ref{fig:EFT_2}, where we compare the the allowed region (shaded in yellow) obtained using  linearized unitarity only and the allowed region (shaded in red)  obtained using full non-linear unitarity. Notice that the axes labels in figures \ref{fig:EFT_1} and \ref{fig:EFT_2} have a relative factor of $(4\pi)^2$. The comparisons in these two plots make it clear how much stronger the bound obtained using full non-linear unitarity is than the one obtained using linearized unitarity only.

The black dashed lines in figures \ref{fig:EFT_1} and \ref{fig:EFT_2} are the positivity bounds obtained in \cite{Caron-Huot:2020cmc}. The relation between the notation of our observables defined in \eqref{eq:description_EFT} and the ones of  \cite{Caron-Huot:2020cmc} can be easily established using appendix \ref{app:polynomials}. Comparing the first line of our \eqref{eq:expansion_amplitude} (there we need to replace $m$ by $M$ and we need to set $m=0$ in \eqref{eq:Mandelstam_hat_1}) with their equation (2.3), we get the relationships between their $g_i$ and our $\lambda_{k,l}^\textrm{EFT}$:
\begin{equation}
	g_2 = \frac{\lambda^\text{EFT}_{2, 0}}{2}M^{-d},\qquad
	g_3 =-\lambda^\text{EFT}_{2,1}M^{-d-2},\qquad
	g_4 =  \frac{\lambda^\text{EFT}_{2,2}}{12}M^{-d-4}, \qquad \ldots
\end{equation}
Thus, the very first bound in equation (4.2) in \cite{Caron-Huot:2020cmc} given by
\begin{equation}
	-10.346 \leq \frac{g_3 M^2}{g_2}\leq 3
\end{equation}
in our notation reads as
\begin{equation}
	\label{eq:positivity_bound_EFT}
	-3/2 \leq \frac{\lambda^\text{EFT}_{2,1}}{\lambda^\text{EFT}_{2,0}}\leq 5.173.
\end{equation}
The positivity bound \eqref{eq:positivity_bound_EFT} is consistent with our numerical bounds given in figures \ref{fig:EFT_1} and \ref{fig:EFT_2}. 

The bound in figure \ref{fig:EFT_1} was already obtained in figure 5 in \cite{Chiang:2022ltp} using a different technique. To match the notation between our and their work, we can compare our equation \eqref{eq:observables_EFT} and their equation (1.4). We conclude that their $g_{i,j}$ are related to our $\lambda_{k,l}^\textrm{EFT}$ by \footnote{In writing these relations, we assumed that the authors use conventions where their coefficients $g_{k,l}$ are dimensionless and they set $M=1$ in their equation (1.4).}
\begin{equation}
	g_{2,0} = \lambda^\text{EFT}_{20},\qquad
	g_{3,1} = \lambda^\text{EFT}_{2,1},\qquad
	g_{4,2} = \lambda^\text{EFT}_{2,2},\qquad\ldots
\end{equation}
We see that our result in figure \ref{fig:EFT_1} is in perfect agreement with their figure 5.

\section{Full unitarity constraints on EFTs} 
\label{sec:EFTs}
We have derived various novel bounds in section \ref{sec:numeric_bounds}. Let us now show how they can be used for bounding effective field theories on a particular example of pseudo-Goldstone bosons.  There will be two key differences in this analysis from the more standard EFT constraints from dispersion relations.  The first is that by imposing the full unitarity constraint, we obtain bounds on the contact term $\sim \phi^4$, parameterized by $\lambda_{0,0}$ or $\Lambda_{0,0}$, which at tree-level contributes only to the real part of the amplitude and therefore is not constrained by positivity or linearized unitarity.  The second is that we can keep track of the finite, nonzero value of the mass $m$, rather than taking the limit $m\rightarrow 0$ from the beginning.  We will see that our full unitarity bounds on the leading derivative interaction $\sim (\partial \phi)^4$ depends in an interesting way on the size of the contact term.\footnote{The 2d version of this argument, which is slightly cleaner, is covered in section \ref{sec:2dEFT}.}

So, consider the effective theory of a pseudo-Goldstone boson, and take $M$ to be the cutoff.  Up to field redefinitions,  there are two quartic  interactions with up to  four derivatives:

\begin{equation}
	\CL_{\rm EFT} = -\frac{1}{2} (\partial\phi)^2 - \frac{1}{2}m^2\phi^2+
	\frac{1}{M^{d}} (-a m^4 \phi^4 + b (\partial \phi)^4 +\ldots ),
	\label{eq:SimpleEFTLag}
\end{equation}
where  $a$ and $b$ are real dimensionless parameters and $m$ is the mass of the pseudo-Goldstone boson.
Dots denote interactions that should be higher order in $1/M$. When $m=0$,  we recover the case of the true Goldstone boson. 

Using the Lagrangian density \eqref{eq:SimpleEFTLag}, it is straightforward to compute the interacting part of the scattering amplitude. At tree level, it reads as
\begin{multline}\label{eq:three_level_amplitude}
	m^{d-4}	\mathcal{T}(s,t,u) = \left(\frac{m}{M}\right)^d \times\\
	 \left( -24 a
	+ 2b m^{-4}\left((s-2m^2)^2 + (t-2m^2)^2 + (u-2m^2)^2\right)+\ldots\right).
\end{multline}
Using \eqref{eq:three_level_amplitude} and the definition \eqref{eq:PA_general_d}, we can also compute the spin-zero EFT partial amplitude. It reads as
\begin{equation}
	m^{d-4}\CT_0(s) =C \left(\frac{m}{M}\right)^d\times
	\left( (-6 a(d-1) + b d) + b(d-2) \left(\frac{s}{m^2}-2\right)+\ldots\right),
\end{equation}
where the constant $C$ is simply defined as
\begin{equation}
	 C\equiv \frac{\pi^{\frac{3-d}{2}}}{2^{d-3}\Gamma(\frac{1+d}{2})}.
\end{equation}

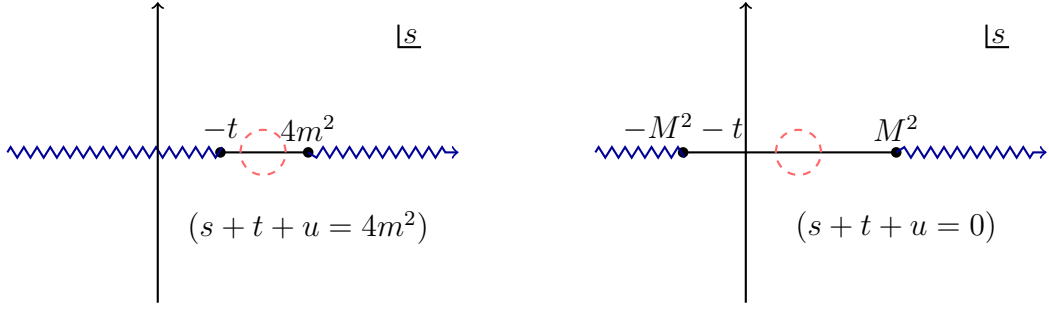
\begin{figure}[tb!]
	\begin{center}
		
		\begin{tikzpicture}[thick]
			
			\node (label) at (2,-1) {$(s+t+u=4m^2)$};
			
			\draw (3.2,1.7) -- (3.2,1.4) -- (3.5,1.4) ;
			\node (label) at (3.36,1.55) {$s$};
			
			\fill(5/6, 0) circle (2pt) node[above] (l_branch) {$-t$} (2, 0) circle (2pt) node[above] (r_branch) {$4m^2$};
			
			
			\draw	[->,decorate,decoration={zigzag,segment length=5,amplitude=2,pre=lineto,pre length=0,post=lineto,post length=3}, blue!60!black]  (-2, 0) -- (l_branch.south) (r_branch.south) -- (4, 0) node [above left, black] {};
			
			\draw(l_branch.south) -- (r_branch.south);
			\draw[->] (0, -2) -- (0, 2) node[below left=0.1] {};
			
			\draw[color=red!60, thick, dashed](1.4,0) circle (0.3);
			
		\end{tikzpicture}
		\qquad        \qquad	
		\begin{tikzpicture}[thick]
			
			\node (label) at (2,-1) {$(s+t+u=0)$};
			
			\draw (3.2,1.7) -- (3.2,1.4) -- (3.5,1.4) ;
			\node (label) at (3.36,1.55) {$s$};
			
			\fill(-5/6, 0) circle (2pt) node[above] (l_branch) {$-M^2-t$} (2, 0) circle (2pt) node[above] (r_branch) {$M^2$};
			
			\draw	[->,decorate,decoration={zigzag,segment length=5,amplitude=2,pre=lineto,pre length=0,post=lineto,post length=3}, blue!60!black]  (-2, 0) -- (l_branch.south) (r_branch.south) -- (4, 0) node [above left, black] {};
			
			\draw(l_branch.south) -- (r_branch.south);
			\draw[->] (0, -2) -- (0, 2) node[below left=0.1] {};
			
			\draw[color=red!60, thick, dashed](0.7,0) circle (0.3);
			
		\end{tikzpicture}
		\caption{Analytic structure in the $s$ complex plane for a fixed value of $t$ of two classes of amplitudes considered in the literature: the left plot is for the nonperturbative amplitude and the right one is for the EFT amplitude. The red dashed circle schematically indicates the region where we evaluate our physical observables $\lambda_{k,l}, \Lambda_{k,l}, \tau_{j;l}$; throughout this region,  the tree level expression \eqref{eq:three_level_amplitude} is a good approximation for the full non-perturbative amplitude.}
		\label{fig:EFT_amp}
	\end{center}
\end{figure}

We will only ever use the EFT amplitude (\ref{eq:three_level_amplitude}) to evaluate observables at the scale $s \sim m^2$, within the region indicated in the left plot of figure \ref{fig:EFT_amp}.  This region is deeply inside the controlled EFT regime if $m\ll M$, where the tree level amplitude \eqref{eq:three_level_amplitude} is a good approximation to the full non-perturbative amplitude.     Consequently, loop corrections to our observables arise at $\CO(\frac{m^{2d}}{M^{2d}})$ and are negligible subleading corrections compared to the tree-level contributions.
We can thus plug (\ref{eq:three_level_amplitude}) into the definitions of our non-perturbative observables \eqref{eq:description_1}, \eqref{eq:description_2} and \eqref{eq:description_3} and relate them with the parameters $a$ and $b$ in the EFT Lagrangian density. We simply get
\begin{equation}
	\lambda_{0,0} = \frac{8}{3} (b-9a) \left(\frac{m}{M}\right)^d, \qquad
    \lambda_{2,0} = 4 b \left(\frac{m}{M}\right)^d,
    \label{eq:EFTlambdas_1}
\end{equation}
\begin{equation}
	\Lambda_{0,0} = 8(b-3a)\left(\frac{m}{M}\right)^d,\qquad
	\Lambda_{2,0} = 4b\left(\frac{m}{M}\right)^d,
	\label{eq:EFTlambdas_2}
\end{equation}
\begin{equation}
	\tau_{0;0} = C \left(\frac{m}{M}\right)^d (bd-6a(d-1)),\qquad
	\tau_{0;1} = C \left(\frac{m}{M}\right)^d b(d-2).
		\label{eq:EFTlambdas_3}
\end{equation}

\begin{figure}[t]
	\begin{center}
		\includegraphics[width=0.42\textwidth]{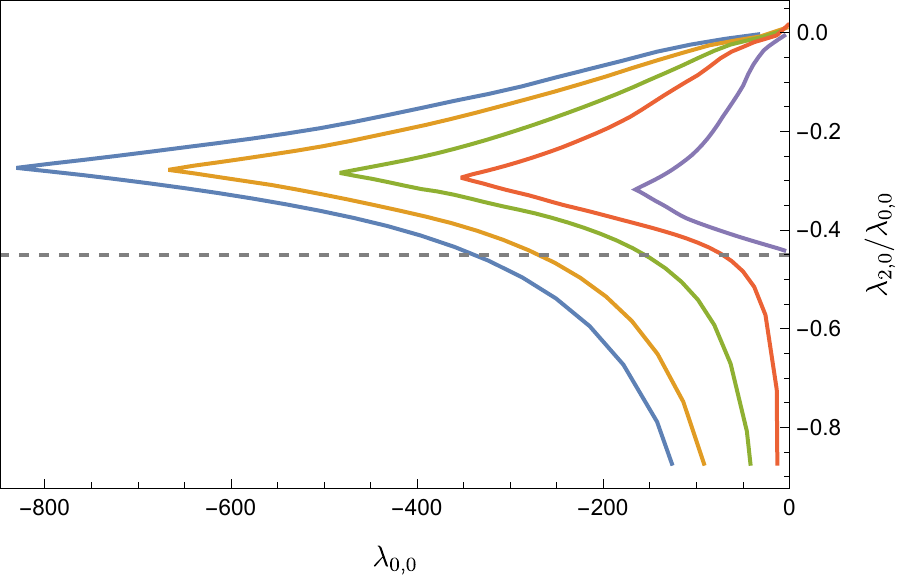}\qquad
		\includegraphics[width=0.515\textwidth]{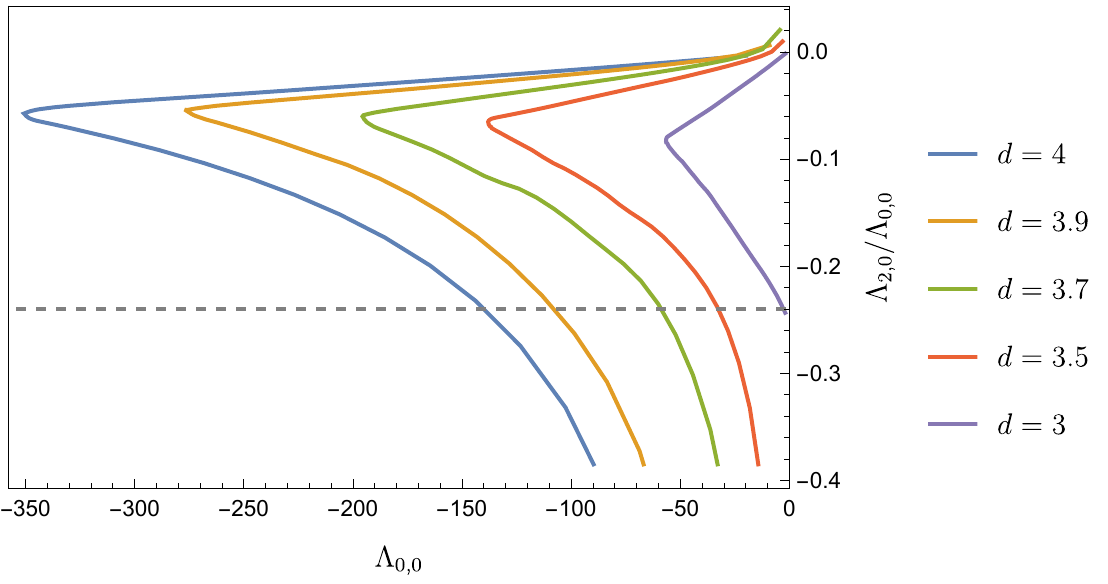}
		\caption{Bounds on $\lambda_{2,0}/\lambda_{0,0}$ vs $\lambda_{0,0}$ (left plot) and bound on $\Lambda_{2,0}/\Lambda_{0,0}$  vs $\Lambda_{0,0}$ (right plot) in various space-time dimensions $d$.  The allowed regions are to the right of the curves. The gray dashed lines at $-0.45$ (left plot) and $-0.24$ (right plot) are added for reference. }
		\label{fig:L21-L20_ratio_Extrapolated}
	\end{center}
\end{figure}

\begin{figure}[t]
	\begin{center}
		\includegraphics[width=0.7\textwidth]{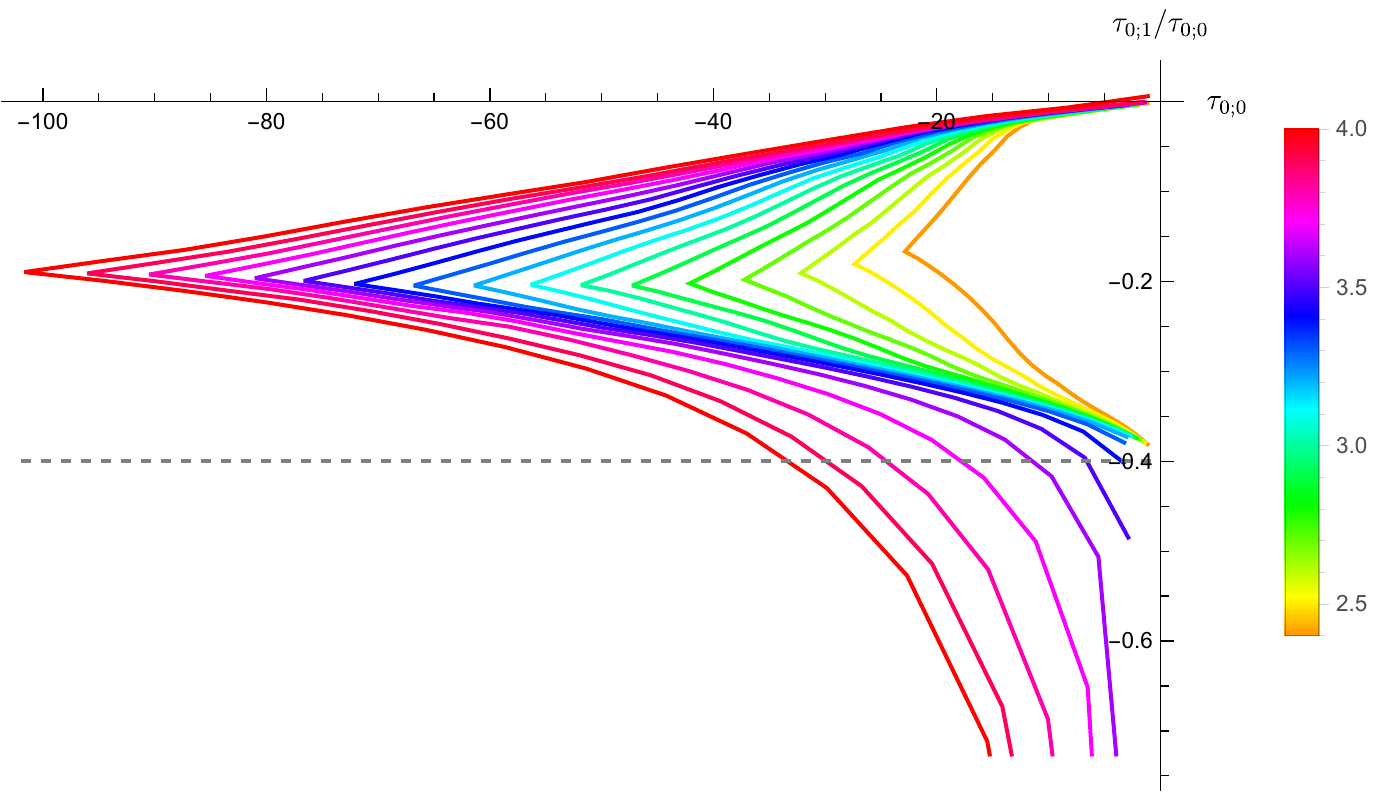}
		\caption{Bounds on the $\tau_{0 ; 1} / \tau_{0 ; 0}$ vs $\tau_{0,0}$.
			Different colors represent different spacetime dimension $d$. The allowed regions are to the right of the curves. The gray dashed line at $-0.4$ is added for reference.}
		\label{fig:PA0_dPA0_ratio_Extrapolated}
	\end{center}
\end{figure}

Let us consider now ratios of these observables, the prefactor $(m/M)^d$ cancels out: 
\begin{equation}
	\label{eq:relations_obs_ab}
	\frac{\lambda_{2,0}}{\lambda_{0,0}} = \frac{3b}{2(b-9a)},\qquad
	\frac{\Lambda_{2,0}}{\Lambda_{0,0}} = \frac{b}{2(b-3a)},\qquad
	\frac{\tau_{0;1}}{\tau_{0;0}} = \frac{b(d-2)}{bd-6a(d-1)}.
\end{equation}
The bounds on these quantities were given in figures \ref{fig:crossing_symmetric}, \ref{fig:plot_forward} and \ref{fig:PA0_dPA0_Extrapolated} respectively. We re-plot them in a more convenient way for our current purposes in figures \ref{fig:L21-L20_ratio_Extrapolated} and \ref{fig:PA0_dPA0_ratio_Extrapolated}. We emphasize that the only assumption about branch cuts that we used in the bounds in these figures was that there is no branch cut below the multi-particle threshold at $4m^2$.  In particular, we did not assume that branch cuts from loops start at $s \sim M^2$.  Nevertheless, we obtain non-trivial bounds on $a$ and $b$. Let us focus on $d=3$ for concreteness. From figures \ref{fig:L21-L20_ratio_Extrapolated} and \ref{fig:PA0_dPA0_ratio_Extrapolated}, we conclude that
\begin{equation}
	d=3:\qquad
	-0.45\lesssim\frac{\lambda_{2,0}}{\lambda_{0,0}} \leq 0,\qquad
	-0.24\lesssim \frac{\Lambda_{2,0}}{\Lambda_{0,0}} \leq 0,\qquad
	-0.40 \lesssim \frac{\tau_{0;1}}{\tau_{0;0}}  \leq 0.
\end{equation}
Assuming that $b\geq 0 $ without loss of generality and taking into account \eqref{eq:relations_obs_ab} we obtain the following bounds from the above inequalities
\begin{equation}
	\label{eq:final_bounds}
	d=3:\qquad
	\frac{b}{a} \lesssim 2.07692,\qquad
	\frac{b}{a} \lesssim 0.972973, \qquad
	\frac{b}{a} \lesssim 2.18182,
\end{equation}
respectively. As a result, we arrive at the final bound in $d=3$ which reads as
\begin{equation}
	d=3:\qquad
	0\leq b\lesssim a.
\end{equation}

Completely analogously, one can derive bounds on $a$ and $b$ in $d<3$ space-time dimensions. In the case of $d>3$, the situation is slightly different. From  figures \ref{fig:L21-L20_ratio_Extrapolated} and \ref{fig:PA0_dPA0_ratio_Extrapolated} we see that the bounds become infinitely weak the closer we approach the origin. In order to put a bound on ratios \eqref{eq:relations_obs_ab}, we need to fix the value of $\lambda_{0,0}$, $\Lambda_{0,0}$ or $\tau_{0;0}$. Let us consider for example figure \ref{fig:PA0_dPA0_ratio_Extrapolated}. For some fixed value of $\tau_{0;0}$ there will always exist a coefficient $\alpha$ such that
\be
\frac{\tau_{0;1}}{\tau_{0;0}}> \alpha.
\ee
Using \eqref{eq:relations_obs_ab}, we conclude that
\begin{equation}
	0 \leq \frac{b}{a} <  \frac{6 \alpha (d-1)}{\alpha d+2-d}.
\end{equation}
This result gives precisely the last inequality in \eqref{eq:final_bounds} if we set $d=3$ and $\alpha=-0.4$.

These bounds require an important caveat.  To understand why, note that in figures  \ref{fig:crossing_symmetric}, \ref{fig:plot_forward} and \ref{fig:PA0_dPA0_Extrapolated}, the location of the tip near the origin changes as a function of $d$, and as $d$ increases from $d=2$ to larger values, the tip moves farther away from the origin. For concreteness, let us focus on figure \ref{fig:PA0_dPA0_Extrapolated}.  In this case, we have shown the location of the tips explicitly in figure \ref{fig:PA0_dPA0_left_tip}.  This means that for any $d>2$, strictly at $m/M=0$, the bound on the ratio $\tau_{0;1}/\tau_{0;0}$ completely disappears; the same is true if $m/M$ is sufficiently small.  So we apparently have an interesting bound that applies even when $m$ is orders of magnitude smaller than $M$, but that disappears in the $m \rightarrow 0$ limit.  

\subsubsection*{Massless Case}

In section \ref{sec:EFTA}, we performed a different numeric bootstrap analysis that tries to take into account more of the expected analytic structure of EFT amplitudes, by assuming that branch cuts in the amplitude can be neglected up to the UV cutoff scale $M$.  This situation is depicted schematically in the right plot of figure \ref{fig:EFT_amp}.  In this case, it is possible to take the limit $m \rightarrow 0$ in the definition of our physical observables, as we did for  $\lambda_{k,l}^{\rm EFT}$ in \eqref{eq:description_EFT}.

Using the massless limit $m=0$ (true Goldstone bosons) of the amplitude \eqref{eq:three_level_amplitude}, one obtains for the observables \eqref{eq:description_EFT} that
\begin{equation}
	\lambda^\text{EFT}_{00} = 0,\qquad
	\lambda^\text{EFT}_{20} = 4b.
\end{equation}
The case $\lambda_{00}^{\rm EFT}=0$ is exactly the case we considered in equation \eqref{eq:l20_EFT}, from which we immediately conclude that in $d=4$
\begin{equation}
	d=4,\;\;m=0:\qquad
	0\leq b \leq 0.01825.
\end{equation}

\section{The $\phi^4$ model}\label{sec:phi4}
It is interesting to place some models in the allowed space of parameters constructed in section \ref{sec:numeric_bounds}. In what follows we will consider the perturbative prediction in $\lambda \phi^4$ theory, in the coupling, and in the $\epsilon$ expansion. In $d=2$, this was done non-perturbatively in  \cite{Chen:2021pgx,chen:2021bmm}.

Let us recall the action of the $\phi^4$ model in general dimensions. The renormalized action reads
\begin{equation}
	S= \int d^d x \left(
	-\frac{1}{2} Z_\phi (\partial \phi)^2-\frac{1}{2}Z_m m^2\phi^2-\frac{1}{4!}Z_\lambda \bar{\lambda} \mu^{\epsilon} \phi^4
	\right),
\end{equation} 
where $m$ is the physical mass and $\bar{\lambda}$ is a renormalized dimensionless coupling constant.

The one-loop computation of the two-to-two scattering amplitude is a textbook result:
\begin{equation}
	\label{eq:amplitude_phi4}
	m^{-\epsilon}\mathcal{T}(s,t,u) = -\left( \frac{\mu}{m} \right)^\epsilon \bar\lambda \times
	\Big(Z_\lambda - \bar\lambda \times \left( \frac{\mu}{m} \right)^\epsilon \left(V(s)+V(t)+V(u)\right)\Big) + \CO(\bar{\lambda}^3),
\end{equation}
where  $V$ in the mostly plus metric reads as
\begin{equation}
	\label{eq:object_V}
	V(s) \equiv \frac{1}{2} \int_0^1 dx\, \frac{\Gamma(\epsilon/2)}{(4\pi)^{d/2}}
	\frac{1}{(1-x(1-x)\frac{s}{m^2})^{\epsilon/2}}.
\end{equation}

Next, we project onto partial waves. Using the definition of the partial amplitude \eqref{eq:Tj} and the explicit one loop expression \eqref{eq:amplitude_phi4}, we can write
\begin{eqnarray}
	m^{-\epsilon}\mathcal{T}_j(s) &= 
	-\left( \frac{\mu}{m} \right)^\epsilon\bar\lambda\left(Z_\lambda - \bar\lambda \left( \frac{\mu}{m} \right)^\epsilon V(s)\right) \frac{2^{2-d} \pi^{(3-d)/2}\delta_{j,0}}{\Gamma\left(\frac{d-1}{2}\right)} \nonumber \\
	&+ \left( \frac{\mu}{m} \right)^{2\epsilon} \bar\lambda^2  \left(
	\mathbf{\Pi}_j\left[ V(t)\right]+\mathbf{\Pi}_j\left[ V(u)\right]
	\right).
	\label{eq:Phi4OneLoop}
\end{eqnarray}
The second line in the above equation is easier to evaluate numerically. 

We still have to renormalize.  First, consider small, but otherwise generic, values of $\bar{\lambda}$ at finite $\epsilon$.  In this case, $\bar{\lambda} \propto \mu^{-\epsilon}$ depends on the RG scale $\mu$, which we have to choose.  A common and convenient choice is $\mu=m$, which we will adopt here as well.  We also have to choose a renormalization scheme.  We will take a physical renormalization scheme, where the counterterms cancel all loop corrections to the physical quantity $\CT_0(2m^2)$:
\begin{equation}
Z_\lambda -  \bar{\lambda} \times V(2m^2)\equiv 1 .
\end{equation}
The idea behind a physical renormalization scheme such as this is to improve the accuracy of the perturbative expansion at a given order, without knowledge of higher order corrections.  By definition, if the counterterm removes all loop corrections to $\CT_0(2m^2)$, then $\CT_0(2m^2)$ is exact at tree-level.  The observable $\partial_s \CT_0(2m^2)$ will still get corrections at all loop orders, but because its leading term is $\CO(\bar{\lambda}^2)$, its dependence on the counterterm shows up at $\CO(\bar{\lambda}^3)$. To the extent that these higher order corrections to $\partial_s \CT_0(2m^2)$ are 
correlated with the loop corrections to $\CT_0(2m^2)$, a physical scheme should minimize them as well.

\begin{figure}[t]
	\begin{center}
		\includegraphics[width=0.9\textwidth]{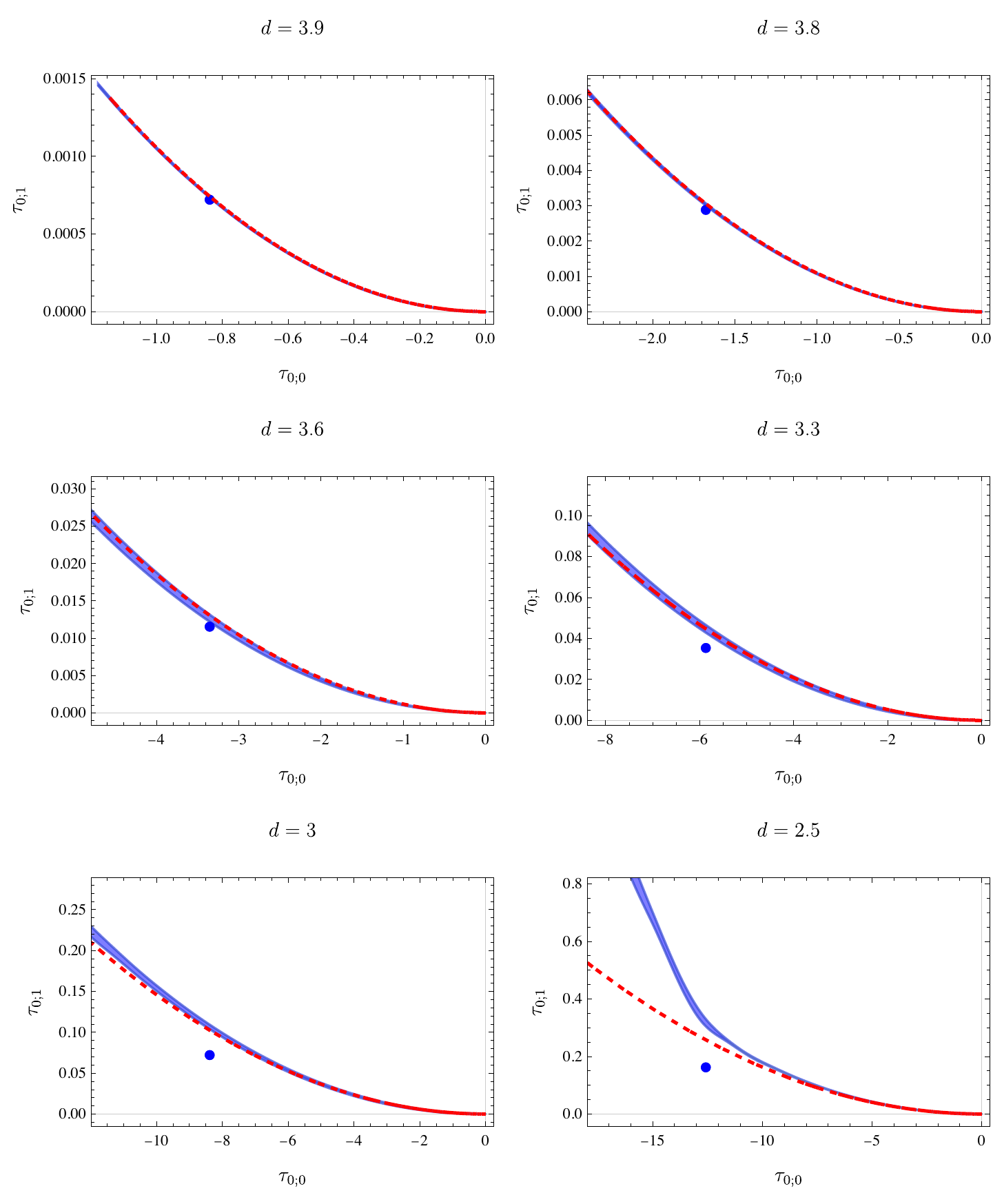}
		\caption{Comparison of the non-perturbative numerical bounds with $\phi^4$ perturbation theory. The blue bands are the numerical S-matrix bootstrap results extrapolated to $N_\textrm{max}\rightarrow\infty$, where the lower boundaries are determined by linear extrapolation $a + b/N_\textrm{max}$ while the upper boundaries are determined by quadratic extrapolation $a + b/N_\textrm{max}^2$. The red dashed lines are the one-loop result of the $\phi^4$ theory given in equation \eqref{eq:phi4OneLoop}. Here, for the $\tau_{0;0}$ values of the red dashed lines, we only include the $\bar{\lambda}$  term, while the $\tau_{0;1}$ values included both terms in \eqref{eq:phi4OneLoop}. The blue dots are the critical point of the $\phi^4$ theory given in equation \eqref{eq:phi4Critical}. }
		\label{fig:phi4Comparison}
	\end{center}
\end{figure}

With these choices, the one-loop amplitude is
\begin{equation}\label{eq:phi4OneLoop}
	m^{-\epsilon}\mathcal{T}_j(s) = 
	- \bar\lambda \frac{ \pi^{(3-d)/2}\delta_{j,0}}{2^{d-2}\Gamma\left(\frac{d-1}{2}\right)} +  \bar\lambda^2  \left(
	\mathbf{\Pi}_j\left[ V(t)\right]+\mathbf{\Pi}_j\left[ V(u)\right]
	\right).
\end{equation} 
In figure \ref{fig:phi4Comparison}, we show the comparison of this one-loop result to the numeric bootstrap bounds on $\tau_{0;0} \equiv m^{-\epsilon}\mathcal{T}_j(2m^2)$ and $\tau_{0;0} \equiv m^{2-\epsilon}\partial_s\mathcal{T}_j(2m^2)$.  Remarkably, at small values of the coupling, the perturbative amplitude lies essentially exactly along the bound, for any value of $d$.  At larger value of the coupling $\bar{\lambda}$, the perturbative prediction eventually deviates from the bound, and actually goes below it, into the non-unitary region, but the perturbative result is not reliable there, and higher-order corrections as well as non-perturbative ones should push the amplitude back up above the bound.\footnote{While it is known \cite{Hogervorst:2015akt} that the Wilson-Fisher fixed point is non-unitary in fractional dimensions, this violation occurs only at very high dimension operators and is unlikely to lead to a non-unitary scattering amplitude. }

We are particularly interested in the scattering amplitude in the vicinity of the critical point. That is, consider the limit where the coupling is tuned very close to its critical value, so that the mass gap is very small compared to the UV scale where the bare parameters are fixed.  This limit is equivalently described by the conformal fixed point theory deformed by the ``$\phi^2$''  operator, defined as the most relevant $\mathbb{Z}_2$ even operator in the conformal theory.  In the IR description, this limit has no free parameters because the coefficient of $\phi^2$ is uniquely fixed in terms of the physical mass $m$; in the UV description, there are no free parameters because $\bar{\lambda}$  is set to its critical value.  In general, this critical point is at strong coupling, but it can be computed perturbatively in an expansion in $\epsilon \equiv d-4$. We take the counterterm to have the form
\begin{equation}
Z_\lambda = \sum_{n=0}^\infty \frac{G_n(\bar{\lambda})}{\epsilon^n}, \qquad G_n(\bar{\lambda}) = \sum_{m=n}^\infty G_{nm} \bar{\lambda}^{m}, \qquad G_{00}\equiv 1.
\end{equation}
Expanding (\ref{eq:Phi4OneLoop}) to $\CO(\bar{\lambda}^2 \epsilon^0, \bar{\lambda} \epsilon,\bar{\lambda}^0 \epsilon^2 )$, we find
\begin{eqnarray}
\label{eq:PWCP}
\tau_{0;0} \equiv  m^{-\epsilon} \CT_0(2m^2) &=&  - \frac{\bar{\lambda}}{2\pi} \nonumber \\
 && + \frac{\bar{\lambda}}{32 \pi^3} \left( \left( - \epsilon + \frac{3 \bar{\lambda}}{(4\pi)^2} \right) \log \frac{\mu}{m} - (4\pi)^2 \bar{\lambda} G_{01} + \dots \right) \nonumber \\
 \tau_{0;1} \equiv m^{2-\epsilon} \partial_s \CT_0(2m^2) &=&   \bar{\lambda}^2 \frac{4+2\pi - \pi^2}{512 \pi^3} \end{eqnarray}
where $\dots$ is easily computed but we suppress it for now for conciseness.  At the critical point
\begin{equation}\label{eq:criticallamba}
\bar{\lambda} = \frac{(4\pi)^2}{3} \epsilon + \CO(\epsilon^2),
\end{equation} 
the dependence on $\mu$ vanishes.  Choosing $G_{01}$ to cancel the remaining terms in the second line of (\ref{eq:PWCP}), we have
\begin{eqnarray}\label{eq:phi4Critical}
\tau_{0;0} &=& -\frac{8 \pi \epsilon}{3}  \approx - 8.38 \epsilon \nonumber \\
\tau_{0;1} &=& -\frac{1}{18} \pi  (\pi^2 -2 \pi -4) \epsilon ^2 \approx 0.072 \epsilon^2.
\end{eqnarray}
In principle, we could continue this calculation to higher orders in $\epsilon$ in order to obtain a more accurate result for the amplitude in the vicinity of the fixed point.  Since the expansion is asymptotic, it would be interesting to try to compute to high orders in $\epsilon$ and Borel resum the result to get a prediction for $\epsilon \sim \CO(1)$.  In practice, such a calculation requires computing high order Feynman diagrams, which is beyond the scope of this paper, and we will have to content ourselves with this low-order result.   At this low order, the only practical difference between this expression and that in  (\ref{eq:phi4OneLoop}) is that we have substituted the perturbative $\CO(\epsilon)$ critical value of the coupling $\bar{\lambda}$, and truncated the loop integrals in an expansion in $\epsilon$ at $\CO(\epsilon)$, rather than using their finite $\epsilon$ values.  The resulting approximation for the  `fixed point' amplitude is indicated by a blue dot in figure \ref{fig:phi4Comparison}.  This result suggests that the amplitude in the vicinity of the fixed point lies on, or nearly on, the boundary of the allowed region. 

In fact, recall that in $d=2$, the fixed point amplitude and the bound can be determined analytically, and the fixed point does indeed lie exactly at the bound.  
Moreover, we  see that at $d$ close to 4, where the fixed point is weakly coupled and therefore under perturbative control, its amplitude still lies along the boundary.  Between these two limits, lacking a strongly coupled calculation of the fixed point amplitude, the best we can say is that figure \ref{fig:phi4Comparison} is consistent with, and suggestive of, the possibility that it continues to saturate the bound.

\begin{center}
\subsection*{Acknowledgments}
\end{center}

We thank Simon Caron-Huot, Andrea Guerrieri, Kelian H\"{a}ring, Ami Katz, Jo\~ao Penedones, Francesco Riva, Marco Serone, David Simmons-Duffin, and Matthew Walters for very helpful conversations. 

ALF and HC were supported in part by the US Department of Energy Office of Science under Award Number DE-SC0015845 and the Simons Collaboration Grant on the Non-Perturbative Bootstrap, and ALF in part by a Sloan Foundation fellowship. The work of DK is supported by the SNSF Ambizione grant PZ00P2\_193411. 

\appendix

\section{Warm-up in $d=2$}
\label{sec:2dwarmup}
In this appendix, we study S-matrix bootstrap in $d=2$. The $d=2$ case is much simpler than the $d>2$ case considered in the main text, and many things can be computed analytically, especially the bounds from the full unitarity constraint. Most of the results in this appendix are analogous to results in higher dimensions in the body of the paper, and we explicitly refer to the higher dimensional results throughout.  So, this appendix should be a useful starting point for the reader who wants to gain more intuition about how the S-matrix bounds work, without the technicalities of the numeric S-matrix bootstrap implementation.

The kinematics of $d=2$ forces $u=0$ and $t=4m^2-s$ for the interacting part of the scattering amplitude $\mathcal{T}(s,t,u)$. As a result, effectively, it is a function of a single variable $s$. Crossing requires
\begin{equation}
	\mathcal{T}(s) = \mathcal{T}(4m^2-s).
\end{equation} 
In $d=2$, there is no spin and as a result, there is only a single partial amplitude denoted by $\CS(s)$ which is simply related to the interacting part of the scattering amplitude as
\begin{equation}
	\label{eq:PA_2d}
	\mathcal{S}(s)=1+i\mathcal{N}_{2}^{-1}\mathcal{T}(s),\qquad \mathcal{N}_2\equiv2\sqrt{s(s-4m^2)}
\end{equation}
contrary to \eqref{eq:PA_general_d} for $d>2$. Due to the simplicity of \eqref{eq:PA_2d} the partial amplitude in $d=2$ has the same analytic and crossing properties as the interacting part of the amplitude $\mathcal{T}(s)$. Unitarity reads as
\begin{equation}
	|\CS(s)|^2 \le 1 \textrm{ for } s>4m^2.
\end{equation}
By a clever change of variables,
\begin{equation}
	z = \frac{ \sqrt{s (4m^2-s)} - 2m^2}{\sqrt{s(4m^2-s)} + 2m^2},
\end{equation}
crossing symmetry of $\CS(z)$ as a function of $z$ is automatic (since $z(4m^2-s) = z(s)$) and the branch cut at $s>4m^2$  is mapped to the boundary of the unit disk.  Since we are assuming that there are no bound states,\footnote{For the case with a bound state pole at $s_i$, one can  construct a function $f(z)$ that is analytic and $|f(z)|\le 1$ inside the unit disk $|z|\le 1$ simply by taking $f(z)\equiv \left( \frac{z-z_{i}}{1-\bar{z}_{i} z}\right) \mathcal{S}\left(z\right)$, where $z_i = z(s_i)$.  
	Since the factor $\frac{z-z_{i}}{1-\bar{z}_{i} z}$ is analytic and $|\frac{z-z_{i}}{1-\bar{z}_{i} z}|\le 1$ inside the unit disk, $f(z)$ is also analytic and $|f(z)|\le 1$  inside the unit disk, and we can now apply the maximum modulus principle and the Schwarz-Pick theorem to $f(z)$. Multiple poles can similarly be removed by including one such factor for each pole.
} $\CS(z)$ has no poles for $|z| < 1$, so analyticity is simply the statement that the function $\CS(z)$ is an analytic function on the disk $|z| < 1$.  Unitarity is simply the statement that $|\CS(z)|^2 \le 1$ on the boundary $|z|=1$.  It is possible to map out the space of functions $\CS(z)$ satisfying these properties using numeric bootstrap methods, by following the strategy explained in section \ref{sec:PNA} by choosing a basis of crossing-symmetric, analytic functions and then imposing unitarity on this space.  This is the strategy that we used in \cite{Chen:2021pgx},  and it is also how we obtained bounds in $d>2$ in this paper.  
However, an advantage of the $d=2$ case is that the problem of understanding the constraints on  $\CS(z)$ is a classic complex analysis problem, and much can be said without invoking numerics.  
As observed in \cite{EliasMiro:2019kyf}, the Schwarz-Pick theorem is a particularly powerful and elegant tool for this purpose.  The Schwarz-Pick theorem and various generalizations are simple applications of the maximum modulus principle for analytic functions (that is, the modulus $|f(z)|$ cannot have strict local maxima except at boundaries) and the fact that M\"obius transformations of the form 
\begin{equation}
	\varphi(z) \equiv \frac{z-z_0}{1-z \bar{z}_0}
	\label{eq:mobius}
\end{equation}
map the unit disk into itself if $|z_0| \le 1$.

\subsection{Nonperturbative bounds}
\subsubsection{Parameterizing the space of amplitudes}
In the $d=2$ case, a natural set of observables to use to parameterize the space of scattering amplitudes is the set of Taylor coefficients around the crossing symmetric point:\footnote{Notice a change of conventions compared to \cite{Chen:2021pgx}. More precisely $\Lambda_0\big|_\text{here}=-\Lambda_0\big|_\text{there}$, instead  $\Lambda_n\big|_\text{here}=\Lambda_n\big|_\text{there}$ for $n\geq 1$.}
\begin{equation}
	\Lambda_n=m^{2(n-1)}\lim _{s \rightarrow 2 m^2} \partial_{s}^{n} \mathcal{T}(s).
\end{equation}
These coefficients uniquely determine $\CT$ since the Taylor series converges in an open set around $s=2m^2$, from which $\CT(s)$ can be analytically continued. Moreover, the Taylor series in $s$ can easily be converted into the Taylor series in $z$, which converges on the `physical sheet' $|z|<1$.  For $d>2$, we need to enlarge the set of coefficients to take into account both $s$ and $t$, and there are more than one natural way to do this.  However, the basic idea is still to do a series expansion around a point that is a finite distance from any poles or branch cuts.

\subsubsection{Imposing bounds}
\begin{figure}[th]
	\begin{center}
		\includegraphics[width=0.4\textwidth]{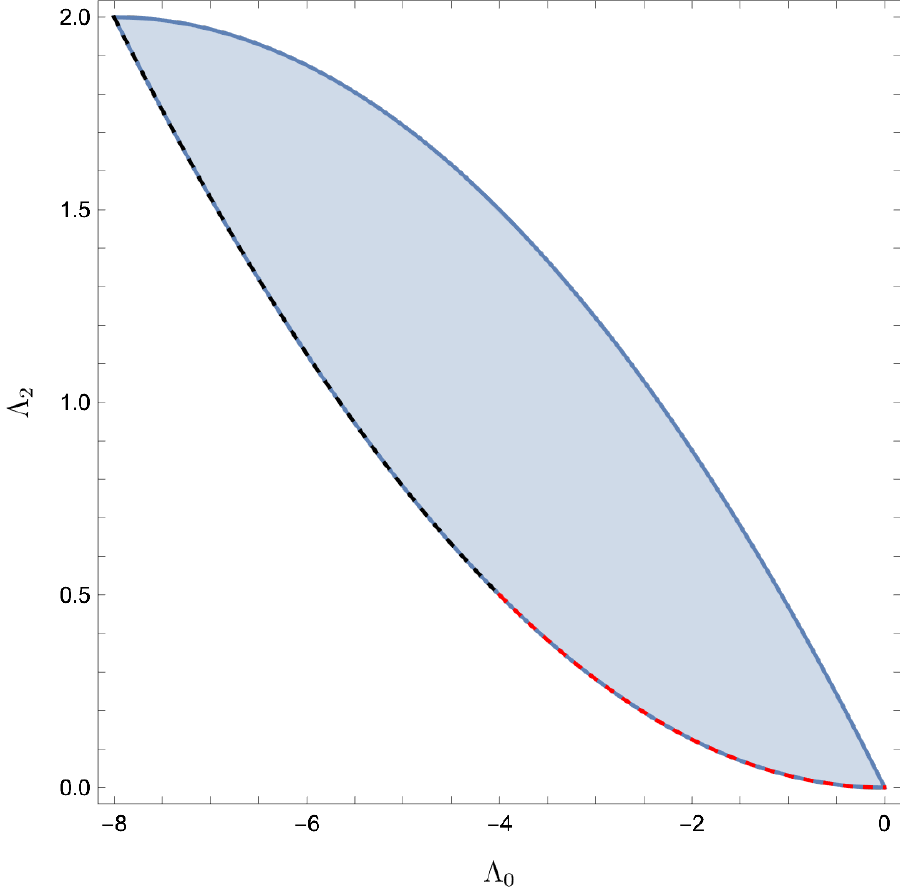}
		\qquad
		\includegraphics[width=0.4\textwidth]{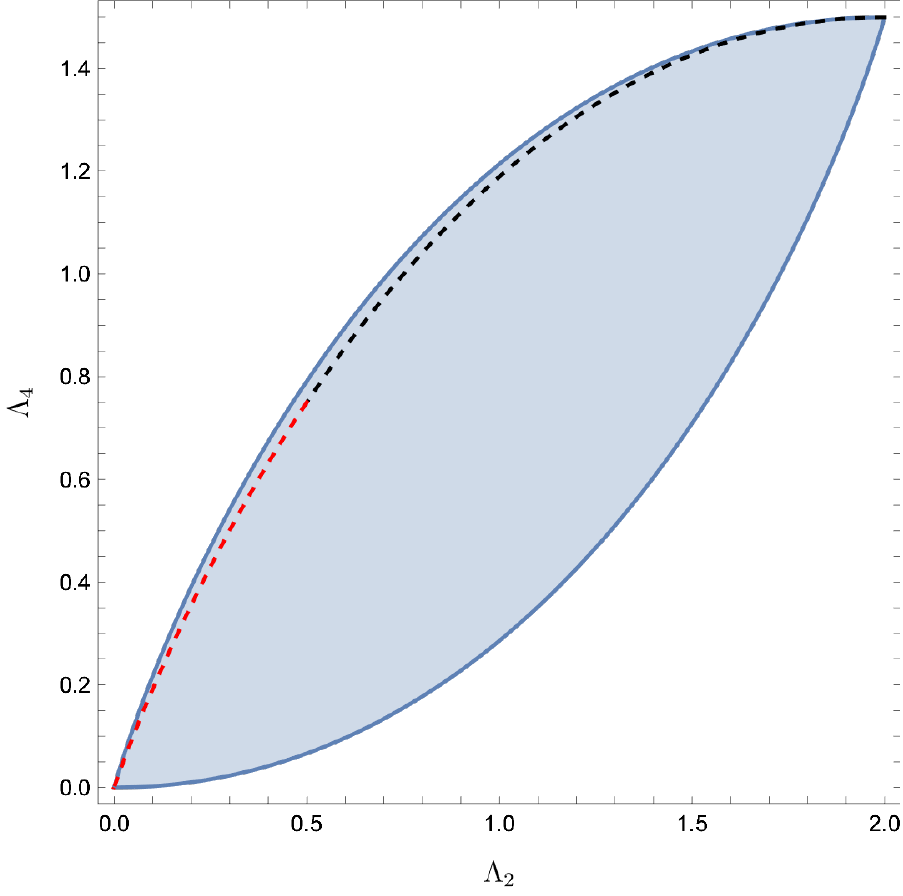}
		\caption{Allowed regions of the S-matrix in the $\Lambda_0-\Lambda_4$ and $\Lambda_2-\Lambda_4$ planes, wihch are also the projections of the 3d plot in figure \ref{fig:Lambda0Lambda2Lambda4_3d} onto these two planes. The plot on the left is precisely what we obtained in \cite{Chen:2021pgx} using numerical S-matrix bootstrap.  
			We have also indicated the locations of the Sinh-Gordon model with red dashed lines, and its analytic continuation, the Staircase model with the black dashed lines.  
		}
		\label{fig:Lambda2Lambda4}
		\vspace{10mm}
		\includegraphics[width=0.44\textwidth]{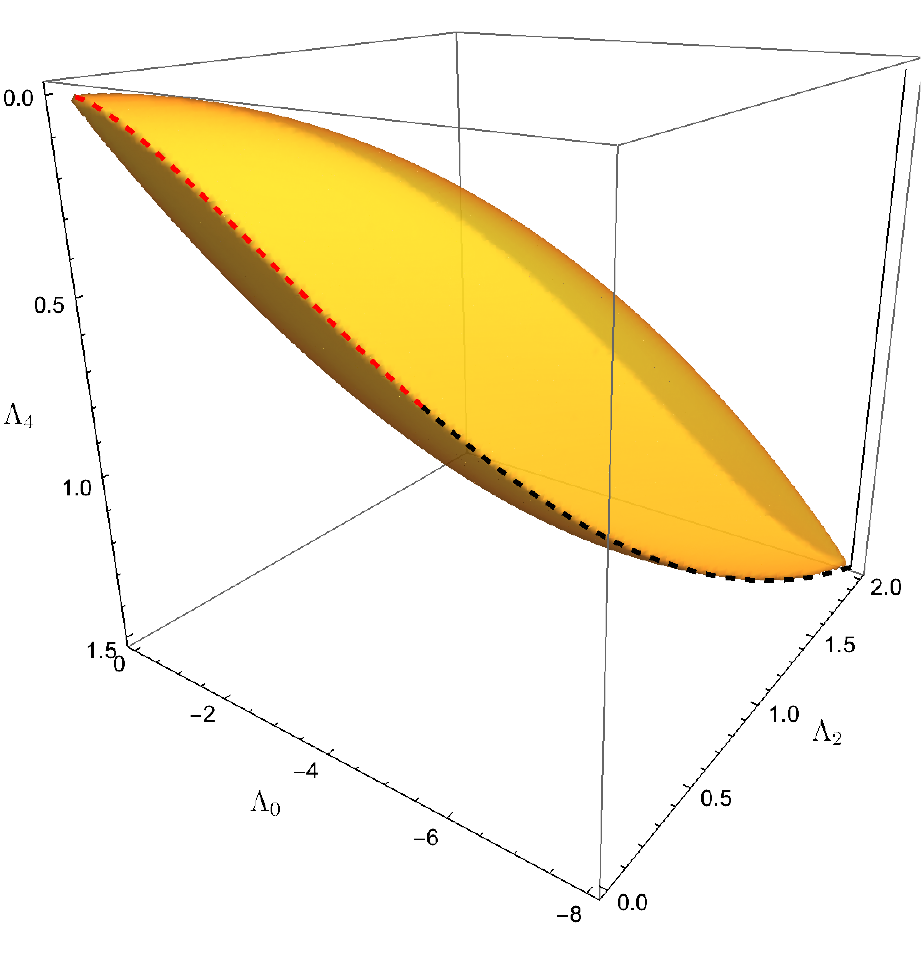}
		\caption{Allowed region of the S-matrix in the $\Lambda_0-\Lambda_2-\Lambda_4$ space. We have indicated the locations of the Sinh-Gordon model with a red dashed line ($-4\le\Lambda_0\le 0$), and its analytic continuation the Staircase model with the black dashed line ($-8\le\Lambda_0\le -4$).}
		\label{fig:Lambda0Lambda2Lambda4_3d}
	\end{center}
\end{figure}

%
On the $z$-plane, $s=2m^2$ is mapped to $z=0$, and expanding $\mathcal{S}(z)$ around $z=0$ we get 
\begin{equation}
	\mathcal{S}(z)=1+\frac{\Lambda_0}{4}+\left(-\frac{\Lambda_0}{2}-2 \Lambda_2\right) z+\left(\frac{8 \Lambda_4}{3}+\frac{\Lambda_0}{2}\right) z^{2}+\mathcal{O}\left(z^{3}\right).
\end{equation}
Applying the maximum modulus principle $|\CS(z)|\le 1$ at $z=0$, we immediately see that the parameter $\Lambda_0$ is restricted to a finite range:
%
\begin{equation}\label{eq:boundLambda0}
	-8\le \Lambda_0\le 0.
\end{equation}
Both endpoints have a simple physical interpretation.  In both cases, $|\CS(0)|=1$, which saturates the maximum modulus and implies that $\CS(z)$ is a constant.  If $\Lambda_0=0$, then $\CS(z)=1$ everywhere and there is no scattering. If $\Lambda_0=-8$ then $\CS(z)=-1$ everywhere, which is the S-matrix for a free fermion.  This is consistent with the fact that there are families of theories, such as the 2d  $\lambda \phi^4$ theory, that interpolate between a free scalar and a free fermion.  
Things become more interesting when we start to look at constraints that simultaneously involve multiple $\Lambda_n$s.  Because $|\CS(0)| \le 1$, we can consider the following function
\begin{equation}
	f(z)=\frac{1}{z} \frac{\mathcal{S}(z)-\mathcal{S}(0)}{1-\overline{\mathcal{S}(0)} \mathcal{S}(z)},
	\label{eq:MMandSP}
\end{equation}
which is also analytic for $|z|\le 1$ and bounded by $|f(z)| \le 1 $ if $|z|=1$.\footnote{The fact that $|f(z)|\le 1$ when $|z|\le 1$ follows from the fact that $z f(z)$ is of the form (\ref{eq:mobius}) with $z_0 \rightarrow \CS(0)$ and $z \rightarrow \CS(z)$.}  Again applying the maximum modulus principle, this time for $|f(z)| \le 1$ at $z=0$, we obtain bounds on $\Lambda_2$:
\begin{equation}\label{eq:boundLambda2}
	\frac{\Lambda_0^{2}}{32} \leq \Lambda_2 \leq -\frac{\Lambda_0}{2}-\frac{\Lambda_0^{2}}{32}.
\end{equation}
This argument is essentially the Schwarz-Pick theorem.   The bounds on $\Lambda_0$ and $\Lambda_2$ given in (\ref{eq:boundLambda0}) and (\ref{eq:boundLambda2}) (shown in the left plot of figure \ref{fig:Lambda2Lambda4}) are exactly the bounds we obtained in \cite{Chen:2021pgx} using a numeric S-matrix bootstrap analysis. Here we see analytically how the bounds are produced by the constraints of crossing, analyticity, and unitarity working in concert. 
We can obtain more bounds by considering additional transformations of $\mathcal{S}(z)$. 
As noted in \cite{EliasMiro:2019kyf}, we can make repeated applications of M\"obius transformations and the maximum modulus principle to make new analytic and bounded functions.  For example,
\begin{equation}
	g(z)=\frac{1}{z} \frac{f(z)-f(0)}{1-\overline{f(0)}f(z)}
\end{equation}
with $f(z)$ from (\ref{eq:MMandSP})
is analytic for $|z| < 1$ and satisfies $|g(z)|\le 1$ for $|z| = 1$. 

Now, $|g(0)| \le 1$ implies
\begin{multline}\label{eq:boundLambda4}
	\left|8\left(3 \Lambda_{0}^{3}+96\left(\Lambda_{0}+4\right) \Lambda_{2}^{2}+48\left(\Lambda_{0}+4\right) \Lambda_{0} \Lambda_{2}-16\left(\Lambda_{0}+8\right) \Lambda_{0} \Lambda_{4}\right)\right|\\
	\leq 3 \left(\Lambda _0^2-32 \Lambda _2\right) \left(\Lambda _0 \left(\Lambda _0+16\right)+32 \Lambda _2\right).
\end{multline}
From (\ref{eq:boundLambda0}), (\ref{eq:boundLambda2}), and the above inequality involving $\Lambda_4$, we can make a 3-dimensional plot of the allowed region in the $\Lambda_0-\Lambda_2-\Lambda_4$ space, which is given in figure \ref{fig:Lambda0Lambda2Lambda4_3d}.

The inequality (\ref{eq:boundLambda4}) involves $\Lambda_0, \Lambda_2$ and $\Lambda_4$. Of course, the full space of allowed S-matrices is an open region in an infinite-dimensional space, which we are parameterizing by the $\Lambda_n$s.  In practice, we always look at finite-dimensional projections of this space.  For instance, we can also project onto the allowed region for $\Lambda_4$ given $\Lambda_2$, for any value of $\Lambda_0$ in $[-8,0]$. This will be given by the projection of the 3d allowed region in figure \ref{fig:Lambda0Lambda2Lambda4_3d} onto the $\Lambda_2-\Lambda_4$ plane. Eliminating $\Lambda_0$ from (\ref{eq:boundLambda4}), we get 
\begin{equation}
	\frac{9}{8}\left(2^{2/3} (\Lambda_2-2)^{4/3}-4 \right) + 
	3 \Lambda_2 \leq \Lambda_4 \leq 3 \Lambda_2-\frac{9}{8} \times 2^{2 / 3}\Lambda_2^{4 / 3}.
\end{equation}
The plot of the above inequality is given in the right part of figure \ref{fig:Lambda2Lambda4}.

The main results of this paper are bounds that generalize figure \ref{fig:Lambda0Lambda2Lambda4_3d} and figure \ref{fig:Lambda2Lambda4}, and related quantities, to higher dimensions. Because of the presence of spin in $d>2$, there are multiple natural higher-dimensional generalizations of the quantities $\Lambda_n$, leading to different ways to present the bounds on the space of scattering amplitudes.  Bounds that directly generalize figure \ref{fig:Lambda2Lambda4} are shown in figure \ref{fig:PA0_dPA0_Extrapolated}, \ref{fig:3dplot}, and \ref{fig:crossing_symmetric}-\ref{fig:FL4}.  
\subsubsection{Positivity and linearized unitarity}
As in the main text of this paper for $d>2$, it is interesting to see how the bounds one obtains using only positivity or linearized unitarity differ from those of the full unitarity. In $d=2$, positivity and linearized unitarity are simply the statements that  
\begin{equation}
	\text { positivity: } 0 \leq \operatorname{Im} \mathcal{T},
\end{equation}
\begin{equation}
	\text { linearized unitarity: } 0 \leq \operatorname{Im} \mathcal{T} \leq 2 \mathcal{N}_{2}.
	\label{eq:LinearizedUnitarity2d}
\end{equation}
The positivity constraint is usually used in combination with dispersion relations.  For instance,
we can pick out $\Lambda_2$ as the following contour integral of $\CT(s)$:
\begin{equation}
	\Lambda_2 = \oint \frac{ds}{2 \pi i} \frac{\CT(s)}{(s-2m^2)^3} = 2\int_{4m^2}^\infty \frac{ds}{2 \pi } \frac{2\textrm{Im}(\CT(s))}{(s-2m^2)^3} \ge 0
	\label{eq:2dDispersion}
\end{equation}
where the second equality is obtained by deforming the contour onto the branch cuts at $-\infty < s <0$ and $4m^2 < s <\infty$.  Moreover, since $2 \textrm{Im}(\CT) \ge \CN_2^{-1} | \CT|^2$, $\Lambda_2$ can only vanish if there is no scattering. Additionally, we expect all $\Lambda_n$s should satisfy a bound like $\Lambda_2 \ge x_n \Lambda_n^2$ for some $x_n$, which we saw explicitly was $x_0 = \frac{1}{32}$ in the case of $n=0$.  The constraint  (\ref{eq:2dDispersion}) versus that of (\ref{eq:boundLambda2}) is a concrete instance of the improvement in the bounds that one can obtain by using the full unitarity constraint. Analogous comparisons on $d>2$ were given in the main text.  In particular, in figures \ref{fig:crossing_symmetric3} - \ref{fig:FL4}, we also showed results from dispersion relations with the positivity condition alongside the results from the full unitarity condition.    
This type of argument can be quite powerful, and although it is weaker than the full unitarity constraint, it has the advantage that it is much more transparent than the full S-matrix bootstrap analysis.  In section \ref{sec:analytic_d>2}, we  analyzed these bounds in more detail in $d>2$, and in particular we applied the method from \cite{Caron-Huot:2020cmc} of constructing null constraints to obtain upper- and lower-bounds on the $d>2$ analogues of $\Lambda_n$.  One of the main such results was (\ref{eq:lambdaTwoSided}). 

\begin{figure}
	\begin{center}
		\includegraphics{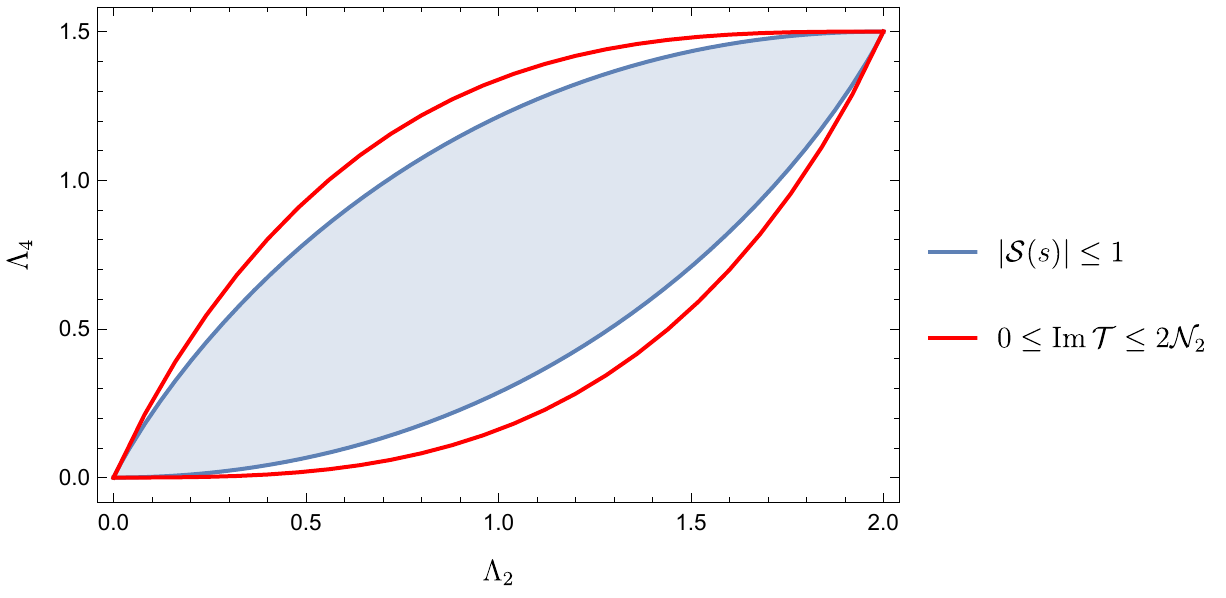}	
		\caption{Comparison of the bounds obtained using the full unitarity constraint (analytic) and using only the linearized unitarity constraint (numerical).}\label{fig:linearVSfullUnivarity2d}
	\end{center}
\end{figure}

Stronger than the positivity constraint but weaker than the full unitarity constraint is the linearized unitarity constraint \eqref{eq:LinearizedUnitarity2d}. In this case, we use the numerical S-matrix bootstrap method to obtain the allowed region in the $\Lambda_2-\Lambda_4$ plane. The comparison with the full unitarity constraint is shown in figure \ref{fig:linearVSfullUnivarity2d}. Interestingly, the tips are the same, while the full unitarity constraint is clearly more constraining. 

\subsection{Relation of bounds to specific theories}
\subsubsection{Weak coupling and integrable models}
\label{sec:Models}
The bounds become most interesting when there are theories of interest either at or very close to the edge of the bounded region.  We have already mentioned the simplest case where $\Lambda_0=0$ or $\Lambda_0=-8$, where the theory becomes a free boson or free fermion, respectively.  The lower bound of the constraint (\ref{eq:boundLambda2}) is particularly interesting.  The maximum modulus principle also says that if the inequality $|f(0)| =1$ is saturated (as it is at the boundaries of (\ref{eq:boundLambda2})), then the function $f(z)$ defined in \eqref{eq:MMandSP} must be a constant: 
\begin{equation}
	\CS(z)-\CS(0) =  cz (1- \overline{\CS(0)} \CS(z)), \qquad |c|=1.
\end{equation}
One can easily infer from this constraint what the S-matrix is along the entire boundary, but to emphasize the connection with higher dimensions, let us first consider the limit where the amplitude is small, $\CT \ll 1$.  Then, if we expand the numerator and denominator of $f(z)$ to linear order in $\CT$, and use the fact that $\CT(0)$ is real, then we have
\begin{equation}
	\CT(z) - \CT(0) = - c z ( \CT(z) - \CT(0) )+ \dots,
\end{equation}
where $\dots$ are higher order in $\CT$. Clearly, the only way to satisfy this equation for all $z$ is for $\CT(z) = \CT(0)$, i.e. $\CT(z)$ is a constant.  But a small, momentum-independent amplitude is just the perturbative interaction $\lambda \phi^4$.  So, although the bounds make no reference to a Lagrangian description, in the limit of weak scattering $\CT \ll 1$ the bound is saturated exactly by the simplest Lagrangian one can think of.  This privileged role of $\lambda \phi^4$ at weak coupling persists in our analysis of higher dimensions as well, as we showed in figure \ref{fig:phi4Comparison}.

More generally, for the lower bound in $(\ref{eq:boundLambda2})$, if $\Lambda_2=\Lambda_0^2/32$, then $|f(z)|=$ const implies that $\mathcal{S}(z)$ is given by 
\begin{equation}
	\mathcal{S}(z)=\frac{z+1+\frac{\Lambda_0}{4}}{1+\left(1+\frac{\Lambda_0}{4}\right) z}.
\end{equation}
One can check that this is exactly the S-matrix of the Sinh-Gordon model when $-4 \le \Lambda_0 \le 0$ (and for $-8\le\Lambda_0\le4$, it is the S-matrix for the Staircase model, which is an analytic continuation of the Sinh-Gordon model). And this is also consistent with what we found in \cite{chen:2021bmm}, that is, the Sinh-Gordon matrix is located at the lower boundary of the allowed region in the $\Lambda_0$-$\Lambda_2$ plane, as also shown in the left plot of the figure \ref{fig:Lambda2Lambda4}.  We also indicate the location of the Sinh-Gordon model by the red dashed lines in figure \ref{fig:Lambda0Lambda2Lambda4_3d}, where it lies on the boundary of the 3d allowed region as expected, and in the right plot of figure \ref{fig:Lambda2Lambda4} in the $\Lambda_2$-$\Lambda_4$ plane, where it is very close to but not exactly at the boundary, due to the projection onto the $\Lambda_2$-$\Lambda_4$ subspace.

Returning to the role of $\lambda \phi^4$ theory, it is interesting to ask whether it continues to have any relation to the bounds when the coupling is strong.  In $d=2$, in \cite{chen:2021bmm}, numerical analysis showed that the line of $(\Lambda_0,\Lambda_2)$ values obtained by dialing the quartic coupling $\lambda$ stayed remarkably close to the allowed lower bound for all values of the coupling from the free theory limit up to the critical coupling $\lambda_*$.  This fact can be understood near the endpoints $\lambda =0$ and $\lambda= \lambda_*$ just from the observation above that the limits $\Lambda=0$ and $\Lambda_0=-8$ are, respectively, the S-matrix of a free scalar and a free fermion.  These are exactly the theories that one obtains in the infrared for $\lambda=0$ and $\lambda=\lambda_*$.  Is it possible that this behavior holds in $d>2$ as well?  We saw in figure \ref{fig:phi4Comparison}  that in fact, an exact analogy cannot hold, because even at $d \sim 4$ where the critical point is weakly coupled and can be computed perturbatively, it does not sit close to the edge of the allowed region for the parameter analogous to $\Lambda_0$.  However, we also saw in figure \ref{fig:phi4Comparison} that the critical point near $d \sim 4$ sits very close to the lower boundary of the analogue of the $\Lambda_0$-$\Lambda_2$ region.  For $d \sim 3$, there is no reliable expansion parameter for the S-matrix near the critical point, and also no strong coupling calculations of this S-matrix that we are aware of, so at best the arguments we can make are merely suggestive.  Nevertheless, because we saw that the S-matrix near the critical point sits along the boundary of the $(\Lambda_0, \Lambda_2)$ allowed region for $d=4-\epsilon$ with $\epsilon \ll 1$ and for $d=2$, we think it is natural to conjecture that it continues to do so for all $d$ from $2\le d < 4$, and we discussed the evidence in favor of this from the $\epsilon$ expansion in section \ref{sec:phi4}.

\subsubsection{Bounds on EFTs}
\label{sec:2dEFT}
A remarkable fact about the combination of analyticity and unitarity is that it can reveal violations of unitarity even in regimes where the EFT is under perturbative control.   Such violations contradict a too-liberal ``anything goes'' philosophy of EFTs that, as long as all irrelevant interactions are suppressed by a UV-cutoff scale $M$ to the appropriate power times at most ${\cal O}(1)$ coefficients, it should be possible to obtain the EFT from a UV-completion.  While it has been known for some time \cite{Adams:2006sv} that the space of EFTs is in fact much more restricted than this, there is likely still much to be learned about the full set of constraints.
As a simple example, consider the application of the $d=2$ bounds to the space of EFT Lagrangians.  Specifically, take the case of a Lagrangian that only has irrelevant interactions, suppressed by a large scale $M$, so that the $S$-matrix is perturbative.  In this limit (i.e., $\Lambda_0\rightarrow0$), the upper bounds of \eqref{eq:boundLambda2} in the two-dimensional space parameterized by $\Lambda_0$ and $\Lambda_2$  reduce to
\be
\frac{-\Lambda_2}{\Lambda_0} \le \frac{1}{2} .
\label{eq:EFTRatioBound}
\ee

For instance, consider a scalar $\phi$ with an approximate shift symmetry $\phi \rightarrow \phi+c$, but a small explicit breaking proportional to $m$.  Assume that the leading interactions in the EFT are
\be
{\cal L}_{\rm eff} \approx \frac{1}{M^2} \left( -a m^4 \phi^4 + b (\partial \phi)^4 \right).
\ee
The amplitude in this case is proportional to ${\cal T} \propto (-3 a+b) m^4 + \frac{b}{2} (s-2m^2)^2$.  Therefore,
\be
\frac{-\Lambda_2}{\Lambda_0} = \frac{b}{3a-b}, 
\ee
and unitarity is violated unless $0< b< a$.  In particular, it is inconsistent to have {\it only} the $(\partial \phi)^4$ interaction when $\phi$ has a mass.\footnote{One might think that, by continuously taking $m$ to zero, this bound would imply that it is inconsistent to have only the $(\partial \phi)^4$ interaction for $m=0$ as well.  However, notice that the coefficient $a$ is defined as $m^{-4}$ times the coefficient of $\phi^4$ in the Lagrangian, and therefore $a$ is undefined in the strict $m=0$ limit. }
A similar, but weaker, bound on EFTs persists as we increase the spacetime dimension $d$, and can be read off from the behavior of the bound near the origin in Fig.~\ref{fig:PA0_dPA0_Extrapolated}.  We discussed this in more detail in section  \ref{sec:EFTs}.

\section{Limit of scattering amplitudes as $d\rightarrow 2$}
\label{app:limit_d2}

In this appendix, we show in what sense the object \eqref{eq:Sj}
continuously approaches a standard 2d $S$ matrix as the limit $d\rightarrow 2$ is taken.  It is because of the subtleties of this limit that we can run our bootstrap analysis efficiently in $d=2$ and in $2.5 < d < 4$, but there is a window $2<d<2.5$ where the numerics become poorly behaved.\footnote{We think it is likely that a more intelligent way of organizing partial waves for $d-2 \ll 1$ could resolve this numerical issue.}   Let us work in the vicinity of $d=2$ and parametrize the deviation from $d=2$ by $\varepsilon$ in the following way
\begin{equation}
d = 2\,(1+\varepsilon).
\end{equation}
Using \eqref{eq:measure}, we define the measure in $d=2$ as the following limit
\begin{align}
\label{eq:measure_d=2}
\mu_{d=2,\,j}(x)\equiv \lim_{\varepsilon\rightarrow 0}
\frac{j!\,\Gamma\left(\varepsilon-\frac{1}{2}\right)
}{4\,\pi^{\varepsilon+\frac{1}{2}}\Gamma(j-1+2\varepsilon)}
\times(1-x^2)^{\varepsilon-1} C_j^{\varepsilon-\frac{1}{2}}(x).
\end{align}
Let us now explore this expression. In the range $x\in(-1,+1)$, we can safely take the limit $\epsilon\rightarrow 0$ which simply leads to
\begin{equation}
\label{eq:measure_d=2_general}
x\neq \pm 1:\qquad
\mu_{d=2,\,j}(x) = -\frac{1}{2}\,C_{j-2}^{3/2}(x).
\end{equation}
Notice, that for $j=0$ the measure \eqref{eq:measure_d=2_general} vanishes. At $x=\pm 1$ we have poles. To treat them correctly we first expand the expression \eqref{eq:measure_d=2} around $x=+1$ and $x=-1$  to the leading order and then take the limit $\epsilon\rightarrow 0$. Independently of the spin value $j$ we obtain
\begin{equation}
\mu_{d=2,\,j}(x) =
\lim_{\varepsilon\rightarrow 0} \frac{\varepsilon/2}{(1-x)^{1-\epsilon}}+
\lim_{\varepsilon\rightarrow 0} \frac{\varepsilon/2}{(1+x)^{1-\epsilon}}+O(1-x^2).
\end{equation}
The latter expression is formally given by the Dirac $\delta$-functions
\begin{equation}
\label{eq:measure_d=2_delta}
\mu_{d=2,\,j}(x) =
\delta(1-x) + \delta(1+x)+O(1-x^2).
\end{equation}

Plugging \eqref{eq:measure_d=2_general} and \eqref{eq:measure_d=2_delta} into \eqref{eq:Tj} and subsequently into \eqref{eq:Sj} we obtain
\begin{equation}
\label{eq:partial_amplitude_2d}
\begin{aligned}
d=2:\qquad
\mathcal{S}_{j}(s)=1&+\frac{i}{2\mathcal{N}_{d=2}}\times
\left(\mathcal{T}(s,0,4m^2-s)+\mathcal{T}(s,4m^2-s,0)\right)\\
&-\frac{i}{2\mathcal{N}_{d=2}}\times
\int_{-1}^{+1} dx\;
C_{j-2}^{3/2}(x) \, \mathcal{T}(s,t(x),u(x)).
\end{aligned}
\end{equation}
In the first line, notice the appearance of the 1/2 factor due to
\begin{equation}
\int_{-1}^{+1}dx\,\delta(1-x)=
\int_{-1}^{+1}dx\,\delta(1+x)=1/2,
\end{equation}

Let us discuss \eqref{eq:partial_amplitude_2d}.
For $j=0$ the last term in \eqref{eq:partial_amplitude_2d} simply vanishes. The $j=0$ partial amplitude consists of two disconnected scattering amplitudes, one with $t=0$ and one with $u=0$. These two pieces are equal to each other. This can be seen by recalling that in $d>2$ the t-u crossing symmetry reads as
\begin{equation}
\mathcal{T}(s,t,u)=\mathcal{T}(s,u,t).
\end{equation}
Focusing on a particular case when $t=4m^2-s$ and $u=0$, we conclude that
\begin{equation}
\mathcal{T}(s,0,4m^2-s)=\mathcal{T}(s,4m^2-s,0).
\end{equation}
As a result, we see that \eqref{eq:partial_amplitude_2d} for $j=0$ is in perfect agreement with (3.30) - (3.32) in \cite{Karateev:2019ymz}.
For $j\geq 2$, the last term in \eqref{eq:partial_amplitude_2d} does not disappear automatically and can only be removed by an additional requirement that $\mathcal{T}$ is independent of $x$. Using the fact that
\begin{equation}
\int_{-1}^{+1} dx\;
C_{j-2}^{3/2}(x) =
\begin{cases}
0,\;j=0,1\\
2,\;j\geq2
\end{cases}
\end{equation}
we get
\begin{equation}
d=2:\qquad
\mathcal{S}_{j=0}(s)=1+\frac{i}{\mathcal{N}_{d=2}}\times
\mathcal{T}(s,4m^2-s,0),\qquad
\mathcal{S}_{j\geq 2 }(s)=1.
\end{equation}
The necessity of invoking this additional assumption in order to recover the $d=2$ result, and the singular nature of the partial waves as this limit is taken, prevent us from obtaining a continuous interpolation between $d=2$ and $d>2$ in this paper.

\section{Symmetric polynomials and amplitudes}
\label{app:polynomials}

Given the series representation \eqref{eq:representation_CS}, it is interesting to notice that it can be cast in an equivalent form in terms of symmetric polynomials.

Let us define our first four symmetric polynomials as
\begin{equation}\label{eq:polinoms_1}
	P_0(s,t,u) \equiv 1, \quad P_1(s,t,u)\equiv0,\quad
	P_2(s,t,u) \equiv \hat{s}^2+ \hat{t}^2+\hat{u}^2,\quad
	P_3(s,t,u) \equiv \hat{s}\ \hat{t}\ \hat{u},
\end{equation}
where we have also defined
\begin{equation}
	\label{eq:Mandelstam_hat_1}
	\hat s  \equiv s - 4m^2/3,\qquad
	\hat t  \equiv  t - 4m^2/3,\qquad
	\hat u \equiv  u - 4m^2/3.
\end{equation}
The rest of the polynomial are built out of the above ones as
\begin{equation}\label{eq:polinoms_2}
	P_4  = P_2^2,\quad 
	P_5  = P_2P_3,\quad
	P_6  = P_2^3,\quad 
	P_6' = P_3^2,\quad 
	P_7  = P_2^2P_3,\quad\ldots
\end{equation}
One can check order by order then that the following holds true
\begin{align}
	\nn
	m^{d-4}\mathcal{T}(s,t,u) &= \lambda_{0,0}\, P_0 + \frac{\lambda_{2,0}}{2}\, m^{-4}P_2 - \lambda_{2,1}\, m^{-6}P_3 +\frac{\lambda_{2,2}}{12}\,m^{-8}P_4-\frac{\lambda_{4,1}}{2}\,m^{-10}P_5 +\ldots\\
	&=	 \sum_{k,l =0}^\infty\lambda_{k,l}m^{-2(k+l)}(s-4m^2/3)^k(t-4m^2/3)^l.
	\label{eq:expansion_amplitude}
\end{align}
where the coefficients of this expansion are precisely the ones defined in \eqref{eq:description_1}.

\bibliographystyle{JHEP}
\bibliography{refs}

\end{document}